\documentclass[%
twocolumn,
nofootinbib,
amsmath,amssymb,
aps,
prd,
floatfix,
nobalancelastpage,
]{revtex4-2}

\usepackage{graphicx}
\usepackage{dcolumn}
\usepackage{bm}
\usepackage{siunitx}
\usepackage{physics}
\AtBeginDocument{\RenewCommandCopy\qty\SI}

\usepackage{easyReview}
\usepackage{slashed}
\usepackage{mathtools}
\usepackage{xspace}
\usepackage{comment}
\usepackage{braket}
\usepackage{float}
\usepackage{fontawesome}
\usepackage{booktabs}
\usepackage{anyfontsize}
\usepackage{listings}
\usepackage{inconsolata}
\usepackage{multirow}
\usepackage{tikz}
\usepackage{subfigure}
\usepackage[dvipsnames]{xcolor}

\tikzset{every picture/.style={line width=0.75pt}}
\usetikzlibrary{calc}
\usetikzlibrary{decorations.pathmorphing}

\definecolor{linkcolor}{rgb}{0.0, 0.28, 0.67}
\definecolor{userInputColor}{HTML}{4287f5}
\definecolor{userCallColor}{HTML}{fcba03}
\definecolor{codeCallColor}{HTML}{039c33}

\usepackage[
   colorlinks=true,
    urlcolor=linkcolor,
   anchorcolor=linkcolor,
    citecolor=linkcolor,
    filecolor=linkcolor,
    linkcolor=linkcolor,
    menucolor=linkcolor,
    linktocpage=true,
    pdfproducer=medialab,
    pdfa=true
]{hyperref}

\DeclareSIUnit\erg{erg}
\DeclareSIUnit\year{yr}
\DeclareSIUnit\jansky{Jy}
\DeclareSIUnit\dex{dex}
\DeclareSIUnit\deg{deg}
\DeclareSIUnit\angstrom{\text{Å}}
\DeclareSIUnit\eV{e\kern-.05em V}
\DeclareSIUnit \parsec {pc}
\DeclareSIUnit[]\msol
{\text{\ensuremath{M_{\odot}}}}
\DeclareSIUnit\beam{beam}
\graphicspath{ {./images/} }

\defcitealias{Baker:2025zhf}{Paper~I}

\DeclareMathOperator{\Si}{Si}
\DeclareMathOperator{\Ci}{Ci}

\newcommand{\githubicon}{\href{https://github.com/bakerem/dark_photons_radio_sky/tree/v1.0.0}{\faGithub}\xspace}
\newcommand{\nblink}[1]{\href{https://github.com/bakerem/dark_photons_radio_sky/blob/v1.0.0/notebooks_for_paper/#1.ipynb}{\faGithub}}

\begin{document}

\title{Dark Photons in the Radio Sky: II. Resonant Conversions in the Intergalactic Medium}

\author{Ethan Baker}
\email{ebaker@bu.edu}
\thanks{ORCID: \href{https://orcid.org/0000-0002-0520-4235}{0000-0002-0520-4235}}
\affiliation{Physics Department, Boston University, Boston, MA 02215, USA}

\author{Hongwan Liu}
\email{hongwan@bu.edu}
\thanks{ORCID: \href{https://orcid.org/0000-0003-2486-0681}{0000-0003-2486-0681}}
\affiliation{Physics Department, Boston University, Boston, MA 02215, USA}

\date{\today}
\begin{abstract} 
    This is the second in a pair of papers forecasting the sensitivity of the Square Kilometre Array (SKA) to dark photons, a highly motivated, simple extension of the Standard Model. Through a kinetic mixing term, visible photons from the cosmic microwave background can resonantly convert into dark photons, generating new temperature anisotropies in the sky. In this work, we detail the entire analysis pipeline that we use to compute SKA's sensitivity, focusing on resonant conversions that occur in the intergalactic medium. We also discuss the sensitivity of 21-cm experiments to dark photons. Our results show that both SKA in combination with galaxy surveys and 21-cm experiments could discover dark photons with masses between $5\times 10^{-15}$ and $5\times 10^{-12}$ eV, and kinetic mixing parameter $\epsilon$ as low as $10^{-8}$.~\githubicon
\end{abstract}

\maketitle

\section{Introduction}
The cosmic microwave background (CMB) is one of the most powerful tools in the search for physics beyond the Standard Model. 
The blackbody spectrum of the CMB and the angular power spectrum of its anisotropies measured in the microwave frequency range are arguably two of the most precisely measured quantities in all of science, and are highly sensitive probes of cosmology, Standard Model physics and beyond. 
At the same time, radio frequency data from a large variety of radio telescopes have already delivered significant new-physics results, including competitive searches for axions~\cite{Foster:2020pgt,Battye:2021yue,Noordhuis:2022ljw,Zhou:2022yxp,Sun:2023gic,Yang:2025yvh}, dark photon dark matter (DM)~\cite{An:2022hhb, An:2023wij} and generic constraints on DM annihilation and decay~\cite{Egorov:2013exa,Natarajan:2013dsa,Chan:2019ptd,Keller:2021zbl,Egorov:2022fkc}.
The remarkable success of these measurements motivates the measurement of the CMB energy spectrum and anisotropy in the radio band as a promising avenue to study the early Universe and search for new physics. 

Current and future radio facilities aiming to detect the cosmological 21-cm signal are, in essence, trying to do precisely that. 
The 21-cm signal can be thought of as a measurement of the distortion to the CMB due to the absorption or emission of 21-cm photons by neutral hydrogen atoms undergoing hyperfine transitions; such techniques promise to open a new window into the cosmic dark ages, and potentially offer exquisite sensitivity to certain beyond the Standard Model (BSM) scenarios. 
In particular, new-physics models that inject excess energy into the intergalactic medium (IGM) or substantially affect structure formation can impact the evolution of neutral hydrogen, leaving unmistakable signatures in 21-cm measurements.
The 21-cm sky-averaged (global) signal---measured at experiments like Experiment to Detect the Global EoR Signature (\textsc{EDGES})~\cite{Monsalve:2016xbk}, Shaped Antenna Measurement of the background Radio Spectrum (\textsc{SARAS})~\cite{Singh:2017syr}, and Probing Radio Intensity at high-Z from Marion (\textsc{PRIZM})~\cite{zotero-item-2534}---has been shown to provide strong sensitivity to e.g.\ DM annihilation and decay~\cite{DAmico:2018sxd,Liu:2018uzy,Mitridate:2018iag,Clark:2018ghm}, DM-baryon scattering~\cite{Barkana:2018qrx,Mondal:2023bxb,Rahimieh:2025fsb}, primordial black holes \cite{Saha:2021pqf}, and an excess radio background~\cite{Pospelov:2018kdh,Caputo:2020avy,Mondal:2023bxb}. 
Other 21-cm experiments measure fluctuations in the 21-cm signal through its power spectrum.
Forecasts have shown that measurements of this power spectrum are highly sensitive to generic scenarios of annihilating and decaying DM, as well as to well-motivated DM models such as axions, millicharged dark matter, and primordial black holes~\cite{Evoli:2014pva,Liu:2018uzy,Barkana:2022hko,Sun:2023acy,Facchinetti:2023slb,Plombat:2024kla,Bae:2025uqa,Rahimieh:2025lbf,Sun:2025ksr,Agius:2025nfz}.  
Experiments such as Giant Metrewave Radio Telescope (GMRT)~\cite{Paciga:2013fj}, Murchison Widefield Array (MWA)~\cite{Dillon:2013rfa}, the LOw Frequency ARray (LOFAR)~\cite{Patil:2017zqk}, Precision Array for Probing the Epoch of Reionization (PAPER) \cite{Kolopanis:2019vbl}, Large-Aperture Experiment to Detect the Dark Ages (LEDA)~\cite{Garsden:2021kdo}, and Hydrogen Epoch of Reionization Array (HERA)~\cite{HERA:2021bsv} have set upper limits on the 21-cm power spectrum and have more data on the way, meaning these experiments could soon be highly sensitive to BSM physics. 

But forthcoming facilities hold even greater promise.
The Square Kilometre Array (SKA) is currently under construction and will be the largest radio experiment ever constructed. 
SKA will produce both images of the radio sky and interferometric data for 21-cm Epoch of Reionization (EoR) intensity mapping studies~\cite{Braun:2019gdo}; both modes will be powerful probes of new physics~\cite{Weltman:2018zrl}. 
Imaging modes will be sensitive to axion models, as well as excess radio emission from dark matter decay and annihilation~\cite{Colafrancesco:2015ola,Kelley:2017vaa,Caputo:2018ljp,Caputo:2018vmy,Safdi:2018oeu, Cembranos:2019noa,Leroy:2019ghm,Ghosh:2020hgd,Chen:2021rea,Wang:2021hfb,Sun:2023wqq,Todarello:2023xuf}.
Other work has shown that SKA images will be sensitive to radio flux from primordial black holes~\cite{Gaggero:2016dpq} and will constrain the properties of neutrinos~\cite{Villaescusa-Navarro:2015cca}. 

In this paper and our preceding paper~\cite{Baker:2025zhf} (henceforth~\citetalias{Baker:2025zhf}), we show that SKA can give us unprecedented sensitivity to one of the simplest renormalizable extensions of the Standard Model: dark photons. 
The dark photon is a hypothetical gauge boson of a new dark $U(1)'$ symmetry with a mass $m_{A'}$ and a parameter $\epsilon$, which controls the strength of their coupling to Standard Model photons through a kinetic mixing term.
Through this kinetic mixing, $\gamma \leftrightarrow A'$ conversions can occur.
In an ionized gas, the probability of these conversions is resonantly enhanced when the dark photon mass $m_{A'}$ is equal to the effective plasma mass $m_{\gamma}(\vec{x})$, which tracks the free-electron number density at each point in space $\vec{x}$.
Here, we focus on $\gamma \to A'$ conversions---our results do not assume that dark photons make up any of the dark matter in the universe. 

At low $m_{A'}$, the Compton wavelength of the dark photon becomes longer than the typical length scales of terrestrial experiments, and so cosmological searches for $\gamma \to A'$ conversions are crucial. 
One way this can be done is by searching for spectral distortions to the sky-averaged CMB blackbody spectrum created by $\gamma \to A'$ conversions in the IGM after recombination, or in the photon-baryon plasma before recombination~\cite{Mirizzi:2009iz, Kunze:2015noa,Caputo:2020bdy,Caputo:2020rnx,Garcia:2020qrp, Chluba:2024wui,Arsenadze:2024ywr}. 
These analyses have grown in sophistication over time since the initial idea was proposed in Ref.~\cite{Mirizzi:2009iz}, and now include inhomogeneities for conversions after recombination~\cite{Caputo:2020bdy,Caputo:2020rnx,Garcia:2020qrp}, and a proper treatment of photon disappearance before recombination~\cite{Chluba:2024wui,Arsenadze:2024ywr}.
A similar formalism has also been used to search for conversions between axions and visible photons using CMB spectral distortions, where the importance of anisotropies was also recognized \cite{Mirizzi:2005nga,Mirizzi:2009nq,Ejlli:2013uda,Schlederer:2015jwa,Tashiro:2013yea,Mukherjee:2018oeb,Mukherjee:2019dsu,Buen-Abad:2020zbd,Goldstein:2024mfp,Mehta:2024wfo,Mehta:2024sye,Mondino:2024rif,Kushwaha:2025yja,Mehta:2025qfr}. 
More recently, searches for anisotropies in the sky sourced by $\gamma \to A'$ conversions have been proposed~\cite{Pirvu:2023lch} and conducted using \textit{Planck} CMB data~\cite{McCarthy:2024ozh}.
Along two different lines of sight, CMB photons will travel through different environments, and different fractions of these photons will convert to dark photons.
Therefore, $\gamma \to A'$ conversions will create patches of ``missing'' photons across the sky, with the patch size being set by the angular scale of inhomogeneities in $m_{\gamma}^2$. 
References~\cite{Pirvu:2023lch,McCarthy:2024ozh} modeled $\gamma \to A'$ conversions in dark matter halos, and searched for the resulting anisotropies in \textit{Planck} data, as well as in a cross-correlation between \textit{Planck} and the number density of galaxies from the \textit{unWISE} survey~\cite{Krolewski:2019yrv,Krolewski:2021yqy}, finding leading limits on $\epsilon$ around $m_{A'} \sim \qty{e-12}{\eV}$.
It is natural to consider this cross-correlation, because the probability of $\gamma \to A'$ conversions occurring is set by the electron number density $n_e(\vec{x})$ through the resonance condition, and the presence of galaxies traces overdensities in $n_e(\vec{x})$.
Also, as explained more fully below, one expects this cross-correlation to scale as $\epsilon^2$, whereas the autocorrelation signal scales as $\epsilon^4$. 
Therefore, the cross-correlation signal is potentially less suppressed at small $\epsilon$, hinting at its promise as an observational tool.

In~\citetalias{Baker:2025zhf} and this paper, we forecast the sensitivity of radio experiments to $\gamma \to A'$ conversions, and show that SKA and a 21-cm global signal experiment could discover dark photons in the mass range $\qty{e-14}{\eV} \lesssim m_{A'} \lesssim \qty{e-11}{\eV}$ with $\epsilon$ as low as $10^{-8}$, surpassing the limits set by \textit{Planck} and CMB spectral distortion measurements significantly. 
In~\citetalias{Baker:2025zhf}, we forecast the sensitivity of SKA in combination with a low-redshift galaxy survey to radio-sky anisotropies generated by $\gamma \to A'$ conversions in gas surrounding dark matter halos, over a range of masses centered on $m_{A'} \sim \qty{e-12}{\eV}$. We find that SKA will have unprecedented sensitivity to this signal, and discuss the relative advantages of SKA over \textit{Planck} in detail.

In this paper, we instead focus on $\gamma \to A'$ conversions occurring within the intergalactic medium. 
We describe in detail how to compute the theoretical dark photon signal, our procedure for producing mock observations of the radio sky and its cross correlation with galaxy surveys, the internal linear combination (ILC) algorithm which we use to extract our signal from radio maps, and our statistical methods that we use to forecast the sensitivity of all of the radio experiments we consider.
To forecast the sensitivity of SKA, we use a multistep analysis pipeline that is the main focus of this paper.
The SKA analysis procedure can be broken into several distinct steps. 
First, we derive a theoretical prediction for the $\gamma \to A'$ signal using analytic and simulation methods. 
We model three cosmological environments where $\gamma \to A'$ conversions could occur: dark matter halos, the IGM during the EoR, and the late-universe IGM. 
Second, in order to properly forecast the sensitivity of SKA to $\gamma \to A'$ conversions, we must model the substantial foregrounds that this experiment will also observe.
We begin by producing nine mock foreground-only maps at SKA frequencies using the \textsc{pysm3} and \textsc{epspy} codes, which then are inputs to the rest of our analysis pipeline.
We then process these null-signal maps using a pipeline based on the ILC algorithm, which constructs the minimum-variance linear combination of radio maps subject to the constraint that the strength of the dark photon signal is not reduced, resulting in a single map with the greatest possible SNR.
We then compute the autocorrelation of this post-ILC map and its cross-correlation with mock galaxy catalogs. 
This would be the same pipeline used to analyze real SKA data; by applying it to our null-signal mock observations, we can forecast the sensitivity of SKA to dark photons in a realistic setting by comparing the results to our theoretical models using a Gaussian likelihood.  

Beyond the SKA imaging experiment, we also discuss the sensitivity of 21-cm global signal and power spectrum experiments to $\gamma \to A'$ conversions.
We find that $\gamma \to A'$ conversions substantially impact both signals. 
However, the search for dark photons in 21-cm power spectrum experiments is significantly complicated by the experimental details.

Overall, we find that SKA and global 21-cm experiments will be more sensitive to conversions within the IGM than searches for $\gamma \to A'$ spectral distortions in the CMB blackbody spectrum measured by Far Infrared Absolute Spectrophotometer (FIRAS) for masses in the range $\qty{5e-15}{\eV} \lesssim m_{A'} \lesssim \qty{e-13}{\eV}$. Together with the results from~\citetalias{Baker:2025zhf}, we find that near-future radio experiments will provide a powerful new avenue to search for ultralight dark photons, highlighting the tremendous potential radio observations have to probe BSM physics more generally. 

The structure of the remainder of this paper is as follows: in Sec.~\ref{sec:signal_modeling}, we review the formalism of $\gamma \to A'$ conversions and describe our analytic and simulation methods for modeling the signal from these conversions. 
Then, in Sec.~\ref{sec:mock_observations}, we first outline our process for modeling the astrophysical foregrounds that complicate the search for the $\gamma \to A'$ signal. Then, we describe the internal linear combination technique that we would use at SKA to extract the $\gamma \to A'$ signal as well as the mock galaxy catalogs that we cross-correlate with the post-ILC map.
We then detail our statistical modeling techniques in Sec.~\ref{sec:stats}. 
Finally, we discuss the feasibility of searching for $\gamma \to A'$ conversions using EoR 21-cm experiments in Sec.~\ref{sec:21cm_expts}. 
Throughout this paper, we use natural units $\hbar = c = k_B = 1$. 

\section{Signal Modeling}
\label{sec:signal_modeling}

The CMB provides a nearly isotropic backlight of photons. 
If dark photons exist, CMB photons can convert into dark photons anywhere along their path from the surface of last scattering to us. 
As these photons propagate through the cosmos, they pass through many different environments with varying effective plasma mass $m_\gamma^2$, leading to changing conditions under which resonant $\gamma \to A'$ conversions can happen. 
As the observer, however, we are only able to measure the integrated effect of all these conversions along each line of sight.
To obtain an accurate prediction for the observable signal, we must therefore model as many environments as possible where these conversions can occur.
This also allows us to search for a larger range of $m_{A'}$, since each of these environments has a different typical value of $m_{\gamma}^2$. 
In principle, we should add up all possible contributions to the signal from all environments where $\gamma \to A'$ conversions occur; however, we will find that conversions in just one of the environments considered here will typically be dominant, accounting for most of the signal.

In this work, we model three environments where $\gamma \to A'$ conversions occur:
\begin{enumerate}
\item \textbf{Environment A}: conversions that occur in dark matter halos from $0.005 \lesssim z \lesssim 4$, using the methods described in Refs.~\cite{Pirvu:2023lch,McCarthy:2024ozh}, which is relevant for $\qty{e-13}{\eV} \lesssim m_{A'} \lesssim \qty{e-11}{\eV}$. This environment is the focus of~\citetalias{Baker:2025zhf}. 
\item \textbf{Environment B}: conversions in the IGM from $5 \lesssim z \lesssim 35$, which occur for dark photons with $\qty{8e-15}{\eV} \lesssim m_{A'} \lesssim \qty{e-13}{\eV}$. 
\item \textbf{Environment C}: conversions that occur in the IGM of the late Universe from $0.005 \lesssim z \lesssim 4$. 
These conversions also occur for dark photons with $\qty{8e-15}{\eV} \lesssim m_{A'} \lesssim \qty{e-13}{\eV}$. 
\end{enumerate}
In this section, we review the formalism of $\gamma \to A'$ conversions and detail our methods for each of these environments. Throughout, we use superscripts to indicate the environment.

\subsection{\texorpdfstring{Formalism of $\gamma \to A'$ Conversions}{Formalism of Gamma to A' Conversions}}

The Lagrangian describing a dark photon $A'_\mu$ with mass $m_{A'}$ that kinetically mixes with the Standard Model photon $A_\mu$ is
\begin{equation}
    \mathcal{L} \supset -\frac{1}{4}F_{\mu\nu}^2 - \frac{1}{4}{F}_{\mu\nu}^{\prime 2} - \frac{\epsilon}{2}F^{\mu \nu}F'_{\mu\nu} + \frac{1}{2} m_{A'}^2 A_\mu^{\prime 2}\, ,
\end{equation}
where $F_{\mu\nu}$ and $F_{\mu\nu}'$ are the field strength tensors for the Standard Model and dark photons, respectively. 
In UV complete models, the kinetic mixing can be highly suppressed by loop effects and therefore can naturally assume a very small value. 
For example, if the kinetic mixing is induced through gravity, $\epsilon$ could be as small as $\sim 10^{-13}$~\cite{Gherghetta:2019coi}. 
The dark photon can acquire a mass either through a spontaneous breaking of the dark $U(1)'$ symmetry or through a Stueckelberg mechanism~\cite{Fabbrichesi:2020wbt}. 

One observable consequence of this kinetic mixing is the oscillation of photons into dark photons or vice versa in a process similar to the Mikheyev--Smirnov--Wolfenstein (MSW) effect. 
In a plasma, conversions occur resonantly at a point $\vec{x}$ when~\cite{Caputo:2020rnx}
\begin{equation}
    m_{A'}^2 = m_{\gamma}^2(\vec{x}) \approx 4\pi \alpha_{\rm EM}\frac{n_e(\vec{x})}{m_e} \, ,
\end{equation}
where $\alpha_{\rm EM}$ is the electromagnetic fine-structure constant, $n_e$ is the electron number density, and $m_e$ is the electron mass. 
$m_{\gamma}^2$ is an effective plasma mass for the Standard Model photon due to a modification of its dispersion relation from electromagnetic interactions in the plasma~\cite{Mirizzi:2009iz}. 
Typical values of $m_{\gamma}$ over the redshift range we consider range from $\sim \qty{e-15}{\eV}$ to $\sim\qty{e-11}{\eV}$, and so the models we consider are sensitive to $\gamma \to A'$ conversions across this same range of $m_{A'}$. 

In the limit where the probability of conversion is small, the probability of $\gamma \to A'$ conversions along a line of sight $\hat{n}$ is given by
\begin{equation}
    P_{\gamma\to A'}(\hat{n}) = \sum_i \frac{\pi \epsilon^2 m_{A'}^2}{\omega(t_{{\rm res},i})} \left|\frac{d \ln m_{\gamma}^2(\hat{n},t)}{dt} \right|^{-1}_{t=t_{{\rm res}, i}}\, ,
    \label{eq:conversion_probability}
\end{equation}
where $i$ indexes the times $t_{{\rm res}, i}$ where resonances occur along a line of sight $\hat{n}$, and $\omega(t_{{\rm res},i})$ is the energy of the photon at $t_{{\rm res}, i}$~\cite{Mirizzi:2009iz}.  

In the sky, these $\gamma \to A'$ conversions lead to the disappearance of some fraction of the backlight CMB photons.  
The probability of disappearance integrated along a line-of-sight $\hat{n}$ at a detected energy $\omega_0$ is denoted by $P_{\gamma\to A'}(\omega_0, \hat{n})$; this leads to a corresponding reduction in intensity of the backlight photons $\Delta I(\omega_0, \hat{n})=-P_{\gamma\to A'}(\omega_0, \hat{n})B(\omega_0)$~\cite{DAmico:2015snf},
where
\begin{equation}
    B(\omega_0) = \frac{\omega_0^3}{2\pi^2}\left[\exp\left(\frac{\omega_0}{T_{\gamma,0}}\right) -1 \right]^{-1}
\end{equation}
is the Planck blackbody specific intensity in frequency, and $T_{\gamma, 0}=\qty{2.73}{\kelvin}$ is the present-day temperature of the CMB~\cite{Fixsen:2009ug}. 
This corresponds to a change in thermodynamic temperature
\begin{equation}
    \Delta T(\hat{n}) = -P_{\gamma \to A'}(\omega_0, \hat{n})T_{\gamma,0}(1-e^{-x})/x
\end{equation}
with $x \equiv \omega_0/T_{\gamma,0}$; this is the temperature unit we use in the rest of this work.
Notice that $(1-e^{-x})/x \approx 1$ for $x \ll 1$, while at $x \gtrsim 1$ the signal becomes exponentially suppressed. 
Since we focus on radio frequencies here, we make the approximation that $\Delta T(\hat{n})=-P_{\gamma \to A'}(\omega_0, \hat{n})T_{\gamma,0}$ throughout and thus the observed signal has an $\omega^{-1}$ frequency dependence. 

The disappearance of CMB photons generates anisotropic cold patches beyond those expected in $\Lambda$ cold dark matter ($\Lambda$CDM) cosmology and astrophysical foregrounds. 
Additionally, because the locations of $\gamma \to A'$ conversions are set by the resonance condition $m_{A'}^2=m_{\gamma}^2(\vec{x}) \propto n_e(\vec{x})$, these conversions are correlated with the distribution of $n_e$, and therefore ultimately of baryons.
Then, since the distribution of galaxies is also correlated with the distribution of baryons, we expect a correlation between the overdensity of galaxies $\delta_g$ and $\Delta T$. 
Cross-correlations have proven to be a particularly effective tool for dark photon searches~\cite{McCarthy:2024ozh}, providing enhanced sensitivity compared to autocorrelation analyses alone.

We model these correlations by first constructing real-space summary statistics. 
We define the two-point autocorrelation function $\xi^{\rm TT}(\mu)$ of $\Delta T$ and the two-point cross-correlation function $\xi^{\rm Tg}(\mu)$ between $\Delta T$ and $\delta_g$, which are each defined as 
\begin{align}
    \xi^{\rm TT}(\mu) &\equiv \left\langle \Delta T(\hat{n})\Delta T(\hat{n}')\right\rangle - \langle \Delta T\rangle^2 \,, \label{eq:auto_xi} \\
    \xi^{\rm Tg}(\mu) &\equiv \left\langle \Delta T(\hat{n})\delta_g(\hat{n}')\right\rangle - \langle \Delta T(\hat{n})\rangle \langle \delta_g(\hat{n})\rangle \, , \label{eq:cross_xi}
\end{align}
where $\langle \dots \rangle$ indicates an average over the sky, and $\mu\equiv \hat{n}\cdot \hat{n}'$. 
The fact that the correlation function only depends on $\mu$ follows from the assumption of statistical isotropy.
In some contexts, it is more useful to consider the angular power spectrum associated with these correlation functions, which is defined as 
\begin{equation}
    \label{eq:xi_and_Cl_relation}
    C_\ell^{\rm TX} = 2\pi \int_{-1}^{1} d\mu \, \xi^{\rm TX}(\mu) P_{\ell}(\mu) \, ,
\end{equation}
where $\rm X\in \{T, g\}$ and $P_\ell(\mu)$ is a Legendre polynomial of order $\ell$. 
These observables describe the angular correlations of the anisotropies generated by the $\gamma \to A'$ conversion process, which itself is related to the angular scales of the underlying $m_{\gamma}^2$ distribution. 

These correlation functions and power spectra are the main observables in this work. 
In the following sections, we describe our analytic and simulation methods that we use to compute the theoretical expectation of these observables in the cosmological environments that we consider. 
These will be used to forecast SKA's sensitivity to $\gamma \to A'$ conversions by comparing these theoretical predictions to their observed values computed from mock null-signal data. 

\subsection{Environment A: Dark Matter Halo Conversions}
\label{sec:envA}
In this section, we compute the $\gamma \to A'$ signal from conversions that occur within dark matter halos, which is also the focus of~\citetalias{Baker:2025zhf}. 
We compute these conversions from $0.005 \lesssim z \lesssim 4$, a range that we adopt for consistency with Ref.~\cite{McCarthy:2024ozh}.
In our modeling below, we assume that all gas is ionized, so this redshift range is a conservative choice that ensures that reionization is complete. 
We model this signal by adopting the halo model formalism outlined in Ref.~\cite{Pirvu:2023lch}. 

In the halo model, all matter is distributed in discrete dark matter halos that are described entirely by their redshift $z$ and mass $m$~\cite{Cooray:2002dia,Yang:2002ww}.
All cosmological observables are then related to the distribution of these halos and their properties. 
For example, under the halo model assumption, the matter power spectrum at all scales can be obtained by specifying a halo density profile for every halo of mass $m$ (e.g., a Navarro-Frenk-White [NFW] profile), and a halo mass function, describing the probability of finding a halo of $m$ at redshift $z$. 
To predict the signals expected from $\gamma \to A'$, we therefore want to compute $C_\ell^{\rm A, TX}$ under the halo model assumption. Throughout, we use the superscript ``A'' to indicate the halo environment.

$C_\ell^{\rm A, TT}$ and $C_\ell^{\rm A, Tg}$ were first obtained in Ref.~\cite{Pirvu:2023lch}; for completeness, we include a full derivation in Appendix~\ref{sec:halo_model_der}, which mostly follows the discussion in Ref.~\cite{Pirvu:2023lch}, up to a few pedagogical improvements in the derivation, and a number of novel analytic expressions that are useful for the numerical computation of these variables.
Here, we will simply state the final result, and give some physical intuition for the form of these expressions.

Within the halo model, the correlation functions can be expressed as the sum of a ``one-halo'' term and a ``two-halo'' term, i.e. 
\begin{equation}
C_\ell^{\rm A, TX} = C_\ell^{\rm A, TX, 1h} + C_\ell^{\rm A, TX, 2h} \,.
\end{equation}
The first term in this expression describes correlations due to $\gamma \to A'$ conversions along two lines of sight passing through the gas in the same halo ($\mathrm{X} = \mathrm{T})$, or the correlation between $\gamma \to A'$ conversions in a halo and the galaxies that occupy the same halo ($\mathrm{X} = \mathrm{g}$). 
The second term describes correlations between $\gamma \to A'$ conversions along two lines of sight in the gas surrounding two different halos for $\mathrm{X} = \mathrm{T}$, or conversions in the gas around one halo and a galaxy in a different halo for $\mathrm{X} = \mathrm{g}$. 

In the autopower spectrum $C_\ell^{\rm A, TT}$, the one-halo term is
\begin{widetext}
\begin{alignat}{1}
    C_\ell^{\rm A, TT, 1h} = T_{\gamma, 0}^2\left(\frac{4\pi}{2\ell+1}\right)^2\int d\chi \, \frac{\chi^2}{(1+z)^2} \left(\frac{2\pi \epsilon^2 m_{A'}^4}{\kappa \omega_0}\right)^2 \int dm \, n(z, m)\left|\frac{d\rho_{\rm gas}}{dr}\right|^{-2}_{r_{{\rm res}}}(z, m)\Theta(r_{{\rm vir}}-r_{{\rm res}}) a^2_\ell\,, 
    \label{eq:Cl_A_TT_1h}
\end{alignat}
\end{widetext}
where
\begin{alignat}{1}
    a_\ell &= \frac{2\ell+1 }{2}\int_0^{\theta_{\rm max}}d\theta \, \theta \left(1-\frac{\theta^2}{\theta_{\rm max}^2}\right)^{-1/2} P_\ell(\cos \theta) \,.
\end{alignat}

Let us walk through the expression for $C_\ell^{\rm A, TT, 1h}$ to build some intuition for its form. 
First, $C_\ell^{\rm A, TT, 1h}$ is a thermodynamic-temperature two-point function, and is therefore directly proportional to $T_{\gamma,0}^2$, the square of the thermodynamic temperature of the CMB backlight. 
It is expressed as an integral over $\chi$, the comoving distance at redshift $z$, integrating the probability of conversion along the line of sight. 
The one-halo term encodes contributions from conversions within the same halo, and hence $C_\ell^{\rm A, TT, 1h}$ only contains one integral over redshift. It also contains an integral over the halo mass function $n(z,m)$, averaging over all halos of mass $m$ at redshift $z$.
Since we are considering two lines of sight through the same halo, the squared probability of conversion appears in this expression, with the term proportional to $\epsilon^2$ and the term $|d\rho_{\rm gas}/dr|^{-2}_{r_{\rm res}}$ coming from Eq.~\eqref{eq:conversion_probability}. 
$\rho_{\rm gas}$ is a spherically symmetric, fully ionized gas density profile within a halo of mass $m$ and at redshift $z$, which we obtain from Ref.~\cite{Battaglia:2016xbi}, following Ref.~\cite{Pirvu:2023lch}, with $\kappa = \qty{5.7e-38}{\eV\squared\per\msol\mega\parsec\cubed}$ being a constant conversion factor between gas density and plasma mass, $\kappa \rho_{\rm gas} = m_\gamma^2$.
$r_{\rm res}$ is the radius within the gas distribution where resonant conversions occur, found by solving $m_\gamma^2(r_{\rm res}) = m_{A'}^2$. 
We then enforce a cutoff for the gas profile given by the virial radius of the halo, denoted by $\Theta(r_{\rm vir}-r_{\rm res})$ where $\Theta$ is the Heaviside step function, so that conversions only occur within the halo. 
Finally, $a_\ell$ encodes the geometry of conversions within the halo, with $\theta_{\rm max} \equiv (1+z)r_{\rm res}/\chi$ being the maximum angle at which conversions occur within the halo.

The two-halo term follows a very similar structure, but contains two lines-of-sight integrals passing through two different halos, as follows:
\begin{widetext}
\begin{alignat}{3}
   & C_\ell^{\rm A, TT,  2h} &&= &&T_{\gamma, 0}^2 \left(\frac{4\pi}{2\ell+1}\right)^3 \int d\chi_1 \int d\chi_2 \left(\frac{2\pi\epsilon^2 m_{A'}^4}{\kappa \omega_0}\right)^2 \frac{\chi_1^2}{1+z_1}\frac{\chi_2^2}{1+z_2} \nonumber \\
   & && && \times \int dm_1 \, n(z_1, m_1) \left|\frac{d\rho_{\rm gas}}{dr}\right|^{-1}_{r_{\rm res, 1}}(z_1, m_1) \Theta(r_{\rm vir,1}-r_{\rm res,1}) a_{\ell}(z_1, m_1) \nonumber \\
    & && &&\times \int dm_2 \, n(z_2, m_2) \left|\frac{d\rho_{\rm gas}}{dr}\right|^{-1}_{r_{\rm res, 2}}(z_2, m_2) \Theta(r_{\rm vir,2}-r_{\rm res,2})  a_{\ell}(z_2,m_2) h_\ell (z_1, z_2; m_1, m_2)  \, .
\end{alignat}
\end{widetext}
The new term $h_\ell$ is defined as 
\begin{align}
    h_\ell &= \frac{2\ell+1}{2}\int_{-1}^1d\mu \,\xi^{\rm hh}(\mu; z_1, m_1, z_2, m_2)P_\ell(\mu) \,, 
\end{align}
where $\xi^{\rm hh}(\mu; z_1, m_1, z_2, m_2)$ is the two-point halo correlation function defined in Eq.~\eqref{eq:xi_hh}, and $\mu$ is the cosine of the opening angle between the centers of two halos.  
Intuitively, $\xi^{\rm hh}(\mu)$ encodes the spatial clustering of halos in the universe.

Next, the one- and two-halo terms of the cross-correlation $C_\ell^{\rm A, Tg}$ are individually given by
\begin{widetext}
\begin{alignat}{3}
    & C_\ell^{\rm A, Tg, 1h} &&= &&-T_{\gamma, 0} \left(\frac{4\pi}{2\ell+1}\right)^2\int d\chi \, \chi^2 \frac{2\pi\epsilon^2m_{A'}^4}{\kappa \omega(z)}\int dm \, n(z,m) \left|\frac{d\rho_{\rm gas}}{dr}\right|^{-1}_{r_{\rm res}}(z,m) a_\ell(z,m) u_\ell^{\rm g}(z,m) \Theta(r_{\rm vir}-r_{\rm res})\, , \label{eq:Cl_A_Tg_1h}\\
    & C_\ell^{\rm A, Tg, 2h} &&= &&-T_{\gamma,0} \int d\chi_1\,  \chi_1^2 \,\frac{2\pi \epsilon^2 m_{A'}^4}{\kappa \omega(z)} \int dm_1 \, n(z_1, m_1) \left|\frac{d\rho_{\rm gas}}{dr}\right|^{-1}_{r_{\rm res}}(z_1, m_1) \Theta(r_{\rm vir}-r_{\rm res}) a_\ell(z_1, m_1) \nonumber \\
    & && &&\times \int dm_2 \, n(z_1, m_2) u^{\rm g}_\ell(z_1, m_2) b(z_1, m_1) b(z_1, m_2) P^{\rm lin}\left(\frac{\ell+\frac{1}{2}}{\chi_1}, z_1\right)\left(\frac{4\pi}{2\ell+1}\right)^2 \,. 
    \label{eq:Cl_A_Tg_2h} 
\end{alignat}
\end{widetext}
The intuition for these expressions is similar to those for $C_\ell^{\rm A, TT}$, with the galaxy multipole kernel $u^{\rm g}_\ell(z,m)$---completely encoding the geometry of how galaxies are distributed within halos---in place of one factor of $a_\ell$. 
Note that we have simplified the two-halo term using the Limber approximation~\cite{Limber:1954zz, LoVerde:2008re}, which allows us to perform an integral over one of the lines of sight, $\chi_2$, and to replace $h_\ell(z_1, z_2; m_1, m_2) \to b(z_1, m_1) b(z_1, m_2) P^{\rm lin}(k_\ell, z_1)$, where $b$ is the halo bias, $P^{\rm lin}$ is the linear matter power spectrum, and $k_\ell \equiv (\ell + 1/2) / \chi_1$. 
Again, full details of this derivation are provided in Appendix~\ref{sec:halo_model_der}.

For all of the power spectra, we compute the halo mass function, bias, and density profile with \textsc{halomod}~\cite{Murray:2013qza,Murray:2020dcd}. 
The $z$ integrals are evaluated with 100 linearly spaced bins from 0.005 to 4, and the $m$ integrals are evaluated with 100 logarithmically spaced bins from \qtyrange{e11}{e17}{\msol}. 
We evaluate $k$ integrals with $10^4$ logarithmically spaced bins from \qtyrange{e-4}{e3}{\per\mega\parsec}. 
We use the 2008 Tinker halo mass function $n(z,m)$~\cite{Tinker:2008ff} and the related bias $b(z,m)$ from Ref.~\cite{Tinker:2010my}. 
We defer a discussion of how the galaxy multipole kernel is modeled to Sec.~\ref{sec:galaxy_modeling}, since we adopt the same approach for our mock observations. 
Further details can also be found in Appendix~\ref{sec:hod}. 

\subsection{Environment B: Epoch of Reionization Intergalactic Medium Conversions}
\label{sec:envB}

\begin{figure*}[t!]
    \centering
    \includegraphics[width=\textwidth]{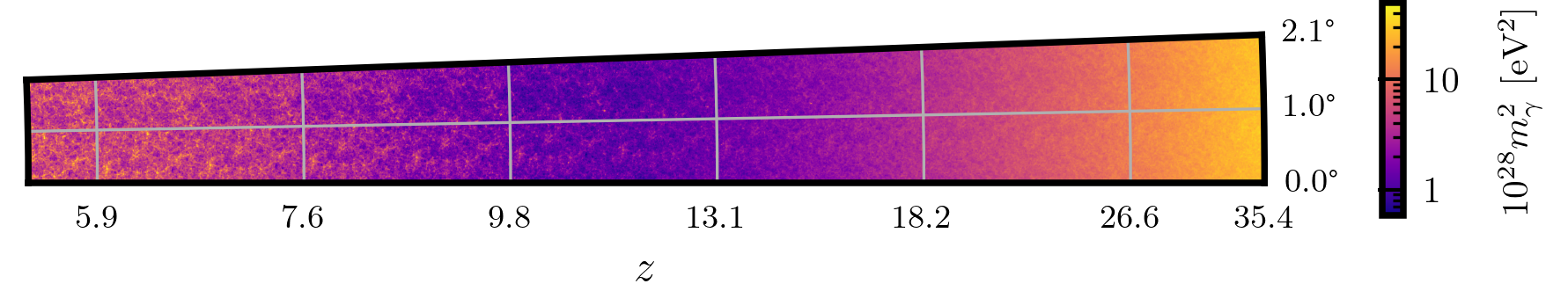}
    \caption{A 2D cross section of the angular lightcone of $m_{\gamma}^2$ that we use in this work, as a function of $z$ and one sky angular coordinate. Lines of sight are drawn horizontally through cells in the light cone. In each cell, we compute $m_{\gamma, i}^2$ and compare it to $m_{A'}^2$ to determine the probability of conversion in that cell.~\nblink{21cm_Lightcone}}
    \label{fig:lightcone}
\end{figure*}

We now move on to considering $\gamma \to A'$ conversions in the IGM during the EoR from $5\lesssim z\lesssim 35$. Throughout, we refer to this as ``environment B.''
This environment is much less dense than the dark matter halos that we previously considered, thereby providing sensitivity to lighter dark photons. 

The evolution of the IGM through the EoR is complicated, and as a result, we rely on simulations to compute the $\gamma \to A'$ signal.
Every $\gamma \to A'$ conversion along a line of sight $\hat{n}$ contributes to the observed temperature deficit $\Delta T$ along that line of sight.
Therefore, to compute $\Delta T$, we need a complete, simulated light cone spanning $5 \lesssim z \lesssim 35$ of the IGM, through which we draw lines of sight and determine where resonant conversions occur along each line.
Then, by summing the contributions from each conversion, we obtain a sky map of the predicted $\Delta T(\hat{n})$ due to $\gamma \to A'$ in environment B. 
We finally compute the angular two-point correlation function of this map to quantitatively analyze the signal.

Much like in environment A, we can also look for the $\gamma \to A'$ signal through cross-correlations between the radio sky and a high-redshift galaxy survey, with some overlap with the EoR redshifts considered here, $5 \lesssim z \lesssim 35$.
To understand why we expect such a correlation, recall that the probability of $\gamma \to A'$ conversions is set predominantly by the free-electron number density at a point in space, and therefore also by the baryon density at that same point.
These same densities are also correlated with galaxy formation: denser regions of the Universe are more likely to form galaxies.
Therefore, the number density of galaxies at a point is correlated with the rate of $\gamma \to A'$ conversions at that same point, and we expect a nonzero cross-correlation between the two signals.

\subsubsection{\texorpdfstring{Simulating $\gamma \to A'$ Conversions in the EoR IGM}{Simulating Photon to Dark Photon Conversions in the EoR IGM}}

We use the seminumerical code \textsc{21cmFAST}~\cite{Mesinger:2010ne, Park:2018ljd, Murray:2020trn} to construct 3D light cones (parametrized by two angular coordinates and $z$) in the range $5 \lesssim z \lesssim 35$. 
The production of the light cone begins by generating initial conditions in a cube of side length \qty{300}{\mega\parsec} at $z=35$, sampled on a $600^3$ grid. 
This is then downsampled to $150^3$ cells to produce the initial coeval box, and evolved in time using second-order Lagrangian perturbation theory.
The light cone is then formed by stitching together these coeval boxes along the redshift direction; in this work, we use a light cone with total angular size \qty{2.1}{\degree} and a transverse resolution of \qty{12.6}{\arcsecond}, giving $600^2 = $ 360,000 lines of sight. 

Our choices of astrophysical parameters in the simulation are the same as the fiducial parameters in Refs.~\cite{Gagnon-Hartman:2025oxd,Davies:2025wsa} (except for $\alpha_\star$, which we take to be the default value in \textsc{21cmFAST}).
Here we fix these parameters and do not consider uncertainties from their variation; a study of the impact of these uncertainties is left to future work. However, we note that changing how reionization proceeds will only shift the times and locations for which a particular mass $m_{A'}$ will resonantly convert, but cannot eliminate conversions altogether for any $m_{A'}$ that lies within the range of plasma masses present in the IGM during the EoR. 

With the simulated light cone, we proceed to compute the $\gamma \to A'$ signal due to conversions along lines of sight through the light cone. 
In order to determine where $\gamma \to A'$ conversions occur, we first compute the plasma mass in each light cone cell $i$ at corresponding redshift $z_i$, given by
\begin{equation}
    m_{\gamma, i}^2 = 4\pi \alpha_{\rm EM}\frac{\overline{n}_{b,0}}{m_e}x_{e,i}(1+z_i)^3(1+\delta_i)\left(1-\frac{3}{4}Y_{\rm He}\right) \,,
\end{equation}
where $\overline{n}_{b,0}$ is the mean present-day baryon number density, $x_{e, i}$ is the free-electron fraction (normalized to the number density of hydrogen, both neutral and ionized), $\delta_i$ is the matter density contrast, and $Y_{\rm He}$ is the helium mass fraction, which we fix to be $Y_{\rm He}=0.245$. 
This expression assumes that the helium ionization fraction is equal to the hydrogen ionization fraction, which is a common assumption in EoR simulations.
A 2D slice of the angular light cone of $m_{\gamma}^2$ that we use in this work is shown in Fig.~\ref{fig:lightcone}.
We can see the global variation of $m_\gamma^2$ from the figure, with the effective plasma mass decreasing with redshift initially due to the expansion of the Universe, but then increasing again once reionization starts to occur in earnest at around $z \sim 10$. 
We can also see local variations in $m_\gamma^2$ due to fluctuations in the matter density and ionization fraction.

\begin{figure*}[t!]
    \centering
    \includegraphics[width=0.45\linewidth]{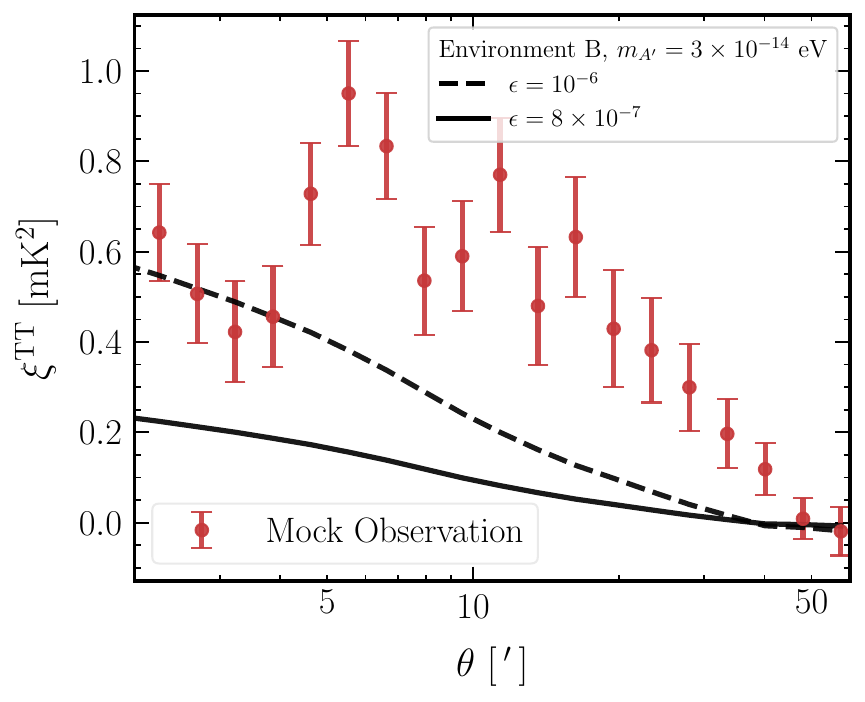}
    \qquad 
    \includegraphics[width=0.47\linewidth]{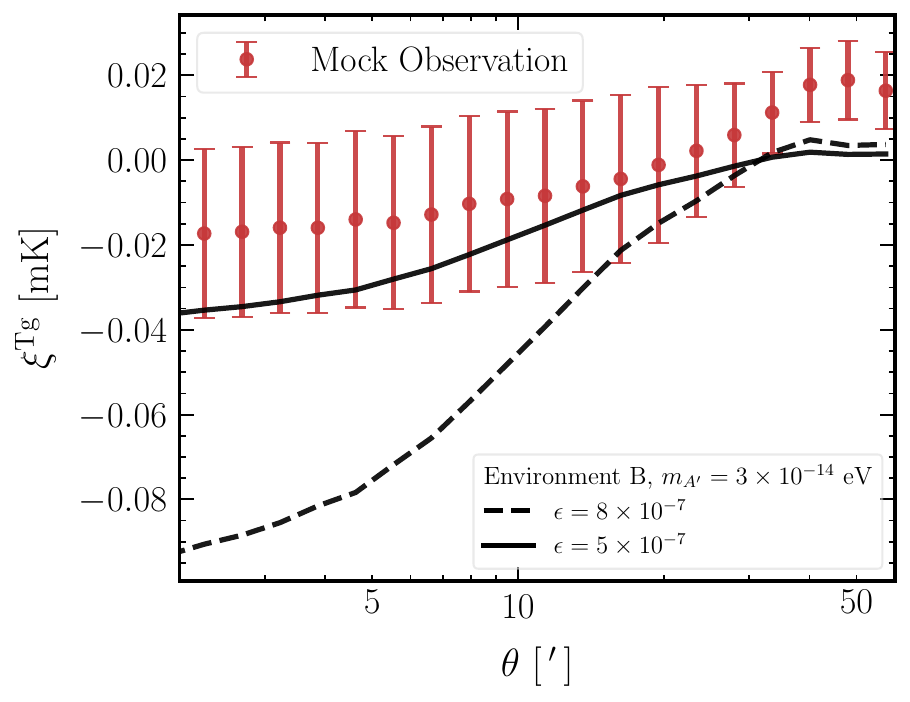}
    \caption{Two-point angular correlation functions from our mock observations, computed in 20 logarithmically spaced angle bins. The error bars shown are the square root of the diagonal elements of the covariance matrix, obtained from our mock observations. These are to be compared to the signal from conversions in environment B. \textit{Left:} $\xi^{\rm TT, obs}$ (red) computed from the post-ILC map, compared to the signal $\xi^{\rm B, TT}$, which is computed from the \textsc{21cmFAST} simulation for $m_{A'}=\qty{3e-14}{\eV}$ at $\omega_0/2\pi = \qty{410}{\mega\hertz}$ for $\epsilon = 8 \times 10^{-7}$ (black, solid) and $\epsilon = 10^{-6}$ (black, dashed).
    \textit{Right:} $\xi^{\rm Tg, obs}$ (red) computed from the cross-correlation of the post-ILC map and the mock \textit{Roman} galaxy catalog. The signal $\xi^{\rm B, Tg}$ is shown for $m_{A'} = \qty{3e-14}{\eV}$, with $\epsilon = 5 \times 10^{-7}$ (black, solid) and $\epsilon = 8 \times 10^{-7}$ (black, dashed).~\nblink{21cmFAST_corr_funcs}}
    \label{fig:21cmfast_corrs}
\end{figure*}

To compute the total probability of resonant conversions along each line of sight, we compare the value of $m_\gamma^2$ in adjacent cells.  
If $m^2_{\gamma,i} < m_{A'}^2 < m^2_{\gamma,i+1}$ between cells $i$ and $i+1$, then we say a conversion has occurred, and assign a probability of conversion to each cell given by  
\begin{equation}
    p_{\gamma \to A', i} = \frac{\pi \epsilon^2 m_{A'}^2}{H(z_{i})\omega_0 (1+z_i)^2}\left[\frac{\ln m_{\gamma,i+1}^2 - \ln m_{ \gamma, i}^2}{z_{i+1} - z_i}\right]^{-1}\!\!\!\!\!\! ,
\end{equation}
where $\omega_0$ is the present-day observed energy of the photon. 
The total probability of conversion along each line of sight is then 
\begin{equation}
 P_{\gamma \to A'} = \sum_{i} p_{\gamma \to A', i} \, . 
\end{equation}
This then gives a two-dimensional map of $\Delta T$, from which we compute the two-point angular correlation function $\xi^{\rm B, TT}$ as defined in Eq.~\eqref{eq:auto_xi} using the default scalar-field correlation function estimator in \textsc{treecorr}~\cite{Jarvis:2003wq}.
We plot $\xi^{\rm B, TT}$ in the left panel of Fig.~\ref{fig:21cmfast_corrs} for $m_{A'}=\qty{3e-14}{\eV}$ at $\omega_0/2\pi = \qty{410}{\mega\hertz}$ for two choices of $\epsilon$. 

\subsubsection{Cross-Correlation with Galaxies}
To obtain the predicted cross-correlation between $\gamma \to A'$ conversions and high-redshift galaxies, we use the latest version of \textsc{21cmFAST} to simulate the formation of high-redshift galaxies in the same light cone~\cite{Davies:2025wsa}.  
The code has a discrete halo sampler which first uses a halo-finder algorithm to locate overdense regions in the simulated matter field that correspond to the largest dark matter halos.
Smaller halos are identified using a conditional halo mass function that constructs merger trees. 
Then, these halos are populated with galaxies using a semiempirical model that is defined by relations such as the stellar-to-halo mass relation and the star formation rate (SFR). 
In particular, the simulation draws from probability distributions of these properties to account for stochasticity in the galaxy population. 

From the simulated galaxies, we can now build a mock galaxy catalog that would be observed by a high-redshift galaxy survey.
We begin by computing the observed luminosity density of each cell. This starts with the SFR $\dot{M}_\star$ in each simulation cell, which is a simulation output and is related to the rest-frame UV luminosity density $L_{\rm UV}$ of the cell by a proportionality
\begin{equation}
    \dot{M_\star} = \kappa_{\rm UV}L_{\rm UV}\, ,
\end{equation}
where $\kappa_{\rm UV}\equiv\qty{1.15e-28}{\msol\per\year \per \erg \second \hertz}$~\cite{Gagnon-Hartman:2025oxd}. 
It is more convenient to express this luminosity density in terms of the absolute AB magnitude $M_{\rm UV}$, given by the usual relation
\begin{equation}
    \log_{10}\left(
        \frac{{L}_{\rm UV}}{\unit{\erg \per\second \per \hertz}}\right) = 0.4 (51.63 - M_{\rm UV})\, .
\end{equation}
To determine if a galaxy at redshift $z$ is observable, we then compute the observed apparent magnitude from $M_{\rm UV}$, as follows:
$$
m_{\rm AB} = M_{\rm UV} + A_{\rm UV} + \mu - 2.5 \log_{10}(1+z) \,,
$$
where $A_{\rm UV}$ is a term that accounts for attenuation from dust and $\mu\equiv 5 \log_{10}(D_L(z)/\qty{10}{\parsec})$ is the distance modulus. 
The final term accounts for the redshift of the flux from the galaxy.
We adopt $A_{\rm UV}=4.43 + 0.79 \ln(10)\sigma_\beta^2 + 1.99\langle \beta\rangle (M_{\rm UV}, z)$ from Ref.~\cite{Tacchella:2012ih}, where $\sigma_\beta=0.34$ and $\langle \beta \rangle$ is the mean UV-continuum slope such that, on average, the flux density of a galaxy, $f_\lambda\propto \lambda^{\langle \beta \rangle(M_{\rm UV}, z)}$. 
If $A_{\rm UV}$ is negative for a given $\langle \beta \rangle$, we set it to 0~\cite{Tacchella:2012ih}.
We take $\langle \beta \rangle(M_{\rm UV}, z)$ from  Refs.~\cite{Trenti:2014hka,Mason:2015cna} for galaxies with $z<8$ and from Ref.~\cite{Cullen:2024abc} for higher redshift galaxies.

\begin{figure}[t!]
    \centering
    \includegraphics[width=\linewidth]{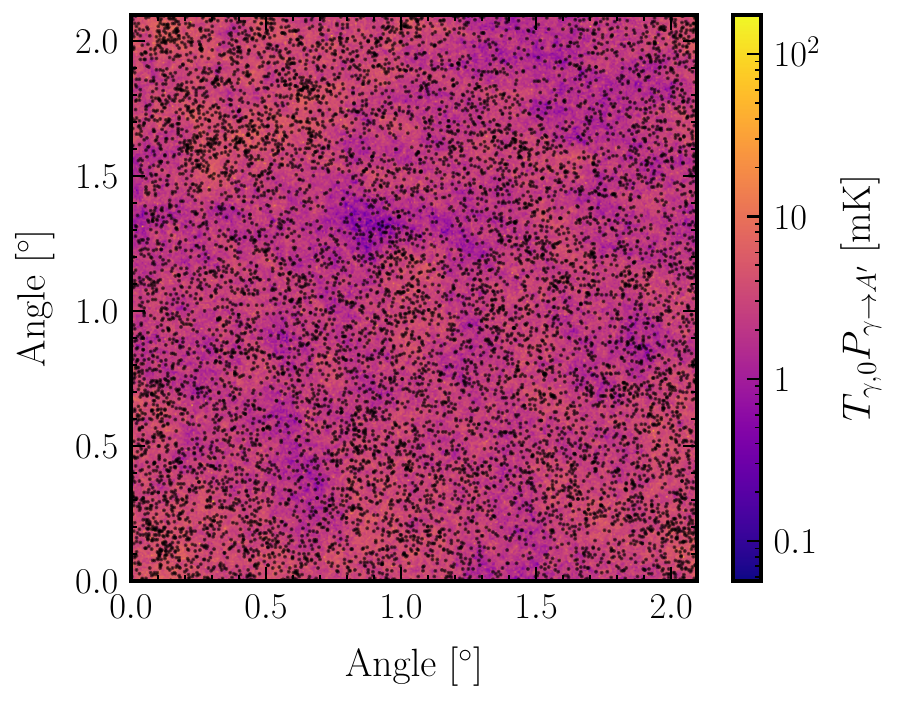}
    \caption{$T_{\gamma,0} P_{\gamma\to A'}$ is shown for $m_{A'}=\qty{3e-14}{\eV}$, $\epsilon=10^{-6}$, and $\omega_0/2\pi =\qty{410}{\mega\hertz}$. Overlaid in black points is a random subset of 10,000 galaxies from the mock \textit{Roman} galaxy survey. $\gamma \to A'$ conversions for dark photons of this mass preferentially occur in regions of the sky with more galaxies.~\nblink{Conversions_Galaxy_Map}}
    \label{fig:conv_gal_map}
\end{figure}

Now that we have the tools to compute $m_{\rm AB}$ in each cell, we construct a mock catalog that would be observed by the upcoming Nancy Grace Roman Space Telescope High Latitude Imaging Survey, or \textit{Roman}~\cite{Wang:2021oec}.
This survey is a \qty{2000}{\deg\squared} Southern Hemisphere survey centered at $l=\qty{260}{\degree}$, $b=\qty{-70}{\degree}$ with limiting magnitude $m_{\rm lim} = 26.7$~\cite{Wang:2005zj}, and overlaps with the all-sky coverage of SKA, making a cross-correlation between these two future surveys possible for the search for $\gamma \to A'$ conversions.
Our mock catalog for environment B is therefore built by flagging a cell as containing an observable galaxy if $m_{\rm AB}$ is less than $m_{\rm lim}=26.7$, and limiting the galaxies considered to $z > 7$, the redshift range most relevant to the dark photons we consider.
Figure~\ref{fig:conv_gal_map} shows a randomly drawn subsample of 10,000 galaxies from this mock catalog, superimposed on the map of $-\Delta T = T_{\gamma, 0}P_{\gamma \to A'}$ for $m_{A'}=\qty{3e-14}{\eV}$, $\epsilon=10^{-6}$, and $\omega_0/2\pi=\qty{410}{\mega\hertz}$.
There are visible overdensities of galaxies in the regions where $P_{\gamma \to A'}$ is largest, which illustrates the expected cross-correlation signal. 
Since $\gamma \to A'$ conversions occur preferentially in regions with many galaxies, we expect a positive correlation between $T_{\gamma,0} P_{\gamma \to A'}$ and galaxy number density, i.e. an anti-correlation between $\Delta T$ and the density of galaxies. 

The output of this modeling process is $\xi^{\rm B, TT}(\theta)$, the two-point autocorrelation function of $\Delta T$, and $\xi^{\rm B, Tg}(\theta)$, the two-point cross-correlation between $\Delta T$ and the mock \textit{Roman} catalog, defined in Eqs.~\eqref{eq:auto_xi} and~\eqref{eq:cross_xi}.
These are more convenient observables than the associated power spectrum because of the nature of the simulation: the \textsc{21cmFAST} simulation has an angular resolution of \qty{12.6}{\arcsecond} but only covers about 4 square degrees, making it difficult to perform the spherical harmonic transform necessary to extract an estimate of the relevant power spectra, since these transforms are defined over the entire sphere. 
It is, however, straightforward to compute correlation functions, which we do with the default estimators of the \textsc{treecorr} package. 
This estimator for $\xi^{\rm B, Tg}$ is defined as the difference between the cross-correlation of the $\Delta T$ field and the galaxy catalog and the cross-correlation of the $\Delta T$ field and a random catalog with $10^6$ points. 
This estimator is necessary because the simulated \textit{Roman} catalog is a set of galaxy coordinates, so we must account for effects such as shot noise. 
The resulting $\xi^{\rm B, Tg}$ then contains the same information as the power spectrum, but is more numerically robust. 
In Fig.~\ref{fig:21cmfast_corrs}, we plot $\xi^{\rm B, Tg}$ for $m_{A'}=\qty{3e-14}{\eV}$ for several choices of $\epsilon$. 
Notably, $\xi^{\rm B, Tg}$ is negative at small angular scales, as expected, since we are looking for photon disappearance.

\subsection{Environment C: Late-Time IGM Conversions}
\label{sec:envC}

So far, we have considered conversions both in dark matter halos at low redshifts (environment A) and the IGM from $5\lesssim z \lesssim 35$ (environment B). 
However, conversions also occur in the IGM at low redshifts ($z \lesssim 5$); these conversions are not captured in the halo model, and are important to consider for several reasons. 
First, since the IGM is much less dense than halos, $m_{\gamma}^2$ is lower than in halos, and thus modeling this environment increases our sensitivity to lower dark photon masses between $\qty{8e-15}{\eV}\lesssim m_{A'} \lesssim \qty{e-13}{\eV}$. 
Second, although we did model dark photon conversions of this mass range in environment B, the probability of conversion favors low redshifts, so the signal from these conversions would dominate over the signal from higher redshift conversions.
To see why, recall that
\begin{equation}
P_{\gamma\to A'} \propto \frac{1}{H(z)(1+z)^2}\left|\frac{d \ln m_{\gamma}^2}{dz}\right|_{z_{\rm res}}^{-1}\, .
\end{equation}
Therefore, after reionization is complete, $P_{\gamma\to A'}$ is larger at low redshifts, since $|d \ln \overline{m_\gamma^2} / dz|^{-1} \sim (1+z)/3$. 
Physically, when $m_{\gamma}^2$ changes more slowly, the resonance timescale is larger, and so the probability of conversion is enhanced. 

Just like the high-redshift IGM (environment B), late-time conversions can only be properly modeled with hydrodynamical simulations. 
In fact, the CMB temperature anisotropies for low-redshift IGM conversions were obtained in Ref.~\cite{Aramburo-Garcia:2024cbz} using simulations and compared directly to the angular power spectrum of the \textit{Planck} maps. 
A more rigorous and potentially stronger limit can be obtained using our signal extraction pipeline applied to CMB data; furthermore, we are also interested in radio surveys in this paper, which was not considered there. 

We therefore leave the complicated task of using hydrodynamical simulations to future work; instead, we estimate this signal by assuming that the one-point and two-point probability distribution functions (PDFs) of baryon density fluctuations follow a lognormal PDF for $m_{\gamma}^2$, following Ref.~\cite{Caputo:2022keo}. 
Ref.~\cite{Caputo:2020rnx} found that treating the one-point PDF as a lognormal distribution agreed well with the one-point PDF of matter fluctuations derived analytically, and argued that a lognormal distribution was a reasonable Ansatz for the one-point PDF of $m_{\gamma}^2$. 
Furthermore, the limits on $\epsilon$ obtained in Refs.~\cite{Caputo:2020bdy,Caputo:2020rnx} using the lognormal Ansatz for the one-point PDF also agree well with limits obtained from simulations~\cite{Garcia:2020qrp}. 
Assuming a lognormal form for the two-point PDF of $m_{\gamma}^2$ has not been studied in detail and is an admittedly crude Ansatz, but in the limit where fluctuations are small, this reduces to a Gaussian two-point PDF, which is the correct limit at high redshifts. This will suffice for an estimate of the signal.

Using these Ans\"{a}tze for the PDFs, we can compute the two-point angular power spectrum of the sky-averaged $\gamma \to A'$ signal, $C_\ell^{\rm C, TT}$. This result was first derived in Ref.~\cite{Caputo:2022keo}, and is recapped in Appendix~\ref{sec:lognormal_derivation} for completeness; the final result is
\begin{equation}
    C_\ell^{\rm C, TT} = T_{\gamma, 0}^2\left[\frac{\pi \epsilon^2 m_{A'}^4}{\omega_0}\right]^2 \int dz \frac{ W^2(z)H(z)}{\chi^2}\widetilde{Q}\left(\frac{\ell}{\chi}, z\right) \, ,
    \label{eq:Cl_envC}
\end{equation}
where $W(z) \equiv [H(z)(1+z)^2]^{-1}$, and 
\begin{alignat}{1}
    \widetilde{Q}(k, z) &= 4\pi \int d\rho \, \rho^2 j_0(k \rho)Q(\rho, z) \, , \\
    Q(\rho, z) &= f_2\left(\rho, m_{A'}^2, m_{A'}^2; z \right)  - [f_1(m_{A'}^2 ; z)]^2 \, . \label{eq:mgamma_corr_func}
\end{alignat}
Here, $f_2(\rho, m_{A'}^2, m_{A'}^2; z)$ is the two-point PDF of $m_{\gamma}^2$, evaluated at two points, both with $m_\gamma^2 = m_{A'}^2$, separated by comoving distance $\rho$ at redshift $z$. 
Similarly, $f_1(m_{A'}^2; z)$ is the one-point PDF of $m_{\gamma}^2$ evaluated at $m_\gamma^2 = m_{A'}^2$ at redshift $z$.
The full expressions for $f_1$ and $f_2$ under the lognormal assumption are given in Eqs.~\eqref{eq:lognormal_mgamma_pdf} and~\eqref{eq:2pt_mgamma_pdf}. 
At large enough values of $\rho$, the distance between the two points in $f_2$ is so large that there is virtually no correlation between the $m_\gamma^2$ values at those two points; in this limit, $Q \to 0$. 
$\widetilde{Q}$ is simply the three-dimensional Fourier transform of $Q$, evaluated assuming isotropy. 
Note that $C_\ell^{\rm C, TT}$ is evaluated using the Limber approximation, which is valid at sufficiently large $\ell$.

\begin{figure}[h]
    \centering
    \includegraphics[width=\linewidth]{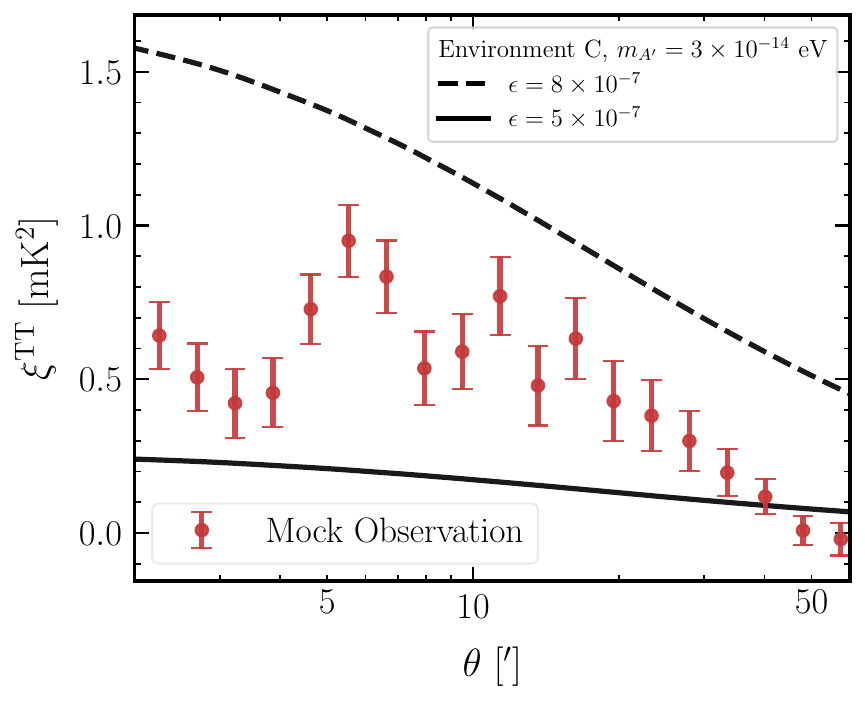}
    \caption{Two-point angular correlation functions from our mock observations, computed in 20 logarithmically spaced angle bins. The error bars are the square root of the diagonal elements of the covariance matrix, obtained from our mock observation. In black, the theoretical prediction $\xi^{\rm C, TT}$ is plotted for $m_{A'}=\qty{3e-14}{\eV}$ at $\omega_0/2\pi = \qty{410}{\mega\hertz}$ for  $\epsilon=5\times 10^{-7}$ (black, solid) and $\epsilon=8\times 10^{-7}$ (black, dashed). This correlation function is compared to the $\xi^{\rm TT, obs}$ from the post-ILC map (red). $\xi^{\rm C, TT}$ is greatest at small angular scales.~\nblink{Analytic_corr_funcs}}
    \label{fig:analytic_xi}
\end{figure}

In practice, we first compute $C_{\ell}^{\rm C, TT}$, and then use Eq.~\eqref{eq:xi_and_Cl_relation} to compute 
\begin{equation}
\xi^{\rm C, TT}(\theta) = \sum_{\ell} \frac{2\ell+1}{4\pi} C_\ell^{\rm C, TT}P_\ell(\cos\theta) \, ,
\label{eq:C_corr_func}
\end{equation} which is the quantity used in our analysis. 
We again use the correlation function instead of the power spectrum because we combine this signal with that from environment B. 
Furthermore, in a more detailed study relying on hydrodynamical simulations, we will likely be limited to producing a simulated patch of sky that is relatively small, as was the case in environment B. 
Then, as in environment B, it is more convenient to consider $\xi$ instead of $C_\ell$, which is more naturally defined over the entire sphere.
For both models, we consider $\gamma \to A'$ conversions over the same range of $m_{A'}$, and so for consistency we add the two signals. 
Note that there is no cross correlation with galaxies for environment C, since the conversions are assumed to happen in the IGM. 
We show the predicted $\xi^{\rm C, TT}$ in Fig.~\ref{fig:analytic_xi} for $m_{A'}=\qty{3e-14}{\eV}$ at $\omega_0 /2\pi = \qty{410}{\mega\hertz}$ and two choices of $\epsilon$. 

\section{Mock Observations}
\label{sec:mock_observations}

Having computed the theoretical predictions for the $\gamma \to A'$ signal, i.e. $C_\ell^{\rm TX}$ and $\xi^{\rm TX}(\theta)$, we now need to understand how we would potentially observe these signals in realistic surveys.
In the radio sky, we need to contend with a substantial astrophysical foreground, coming from both galactic and extragalactic sources.  
As such, it is crucial that we test our ability to extract the $\gamma \to A'$ signal in the presence of these foregrounds by generating realistic, mock observations of the radio sky, as well as a mock galaxy catalog for the cross-correlation analysis.

We begin by creating mock radio maps across a range of frequencies that simulate those that would be observed at an experiment such as SKA.
This involves simulating realistic astrophysical foregrounds, and experimental details such as noise and beam size.
These mock null-signal maps are then processed with the rest of the analysis pipeline that would be used to extract the dark photon signal from actual SKA data.
In particular, the radio maps at different frequencies are optimally combined using the ILC method to produce a single post-ILC map. 
From this map, we can obtain the autocorrelation function and autopower spectra that can then be compared to the theoretical predictions from Sec.~\ref{sec:signal_modeling} to estimate our sensitivity to $\gamma \to A'$ conversions.

For environments A and B, we additionally create mock galaxy catalogs, and use the galaxy overdensity of these maps to perform a cross-correlation with the post-ILC map, which are then once again compared to the theoretically predicted cross-correlation to obtain an estimate of our sensitivity to $\epsilon$ as a function of $m_{A'}$.
We use different techniques to form these mock catalogs in environments A and B.
In environment A, we use the halo occupation distribution model from Ref.~\cite{Kusiak:2022xkt}, which describes galaxies in the ``blue'' subsample of the \textit{unWISE} catalog from Refs.~\cite{Krolewski:2019yrv,Krolewski:2021yqy} to draw a realization of the galaxy survey map. 
In environment B, we make use of the mock catalog of high-redshift galaxies produced by the \textsc{21cmFAST} simulation, described in Sec.~\ref{sec:envB}.

\subsection{Mock Foreground SKA Radio Maps}
\label{sec:foreground_modeling}

We produce mock maps that would be observed by SKA-Mid, which will operate at frequencies between \qtyrange{0.35}{15}{\giga\hertz} \cite{Braun:2019gdo}. 
We therefore generate nine maps at \{0.41, 0.56, 0.77, 1.05, 1.43, 4.94, 6.74, 9.19, 12.53\}~\unit{\giga\hertz}, which are the centers of the SKA-Mid continuum observing bands~\cite{Braun:2019gdo}. 
SKA will be capable of operating in several distinct modes, each with particular advantages and disadvantages. 
Here, we consider a continuum broadband observing mode that sacrifices spectral resolution for angular resolution.
This mode will operate  with spectral resolution of $\Delta \omega /\omega \approx 0.3$ and a beam size of several arcseconds~\cite{Braun:2019gdo}. 
As we explain more fully below, we find that this fine angular resolution is crucial to the improved sensitivity of SKA relative to earlier cosmological dark photon searches.
Previous work that uses similar methods to our pipeline considered a single-dish mode of SKA or other radio experiments that have fine spectral resolution but low angular resolution~\cite{Dai:2024bfa,Joseph:2024ush,DeCaro:2025qly}.
This is important in the intensity mapping applications that those authors consider, but would be less ideal for the discovery of dark photons.

At each frequency that we consider, we model both galactic and extragalactic foregrounds, which we discuss in more detail in Secs.~\ref{sec:galactic_foregrounds} and \ref{sec:extragalactic_foregrounds} below.
We produce maps in the HEALPix format\footnote{\url{https://healpix.sourceforge.io/}} with $N_{\rm side}=2048$, which corresponds to an angular resolution of about \qty{2.6}{\arcminute}. 
The maximum multipole moment we can reach with this resolution is $\ell_{\rm max}=4096$ \cite{Sullivan:2024jim}. 
This choice of $\ell_{\max}$ corresponds to the highest resolution maps that are computationally feasible in our analysis, although radio surveys operate with much finer resolution than this. 
Because of the limited resolution, we only consider correlations on scales larger than \qty{2.6}{\arcminute} or $\ell < 4096$ in this work, keeping in mind that much smaller scales are accessible in realistic experiments such as SKA. 

Besides the astrophysical foregrounds, we also add thermal noise to all the mock foreground maps to account for additional experimental noise. 
We assume that the thermal noise is randomly distributed across the map with a magnitude drawn from a Gaussian distribution with standard deviation $\sigma_{\rm th}$. 
For simplicity, we set the thermal noise standard deviation at the level found by the LOFAR Two Meter Sky Survey Deep Fields Project (LoTSS Deep), $\sigma_{\rm rms}= \qty{7.5}{\micro\jansky\per\beam}$~\cite{Shimwell:2025tui}, which is above the expected level of thermal noise after several hours of observing at SKA~\cite{Braun:2019gdo}.
To convert $\sigma_{\rm rms}$ to a temperature, we use the relation
\begin{equation}
    \sigma_{\rm th} = \frac{\nu^{-2}}{2 \Omega_{\rm pix}}\sigma_{\rm rms} \, ,
\end{equation}
where $\Omega_{\rm pix}$ is the solid angle of a pixel in our foreground map. 
The thermal noise as a function of temperature then varies as a function of frequency, but is constant as a function of flux density, which is the quantity measured by the experiment. 
For the frequencies of interest here, $\sigma_{\rm th} \sim \qtyrange{1}{10}{\milli\kelvin}$. 
We additionally subtract the monopole from each map since we are only interested in fluctuations. 

Finally, since the galactic foregrounds are most substantial near the galactic plane and get less bright away from it, we can improve our SNR by masking out the brightest areas of the sky. 
We therefore choose to mask all of the sky except a \qty{20}{\degree} patch around the center of the \textit{Roman} galaxy survey in all of our analyses; this choice preserves our ability to perform a cross-correlation analysis in environment B with \textit{Roman}, while removing the brightest galactic foregrounds.

\subsubsection{Galactic Foregrounds}
\label{sec:galactic_foregrounds}

\begin{figure}
    \centering
    \includegraphics[width=0.89\linewidth]{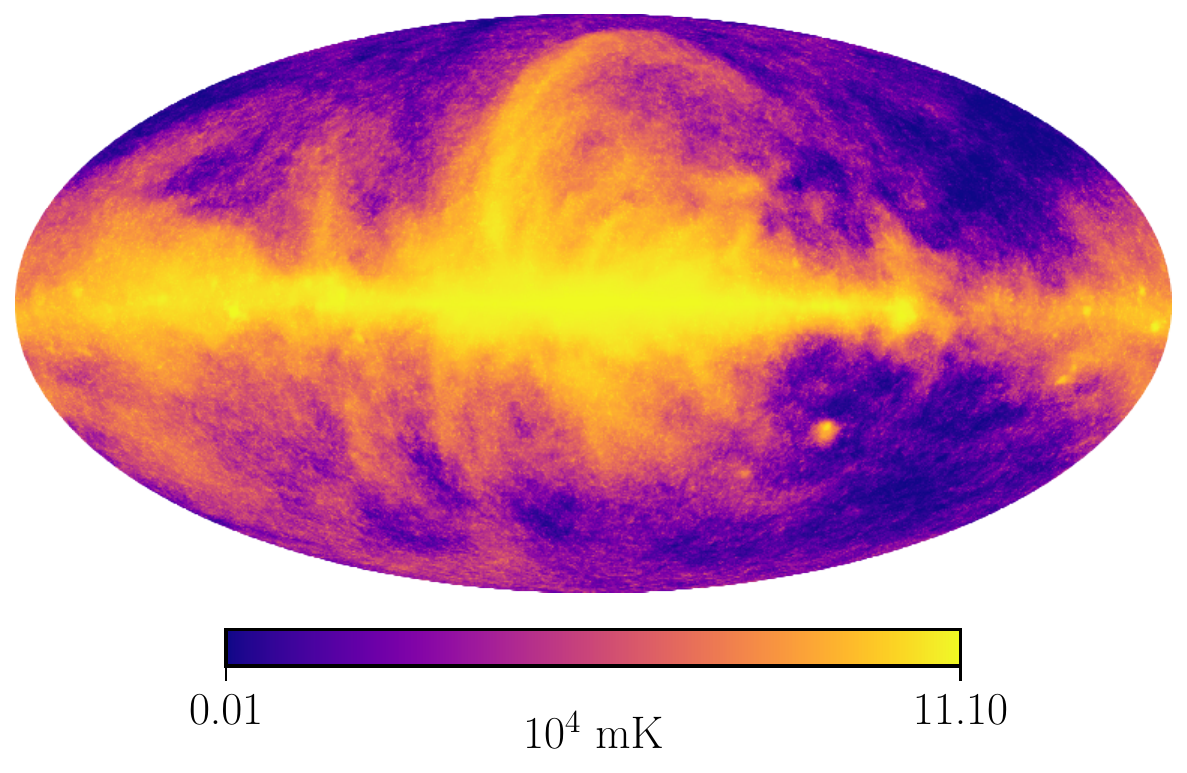}
    \medskip
    \includegraphics[width=0.89\linewidth]{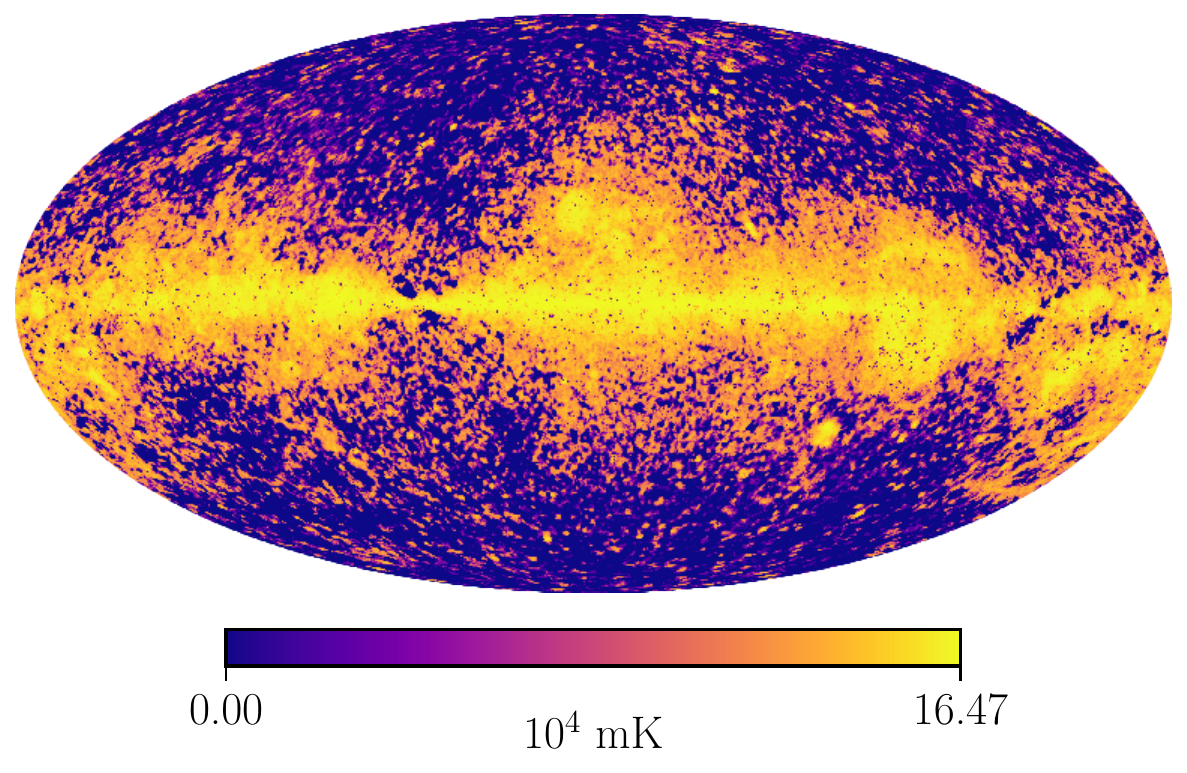}
    \medskip
    \includegraphics[width=0.89\linewidth]{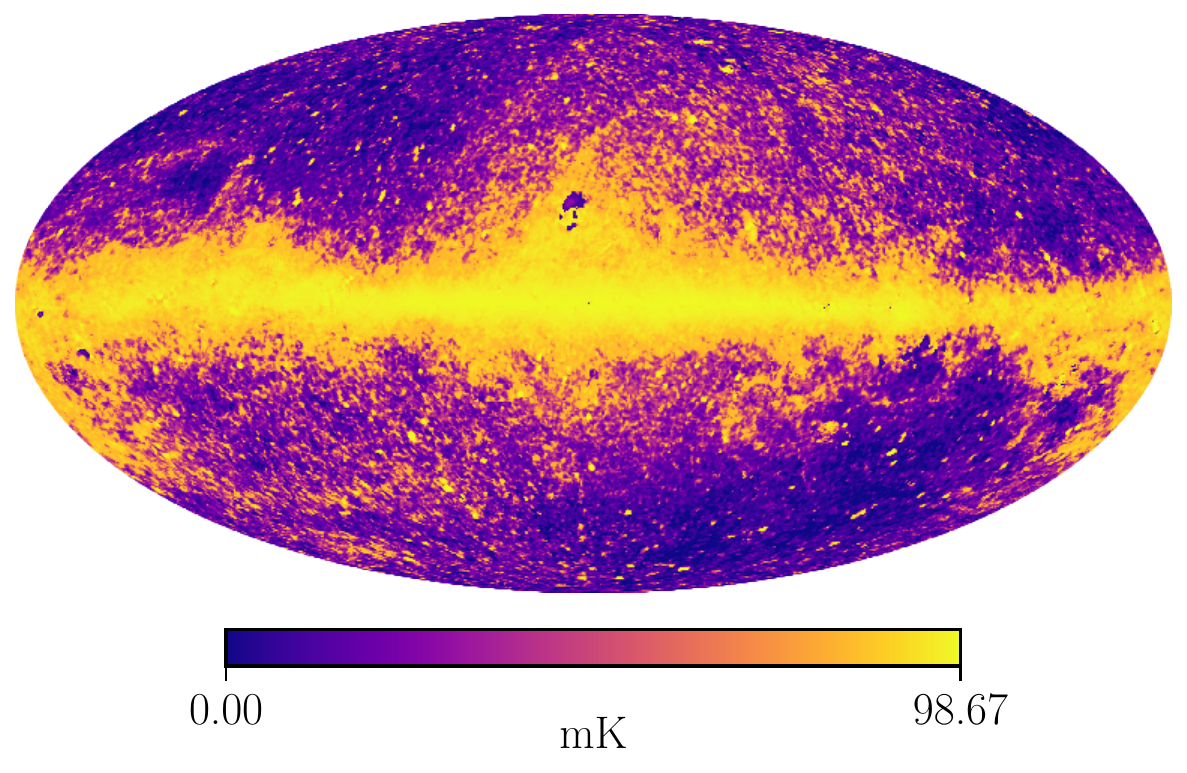}
    \medskip
    \includegraphics[width=0.89\linewidth]{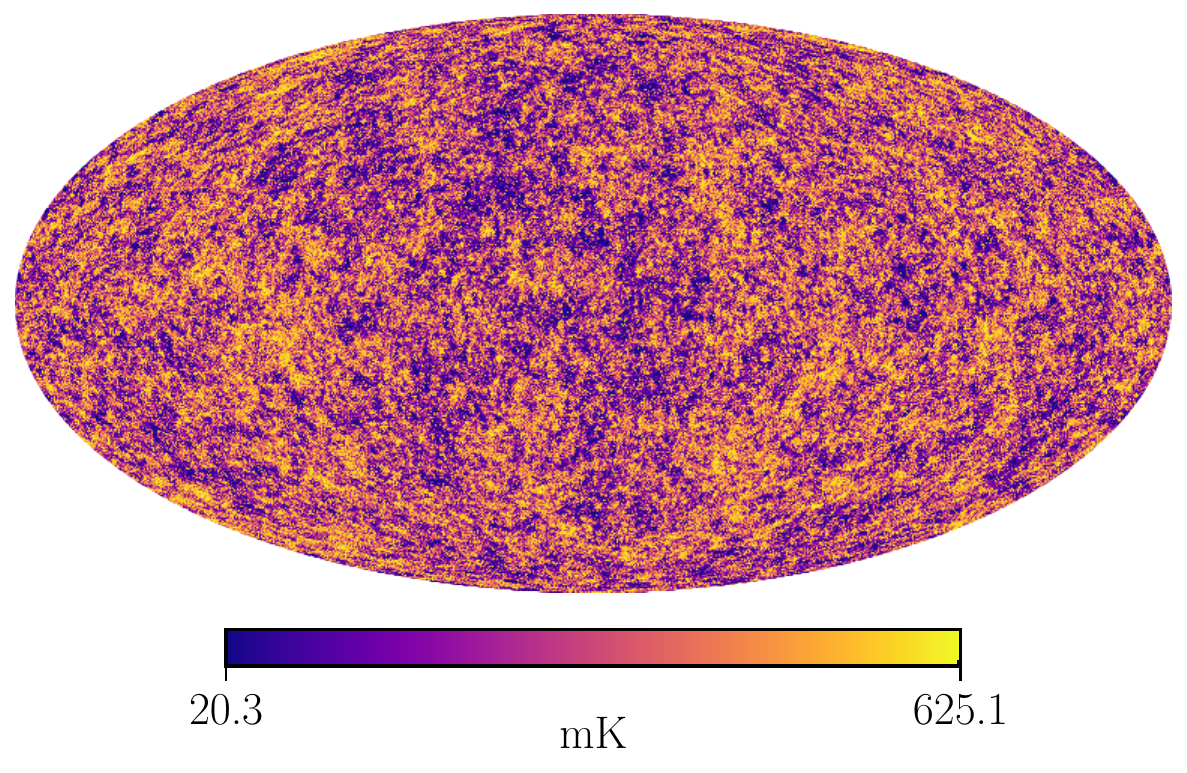}
    \caption{Thermodynamic temperature of the main components of our mock radio foreground, namely (from top to bottom) synchrotron radiation, free-free radiation, spinning dust, and the diffuse background of extragalactic point sources. All maps are made at $\omega_0/2\pi=\qty{1.43}{\giga\hertz}$ in the Mollweide projection, with $N_{\rm side}=2048$.~\nblink{Foregrounds}}
    \label{fig:foregrounds}
\end{figure}

The most important galactic foregrounds at the frequency range that we consider are galactic synchrotron radiation and galactic free-free radiation. 
Spinning dust, known also as anomalous microwave emission (AME), is also important above \qty{1}{\giga\hertz}. 
We model each of these foregrounds using \textsc{pysm3}~\cite{Thorne:2016ifb,Zonca:2021row,Pan-ExperimentGalacticScienceGroup:2025vcd} and provide a brief description of each model for completeness.

We model the synchrotron radiation using the \texttt{s1} model of the \textsc{pysm3} code. 
This model is based on two templates: the reprocessed Haslam \qty{408}{\mega\hertz} map from Ref.~\cite{Remazeilles:2014mba} and the WMAP 9-year \qty{23}{\giga\hertz} Q/U maps~\cite{WMAP:2012fli}. 
The Haslam map template is produced with a resolution of \qty{57}{\arcminute} and therefore does not provide information about synchrotron radiation on the smaller scales that are highly relevant in this work.
To overcome this, \textsc{pysm3} assumes that the angular power spectrum of the synchrotron radiation follows a power law $\ell(\ell+1)C^{\rm ss, sync}_\ell \propto \ell^{-\gamma}$, where $\gamma =0.55$. 
This choice agrees with other observations of diffuse galactic synchrotron radiation at the scales of interest \cite{Chakraborty:2019ecc}.  
The map of small-scale synchrotron radiation is drawn assuming a lognormal distribution obtained from this power spectrum~\cite{Thorne:2016ifb}.
The resulting final map is then the sum of the largescale template map and this small-scale map, which is further rescaled to the appropriate frequency using a spatially varying power law from `Model 4' in Ref.~\cite{Miville-Deschenes:2008lza}.

We model the free-free radiation with the \texttt{f1} model of \textsc{pysm3}, which is based on the fit to the \textit{Planck} 2015 data~\cite{zotero-item-2388} with additional small-scale features added using a similar procedure to that for galactic synchrotron radiation, described more fully in Ref.~\cite{Thorne:2016ifb}. 

Spinning dust is modeled with the \texttt{a1} model of \textsc{pysm3}. 
This model combines two populations of spinning dust based on spatial templates from the \textit{Planck} 2015 characterization of diffuse foregrounds~\cite{Planck:2015mvg}, which are then rescaled according to an emission law from Ref.~\cite{Ali-Haimoud:2008kyz}. 
One population has a peak frequency in this emission law that is spatially varying, while the other has a constant peak frequency. 
At high resolutions, spinning dust is modeled using a thermal dust template \cite{Thorne:2016ifb}. 

Example maps of these foregrounds at one representative frequency are shown in Fig.~\ref{fig:foregrounds}. 
For completeness, we also include a population of thermal dust, and galactic CO emission in our foreground maps at frequencies greater than \qty{1}{\giga\hertz}. We model these using the \texttt{d1} and \texttt{co1} models of \textsc{pysm3}, respectively, and note that their magnitude is much smaller than that of synchrotron radiation, free-free radiation, and spinning dust. 

\subsubsection{Extragalactic Foregrounds}
\label{sec:extragalactic_foregrounds}
Extragalactic point sources are the most significant radio foreground outside the Milky Way. 
We classify these point sources into two categories based on their brightness.

First, the brightest sources are resolved and cataloged by radio surveys such as LoTSS Deep, which recently characterized extragalactic point sources down to fluxes of approximately \qty{0.05}{\milli\jansky} in the European Large Area ISO Survey-North 1 (ELAIS-N1) deep field~\cite{Shimwell:2025tui}.
These sources can be easily masked out in realistic radio surveys and therefore would not substantially impact a search for $\gamma \to A'$ conversions.
In this work, we assume that every point source with flux greater than $S_{\rm max}=\qty{0.1}{\milli\jansky}$ can be resolved and masked in a realistic experiment. 
This threshold is above the minimum flux that LoTSS Deep successfully resolved and is therefore a conservative choice.
In our mock point source maps, we account for this masking by simply not including any point sources with flux greater than $S_{\rm max}$~\cite{Wang:2005zj,Liu:2011hh}, since the beam size of SKA is much smaller than the pixel size of our maps. 

Next, there is a diffuse background of point sources that are not resolved by a radio experiment.
Since they cannot be resolved, it is impossible to mask these sources, and they therefore are the most significant extragalactic foreground. 
This background can be characterized entirely in terms of 1) the normalized source count distribution $dN/dS$, which describes the number of point sources per steradian $N$ per unit flux density $S$~\cite{Mandal:2020zor} normalized to the number density of sources in a survey, and 2) the two-point angular correlation function of the point sources, $\xi_{\rm PS}(\theta)$~\cite{Hale:2023ust}.

We take $dN/dS$ from Ref.~\cite{Gervasi:2008rr}, which is fit to the form
\begin{equation}
    \frac{dN}{dS} = S^{-2.5}\left[\frac{1}{A_1 S^{a_1}+B_1 S^{b_1}}+\frac{1}{A_2 S^{a_2}+B_2S^{b_2}}\right] \,,
\end{equation}
where the coefficients are fit to the $6C$ and $7C$ \qty{150}{\mega\hertz} surveys \cite{zotero-item-2905,zotero-item-2951,zotero-item-2953, zotero-item-2955, zotero-item-2956,zotero-item-2962, zotero-item-2959}. 
This source count distribution agrees with that measured by the LoTSS Deep Survey at brightnesses above \qty{0.1}{\milli\jansky}~\cite{Mandal:2020zor}.
Below this threshold, our adopted $dN/dS$ is larger than that from LoTSS Deep and thus is a more conservative choice.
However, we have checked that our final results are largely insensitive to either choice.

We take the angular correlation function from the LoTSS Data Release 2 (LoTSS-DR2), which measured the clustering of radio sources from a catalog collected from several fields across the Northern Hemisphere~\cite{Hale:2023ust}. 
The two-point correlation function follows a power law 
\begin{equation}
    \xi_{\rm PS}(\theta) = 7.8 \times 10^{-3} \, \theta\,^{-0.821} \, ,
\end{equation}
where $\theta$ is in degrees. 

Using $dN/dS$ and $\xi_{\rm PS}(\theta)$, we construct a mock field of the diffuse point source background. 
We include sources with a minimum flux of $S_{\rm min} = \qty{1}{\micro\jansky}$, the lowest flux for which data exists, and a maximum flux of $S_{\rm max}$, using the code \textsc{epspy}~\cite{Mittal:2024mzv}. 
We can estimate the contribution from dim sources that we exclude from our analysis with our choice of $S_{\rm min}$ by extrapolating $dN/dS$ to lower flux values and integrating $S (dN/dS)$ to get the total flux per steradian due to these sources. 
Using our adopted $dN/dS$ from Ref.~\cite{Mandal:2020zor}, we find that the contribution from sources dimmer than $S_{\rm min}$ is negligible compared to the sources that we do include. 
The final unresolved point sources map is shown in the bottom panel of Fig.~\ref{fig:foregrounds} at $\omega_0/2\pi = \qty{1.43}{\giga\hertz}$.

Besides extragalactic point sources, we also include the lensed CMB and thermal Sunyaev-Zeldovich emission for completeness, using the \texttt{c1} and \texttt{tsz1} models of \textsc{pysm3}.  
However, because the CMB fluctuations are $\mathcal{O}(\unit{\micro\kelvin})$, the CMB fluctuations are not a significant source of contamination compared to the other foregrounds we consider.
The thermal Sunyaev-Zeldovich effect is also negligible at radio frequencies.  

\subsubsection{Needlet ILC Pipeline}
\label{sec:needlet_ilc}

Having constructed our mock radio foreground maps, we are now ready to construct a pipeline that we would use at an SKA-like experiment to extract a dark photon signal. 
The ILC method is a robust technique used extensively in the CMB community to separate the CMB from foregrounds~\cite{WMAP:2003cmr,Tegmark:2003ve,Delabrouille:2008qd,ACT:2023wcq,Dai:2024bfa,Joseph:2024ush,DeCaro:2025qly}. This method leverages the fact that the foregrounds have a different spectrum than the signal of interest, and by taking a linear combination of maps at different frequencies, we can minimize the variance of the resulting map subject to the constraint that the strength of the signal of interest is not reduced.

In our pipeline, this algorithm would extract a component with a spectrum that is proportional to $\omega^{-1}$, i.e. a dark photon-like component. 
To perform a forecast of SKA's sensitivity to $\gamma \to A'$ conversions, we perform the ILC algorithm on our mock, foreground-only maps to obtain a single minimum-variance map.
The two-point correlation functions and angular power spectra of this post-ILC map---either autocorrelation, or in cross-correlation with a mock galaxy survey---are then computed following Eqs.~\eqref{eq:auto_xi}--\eqref{eq:xi_and_Cl_relation}, and compared to the theoretical prediction for the dark photon signal. 
This simulates how the algorithm would perform on SKA data assuming the null-signal hypothesis, and therefore can be used to set an upper limit on our sensitivity to $\epsilon$ as a function of $m_{A'}$. 
The ILC procedure could never perfectly extract the dark photon signal from SKA maps: there would always be some level of residual noise in the map that sets the sensitivity of the experiment.
This pipeline allows us to determine exactly what this residual noise level is likely to be.

In this section, we explain the most basic, real-space ILC procedure to convey the main ideas of the algorithm, closely following the explanation from Ref.~\cite{McCarthy:2023hpa}, before providing a brief overview of the more sophisticated needlet ILC (NILC) procedure~\cite{Delabrouille:2008qd} that we use in our analysis; a more detailed explanation is left to Appendix~\ref{sec:needlet_ilc_appendix}.
We implement this algorithm with the \textsc{pyilc} code \cite{McCarthy:2023hpa} in our analysis pipeline. 

The basic real-space ILC procedure proceeds as follows. 
First, begin by writing an observed map $T_i$ at a pixel $\hat{n}$ as 
\begin{equation}
    T_i(\hat{n})=a_i s(\hat{n}) + n_i(\hat{n})\, ,
\end{equation}
where $i$ indexes each frequency bin, $s(\hat{n})$ is the signal of interest, and $n_i(\hat{n})$ is the sum of the foregrounds and noise. 
$a_i$ is a coefficient that contains the frequency dependence of the signal, which is proportional to $\omega^{-1}$ in our case, leaving $s(\hat{n})$ independent of frequency. 
Then, we can take a linear combination of the maps with weights $w_i$ subject to the constraint that
\begin{equation}
    \label{eq:ilc_constraint}
    \sum_i w_i a_i =1\, ,
\end{equation}
which gives a single map, given by 
\begin{align}
    T_{\rm ILC}(\hat{n}) &= \sum_i w_i T_i(\hat{n}) = \sum_i w_i(a_i s(\hat{n}) + n_i(\hat{n})) \nonumber \\
    &= s(\hat{n}) + \sum_i w_i n_i(\hat{n})\, .
\end{align}
The ILC procedure works by finding a choice of weights $w_i$ that minimizes the variance of $T_{\rm ILC}(\hat{n})$, i.e.\ $\sum_{\hat{n}}[\Delta T_{\rm ILC}(\hat{n})]^2$, where $\Delta T_{\rm ILC}(\hat{n}) \equiv T_{\rm ILC}(\hat{n}) - \langle T_{\rm ILC}(\hat{n}) \rangle$. 
Since the signal strength is not affected by the choice of $w_i$, this procedure effectively minimizes all other components relative to the signal of interest. 
The optimal weights can be found with a Lagrange multiplier that enforces the constraint in Eq.~\eqref{eq:ilc_constraint}, which gives~\cite{Eriksen:2004jg}
\begin{equation}
    w_i =  \frac{a_j \mathcal{C}^{-1}_{ij}}{a_k \mathcal{C}^{-1}_{kl} a_l} \label{eq:clean_coeff}\, ,
\end{equation}
where $\mathcal{C}_{ij} = \sum_{\hat{n}} \Delta T_i(\hat{n}) \Delta T_j(\hat{n})$ is proportional to the covariance matrix of the maps at different frequencies and repeated indices are summed over. 

It is important to note that the normalization of the resulting map is arbitrary and depends on what factors are absorbed into $a_i$ or $s$.  
In this work, we choose $a_i=\omega_i^{-1} / \omega_{0,\rm{ref}}^{-1}$, where $\omega_i$ is the energy of the $i$-th map, and $\omega_{0,\rm{ref}} / 2\pi \equiv \qty{410}{\mega\hertz}$. 
This normalizes the resulting map $T_{\rm ILC}(\hat{n})$ to the magnitude of a signal-only map at energy $\omega_{0,\rm{ref}}$.

The standard ILC method is not effective if there are highly anisotropic foregrounds, since we have not considered any spatial or angular information.
A more sophisticated strategy that does take both spatial and angular information into account is the NILC algorithm~\cite{Delabrouille:2008qd}.
The first step in this process involves filtering each $T_i$ in harmonic space with $N_I$ ``needlet'' filters that separate out different angular scales.
This gives $N_I$ maps for each frequency bin, each of which contains information at different characteristic angular scales. 
One simple example of needlet filters is $N_I$ top hats that are nonzero over a range $\Delta \ell_I$.
Then, the $j$-th filtered map would only contain information about angular scales in the range $\Delta \ell_j$.

Next, each needlet-filtered map is smoothed with a Gaussian kernel that depends on the needlet scale and the covariance of each pixel is computed across frequency bins.
The ILC weights are still given by Eq.~\eqref{eq:clean_coeff}, but now there is one weight for each pixel, needlet scale, and frequency.
This is the key difference from the standard algorithm that greatly improves the NILC's performance. 
The needlet-filtered maps are then summed with these weights and combined to give a single post-ILC map, $T_{\rm ILC}(\hat{n})$. 
More detail about this algorithm is given in Appendix~\ref{sec:needlet_ilc_appendix}.

\begin{figure}
    \centering
    \includegraphics[width=\linewidth]{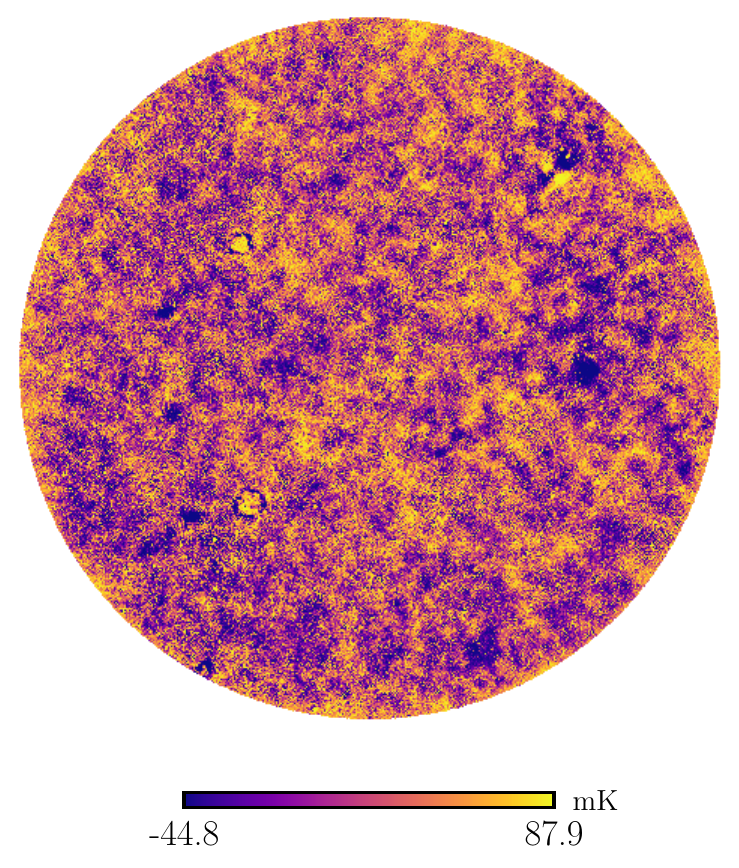}
    \caption{Post-ILC map used in our fiducial analysis in the Gnomonic projection. This map has been enlarged to only show the entire \qty{20}{\degree} patch of the sky that we leave unmasked. Its magnitude is normalized to the dark photon signal at $\omega_0/2\pi=\qty{410}{\mega\hertz}$.~\nblink{ILC_image}}
    \label{fig:post_ilc_map}
\end{figure}

With $T_{\rm ILC}(\hat{n})$, we can now compute $C_\ell^{\rm TT, obs}$ and $\xi^{\rm TT, obs}$, which we will later compare to the theoretical signal. 
We compute the autopower spectrum $C_\ell^{\rm TT, obs}$, used in our environment A analysis, with \textsc{pymaster}~\cite{Alonso:2018jzx,Garcia-Garcia:2019bku,Nicola:2020lhi}.
The \textsc{pymaster} code accounts for mixing between different multipoles introduced by the fact that we mask out most of the sky, ensuring that $C_\ell^{\rm TT, obs}$ is accurate.
To understand the need for this step, one can intuitively understand a mask in pixel space as a convolution between the true power spectrum and the power spectrum of the mask in harmonic space.
This creates an unphysical coupling of different multipoles, which must be algorithmically removed \cite{Alonso:2018jzx}.  
Our fiducial $C_\ell^{\rm TT, obs}$, obtained using the NILC procedure applied to the SKA foreground-only maps, is shown in orange in the left panel of Fig.~1 of \citetalias{Baker:2025zhf}. 
Note that we bin our $C_\ell^{\rm TT, obs}$ with $\Delta \ell = 150$ in our analysis.

An additional foreground processing step is needed to compute $\xi^{\rm TT, obs}$, which is used in environments B and C. 
Since the angular resolution of the \textsc{21cmFAST} simulation is much finer than the \qty{2.6}{\arcminute} resolution of the mock foreground maps, we interpolate the post-ILC map onto a grid with the finer resolution from the \textsc{21cmFAST} simulation. 
This is not strictly what would be observed in a realistic experiment, but it does allow us to estimate the impact of foregrounds without a prohibitive computational cost. 
To limit the impact of this interpolation, we do not consider any correlations below \qty{2.6}{\arcminute} in our analysis. 
Furthermore, we do not consider correlations on scales above \qty{63}{\arcminute}, i.e. half of the simulation box size, since there are very few points included in the computation of the correlation function.
We bin the remaining range of angles with 20 logarithmically spaced bins. 
Then, we compute $\xi^{\rm TT, obs}$ of the fine-resolution map using the default scalar-field correlation function estimator of \textsc{treecorr}; details of this procedure can be found in Sec.~\ref{sec:stats}. 
The final result for $\xi^{\rm TT, obs}$ is shown in the left panel of Fig.~\ref{fig:21cmfast_corrs}, which can then be compared to the signal predictions described in Sec.~\ref{sec:signal_modeling} to estimate the sensitivity to $\epsilon$. 

We validate this entire foreground generation and ILC procedure by comparing our autocorrelation results with the results from Ref.~\cite{McCarthy:2024ozh}.
First, we process real \textit{Planck} maps with the ILC pipeline and find that the results agree extremely well with Ref.~\cite{McCarthy:2024ozh}.
Next, we generate mock \textit{Planck} maps using the same \textsc{pysm3} models as in our fiducial analysis and then process these mock maps with the ILC pipeline. 
We also find good agreement with Ref.~\cite{McCarthy:2024ozh}, validating the entire mock foreground procedure.
More detail about our validation procedure is given in~\citetalias{Baker:2025zhf}.

\subsection{Mock Galaxy Catalogs}
\label{sec:galaxy_modeling}

We also generate mock galaxy catalogs in order to obtain a mock radio-galaxy cross-correlation measurement that we would observe with no $\gamma \to A'$ conversions, in order to estimate the sensitivity of such an analysis to the $\gamma \to A'$ signal.

In environment A, we build a mock catalog based on the halo occupation distribution (HOD) model in Ref.~\cite{Kusiak:2022xkt}, which is also the same HOD model used to compute the galaxy multipole kernel $u_\ell^g$ found in the predicted angular cross-power spectrum $C_\ell^{\rm A, Tg}$, given in Eqs.~\eqref{eq:Cl_A_Tg_1h} and~\eqref{eq:Cl_A_Tg_2h}. 
The output of this model is the autopower spectrum of the overdensity of galaxies $C_\ell^{\rm gg}$, from which we draw a mock galaxy map using the \texttt{synfast} routine of \textsc{healpy}~\cite{Zonca:2019vzt}. 
We then cross-correlate this mock galaxy map with the post-ILC map to get $C_\ell^{\rm Tg, obs}$, again using \textsc{pymaster} to account for mask effects.
In our analysis, $C_\ell^{\rm Tg, obs}$ is shown in the right panel of Fig.~1 of \citetalias{Baker:2025zhf} and is consistent with zero across all multipoles, since by construction, there is no correlation between our mock radio maps and the galaxy maps generated here. 
In actual data, some care must be taken, since the galaxies in the galaxy survey may also be radio emitters, which would introduce a positive correlation between radio maps and the galaxy catalog. 
Such an effect was reported in the search for $\gamma \to A'$ anisotropies in \textit{Planck} data in cross-correlation with the \textit{unWISE} galaxy catalog, and had to be subtracted off~\cite{McCarthy:2024ozh}. 
We comment briefly on this point in Sec.~\ref{sec:stats}. 

The HOD model that we adopt, which we refer to as ``K22,'' describes galaxies in a subsample of the \textit{unWISE} survey, known as the ``blue'' sample~\cite{Krolewski:2019yrv,Krolewski:2021yqy}.
This sample includes galaxies with a mean redshift of 0.6 over a redshift range of $0.3 \lesssim z \lesssim 0.9$. 
The model in Ref.~\cite{Kusiak:2022xkt} was fit to this sample using the autocorrelation of the sample and the cross-correlation of this sample with \textit{Planck} CMB lensing measurements. 
Although the minimum angular scale in this fit is \qty{10}{\arcminute}, we use the model to numerically compute $C_\ell^{\rm gg}$ down to scales of \qty{2.6}{\arcminute}, the resolution of our mock maps. 
In Ref.~\cite{Kusiak:2023hrz}, the authors fit the same HOD model to data at smaller angular scales; in principle, this latter HOD model (which we call ``K23'') would therefore be more appropriate in this study.
However, it was argued in a similar context to what we consider here that the K22 model is more physically robust~\cite{Goldstein:2024mfp}.
Moreover, in Ref.~\cite{McCarthy:2024ozh}, the authors derived constraints on $\epsilon$ using both models and found that their final results were largely insensitive to the HOD choice.
We also explore the impact of this modeling choice in Appendix~\ref{sec:halo_modeling_choices} and find that our final results are insensitive to this choice.
Therefore, we use the K22 model in our fiducial analysis.  

In environment B, our mock catalog is identical to the galaxy catalog constructed in Sec.~\ref{sec:envB} for the predicted signal using \textsc{21cmFAST}. 
There, we described how the theoretical prediction $\xi^{\rm B, Tg}$ is computed by cross-correlating the simulated map of $-T_{\gamma, 0}P_{\gamma \to A'}$ and this mock \textit{Roman} catalog.
For the mock observation, we instead cross-correlate the mock \textit{Roman} catalog with the post-ILC, foreground-only map to obtain $\xi^{\rm Tg, obs}$.
Once again, $\xi^{\rm Tg, obs}$ and the covariance between different angles is computed using the default estimator in \textsc{treecorr}, a procedure we describe in more detail in Sec.~\ref{sec:stats}. 
The final cross-correlation function is shown as red points with error bars in the right panel of Fig.~\ref{fig:21cmfast_corrs}; we can see that it is consistent with zero across all angles, since there is no correlation between the foregrounds and the mock galaxy catalog by construction. 

\section{Statistical Analysis and Results}
\label{sec:stats}

We finally forecast the sensitivity of SKA and galaxy surveys by comparing the theoretical $C_\ell^{\rm TX}$ and $\xi^{\rm TX}$ with the observed $C_\ell^{\rm TX, obs}$ and $\xi^{\rm TX, obs}$, which are computed from the post-ILC, foreground-only map and our mock galaxy survey. 

For environment A, for a fixed value of $m_{A'}$, the likelihoods are 
\begin{widetext}
    \begin{align}
        -2\ln \mathcal{L_{\rm A, TT}}(\epsilon^4) &= (C_\ell^{\rm A, TT}(\epsilon^4) - C_\ell^{\rm TT, obs}) \left(\mathcal{C}^{A, TT}_{\ell \ell'}\right)^{-1} (C_{\ell'}^{\rm A, TT}(\epsilon^4) - C_{\ell'}^{\rm TT, obs})\, , \\
        -2\ln \mathcal{L_{\rm A, Tg}}(\epsilon^2) &= ( C_\ell^{\rm A, Tg}(\epsilon^2) - C_\ell^{\rm Tg, obs}) \left(\mathcal{C}^{A, Tg}_{\ell \ell'}\right)^{-1} (C_{\ell'}^{\rm A, Tg}(\epsilon^2) - C_{\ell'}^{\rm Tg, obs})\, ,
    \end{align}
\end{widetext}
where repeated indices are summed over. 
Note that $C_\ell^{\rm A, TT} \propto \epsilon^4$ and $C_\ell^{\rm A, Tg} \propto \epsilon^2$, as described in Sec.~\ref{sec:signal_modeling}.
$\mathcal{C}_{\ell \ell'}$ is the Gaussian covariance between different $\ell$ bins of the mock-observed power spectrum, also computed using \textsc{pymaster}.
The Gaussian covariance of the power spectrum between two fields $X$ and $Y$ (where $X$ and $Y$ are either $\mathrm{T}$ or $\mathrm{g}$) is given by 
\begin{equation}
    \mathcal{C}^{XY}_{\ell \ell'} = \delta_{\ell \ell'} \frac{C_\ell^{XY} C_\ell^{XY}+ C_{\ell}^{XX}C_\ell^{YY}}{2\ell+1} \, .
\end{equation}
However, if the fields are masked, a more sophisticated approach is needed because the mask introduces mode-mixing effects. 
This mode mixing in turn causes non-diagonal elements to appear in the covariance matrix. 
To handle this complication, we use the pseudo-$C_\ell$ framework of \textsc{pymaster}~\cite{Garcia-Garcia:2019bku,Nicola:2020lhi}. 

We compute the likelihood for $200 \leq \ell \leq 4096$ using bins of $\Delta\ell = 150$. The lower limit is a conservative choice set such that we can be confident in the Limber approximation (which was used to compute $C_\ell^{\rm A, Tg}$)~\cite{LoVerde:2008re}. 
The upper limit is set at $2N_{\rm side}$, the recommended maximum $\ell$ for the resolution of our mock radio maps~\cite{Sullivan:2024jim}. 

The likelihoods for environments B and C are similarly
\begin{widetext}
    \begin{align}
        -2\ln \mathcal{L_{\rm B,TT}}(\epsilon^4) &= (\xi_i^{\rm B, TT}(\epsilon^4) - \xi_i^{\rm TT, obs}) \left(\mathcal{C}^{B,TT}_{ij}\right)^{-1} (\xi_j^{\rm B,TT}(\epsilon^4) - \xi_j^{\rm TT, obs}) \, , \label{eq:Ba_like} \\
        -2\ln \mathcal{L_{\rm B+C, TT}}(\epsilon^4) &= (\xi_i^{\rm B+C, TT}(\epsilon^4) - \xi_i^{\rm TT, obs}) \left(\mathcal{C}^{B+C,TT}_{ij}\right)^{-1} (\xi_j^{\rm B+C, TT}(\epsilon^4) - \xi_j^{\rm TT, obs})\, , \label{eq:BCa_like}\\
        -2\ln \mathcal{L_{\rm B, Tg}}(\epsilon^2) &= (\xi_i^{\rm B, Tg}(\epsilon^2) - \xi_i^{\rm Tg, obs}) \left(\mathcal{C}^{B, Tg}_{ij}\right)^{-1} (\xi_j^{\rm B, Tg}(\epsilon^2) - \xi_j^{\rm Tg, obs})\, , \label{eq:Bc_like}
    \end{align}
\end{widetext}
where $i,j$ index the $\theta$ bins at which $\xi$ is computed, and repeated indices are summed over. 
Once again, all theoretical predictions for autocorrelation scale as $\epsilon^4$ and cross-power correlations scale as $\epsilon^2$.
$\xi_i^{\rm B+C, TT}=\xi_i^{\rm B,TT} + \xi_i^{\rm C, TT}$, since the two environments are independent of each other.
We consider the sum of the two signals since they include $\gamma \to A'$ conversions over the same range of $m_{A'}$. 
The covariance matrices $\mathcal{C}_{ij}$ here encapsulate the correlation between different angular bins of the mock-observed correlation functions. 

We compute the correlation functions and estimate their covariances using the default \textsc{treecorr} estimators and the bootstrap resampling method. 
In the bootstrap estimate, we first divide the map into $N_{\rm patch}=50$ patches and resample $N_{\rm patch}$ of these patches with replacement.
Then, we compute the correlation function using this resampled field, with each patch weighted by the number of times that it appears.
We then repeat this process 500 times.
This bootstrap process then provides an estimate of the true covariance by resampling the original dataset many times to estimate the statistics of the true underlying distribution \cite{Mohammad:2021aqc}. 
Plots of our observed and predicted correlation functions for environments B and C are shown in Figs.~\ref{fig:21cmfast_corrs} and~\ref{fig:analytic_xi}, respectively, where the error bars are the square root of the diagonal entries of the covariance matrix. 
Note that $\xi^{\rm TT, obs}$ is the same in both figures. 

From the relevant likelihoods, we construct the test statistic $\lambda$ based on the likelihood ratio~\cite{Cranmer:2014lly}
\begin{equation}
    \lambda(\epsilon) \equiv \begin{cases} 
        1 \,, & \epsilon < \hat{\epsilon} \,, \\
        \mathcal{L}(\epsilon) / \mathcal{L}(\hat{\epsilon}) \,, & \epsilon \geq \hat{\epsilon} \, ,
    \end{cases}
\end{equation}
where $\hat{\epsilon}$ maximizes the likelihood. 
The 95\% confidence upper limit on $\epsilon$ is then found from a $\chi^2$ distribution with 1 degree of freedom by computing where $-2\ln \lambda(\epsilon)=2.71$~\cite{ParticleDataGroup:2024cfk}. 

For models B and C, one final step is needed. 
By the nature of the bootstrap covariance estimation process, there is some variation in the calculation of $\mathcal{C}_{ij}$, which enters into the likelihoods in Eqs.~\eqref{eq:Ba_like}--\eqref{eq:Bc_like}. 
Therefore, the inferred upper limits on $\epsilon$ can also vary upon repeated calculation of $\mathcal{C}_{ij}$.
To account for this variation, we compute $\xi$ and the covariance 25 times from the same map and compute an upper limit for $\epsilon$ using the likelihood ratio.
Then, we take the median of these limits and report the 90\% containment band around this median to demonstrate the spread in the resulting limit. 

\begin{figure*}[t!]
    \centering
    \includegraphics[width=0.6\linewidth]{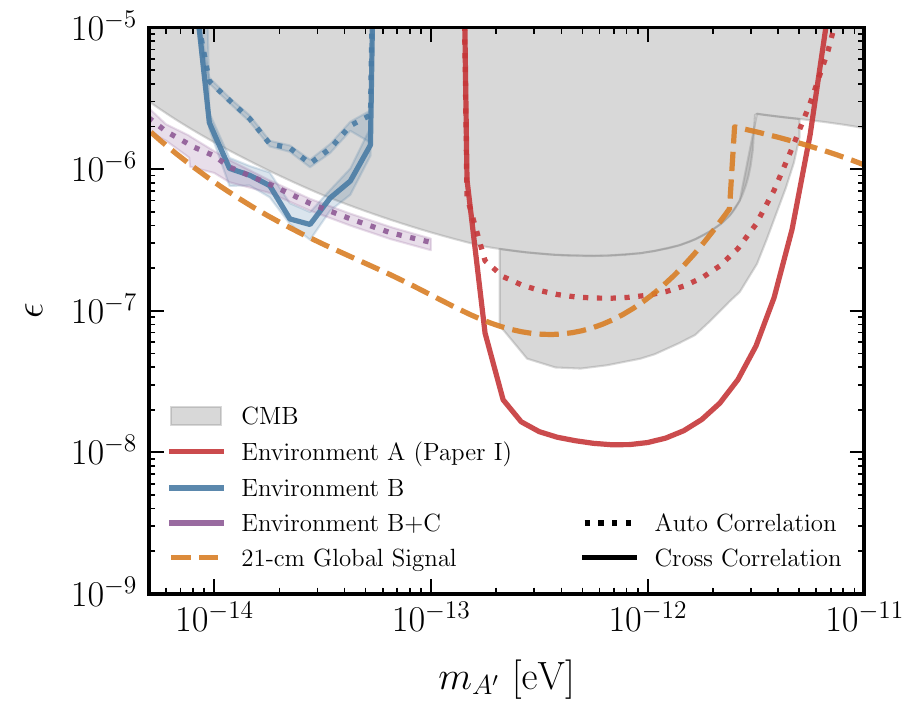}
    \caption{Forecasted sensitivities to the dark photon kinetic mixing parameter $\epsilon$ expected using autocorrelation (dotted lines) and SKA in cross-correlation with various galaxy surveys (solid lines), including \textit{(i)} environment B $\gamma \to A'$ conversions in the IGM during the EoR, using the SKA autocorrelation function (blue, dotted) and the cross-correlation function between SKA and the \textit{Roman} galaxy survey (blue, solid); \textit{(ii)} the estimated limit from environment C $\gamma \to A'$ conversions in the late Universe IGM, using the SKA autocorrelation function (purple, dotted); and \textit{(iii)} environment A $\gamma \to A'$ conversions in dark matter halos, using the SKA autopower spectrum (red, dotted) and the cross-power spectrum between SKA and a low-redshift galaxy survey (red, solid) from \citetalias{Baker:2025zhf};
    90\% expected limit bands are shown for limits due to conversions in environments B and C. 
    Also shown is the estimated sensitivity of a 21-cm global signal experiment to $\gamma \to A'$ conversions (orange, dashed) by determining the minimum value of $\epsilon$ such that $T_{\gamma, 0}P_{\gamma \to A'}\gtrsim \qty{100}{\milli\kelvin}$. Existing limits from spectral distortions of the CMB~\cite{Caputo:2020bdy} and the anisotropy power spectrum~\cite{McCarthy:2024ozh} are shaded in gray.~\nblink{Limits}}
    \label{fig:ska_lims}
\end{figure*}

Using this statistical procedure, we project that SKA will be highly sensitive to $\gamma \to A'$ conversions, as shown in Fig.~\ref{fig:ska_lims}, demonstrating the power of radio telescopes to search for dark photons. 
Our results show that SKA will be most sensitive to conversions in environment A when cross-correlated with a low-redshift galaxy survey, improving on CMB limits \cite{McCarthy:2024ozh} by a factor of 4; this result is discussed in greater detail in \citetalias{Baker:2025zhf}. 

Additionally, we find that SKA is likely to have leading sensitivity to dark photons with $10^{-14} \lesssim m_{A'} \lesssim 4 \times 10 ^{-14}$, which undergo conversions in environment B.
When cross-correlated with the \textit{Roman} galaxy survey, SKA  will be sensitive to $\epsilon$ as low as $5\times 10^{-7}$ over this mass range, a sensitivity that slightly exceeds current limits from spectral distortions of the cosmic microwave background. 
Finally, we find that SKA is likely to be highly sensitive to $\gamma \to A'$ conversions in environment C from our preliminary estimate, which is relevant for $5\times 10^{-15} \lesssim m_{A'} \lesssim 10^{-13}$. 
Again, we caution that this signal was estimated using a simple Ansatz for the distribution of $m_{\gamma}^2$; however, this result motivates the use of a hydrodynamical simulation to better model $\gamma \to A'$ conversions in the late Universe IGM.

The major improvement in SKA's performance is due to two main effects.
First, SKA will operate with much smaller beams than the CMB, meaning it will have much more information about small-scale physics.
This new information greatly improves SKA's sensitivity to $\gamma \to A'$ conversions.
We provide more detail about this point in~\citetalias{Baker:2025zhf}.
A second, related advantage is improved point source subtraction at SKA. 
As noted in Sec.~\ref{sec:extragalactic_foregrounds}, radio telescopes can resolve much fainter radio sources than \textit{Planck}. 
For example, LoTSS Deep resolved sources with flux densities as low as \qty{0.05}{\milli\jansky}, whereas the faintest sources resolved in the \textit{Planck} Catalog of Compact Sources have flux densities of $\sim \qty{100}{\milli\jansky}$~\cite{Planck:2015bin}. 
SKA is expected to improve on LoTSS's performance, and we find that this improved point source subtraction modestly improves SKA's sensitivity as well. 

For environment C, we argued in Sec.~\ref{sec:envC} that the angular correlation function of the $\gamma \to A'$ signal in environment C, $\xi^{\rm C, TT}$ was a more natural observable.
However, the theoretical prediction was made in the form of the angular power spectrum $C_\ell^{\rm C,TT}$, and so we could also have performed a likelihood analysis with $C_\ell^{\rm TT,obs}$. 
We find $\mathcal{O}(1)$ agreement between these two methods in terms of the final sensitivity to $\epsilon$, despite the drastic differences in how the analyses proceed.

As we have noted several times in this paper so far, our fiducial model for mock observations cannot capture any potential cross-correlation between galaxies in a galaxy survey and unresolved extragalactic point sources that would contribute to the foreground, since galaxies themselves source radio emission. 
In a realistic experiment, this could potentially show up as a positive cross-correlation between the post-ILC map and the galaxy overdensity map. 
If this bias is not removed in real data, we would tend to overestimate the limit on dark photons, since the signal is a \textit{negative} cross-correlation, and so it is important to remove it when obtaining an actual limit.
In Ref.~\cite{McCarthy:2024ozh}, for example, the authors found that this same correlation biased their final results slightly, and would have led to overestimated limits on $\epsilon$ from a cross-correlation between \textit{Planck} and \textit{unWISE} data. 
However, Ref.~\cite{McCarthy:2024ozh} also demonstrated that the main impact was at large angular scales, and showed that it was possible to successfully eliminate this bias by estimating it from cross-correlations between actual radio maps and galaxy catalogs. 
We also note that our results correspond to the successful removal of this bias and therefore do not overestimate the derived sensitivity to $\epsilon$. 
In Appendix~\ref{sec:radio_bias}, we also perform a crude estimate of the magnitude of this bias and find that it corresponds roughly to the magnitude of a dark photon signal for $m_{A'}\sim \qty{5e-13}{\eV}$ and $\epsilon \sim 5\times 10^{-8}$. 
Therefore, it is highly likely that even after the removal of this bias, the sensitivity of SKA will exceed current constraints.

Beyond this, following Refs.~\cite{McCarthy:2023hpa,McCarthy:2024ozh}, one might also consider explicitly deprojecting all other foregrounds in real data, a more conservative approach that adjusts the ILC weights to explicitly remove known foregrounds from the post-ILC map while optimally preserving the signal. 
Performing such a procedure could modestly weaken our results.

In any case, since our analysis shows that a cross-correlation between SKA and low-redshift galaxies will improve on the CMB limits on $\epsilon$ by a factor of 4, both deprojection and the radio bias subtraction would need to reduce the precision of the observed cross-correlation by a factor of 16, in order for radio surveys to not be competitive with CMB searches. 
We therefore have reason to be optimistic about our analysis being competitive with CMB limits. 

\section{\texorpdfstring{$\gamma \to A'$ Conversions in 21-cm Experiments}{ Gamma to A' Conversions in 21-cm Experiments}}
\label{sec:21cm_expts}

The goal of 21-cm cosmology is to measure the spectral distortion to the CMB backlight due to the absorption or emission of 21-cm photons from neutral hydrogen hyperfine transitions during the EoR. 
Since $\gamma \to A'$ conversions would act as an additional source of distortion, it is natural to consider the outlook of the 21-cm signal for studying $\gamma \to A'$ conversions. 

The main observable is the brightness temperature of 21-cm photons relative to the CMB backlight, which for simplicity we can think of as being a function of comoving position $\vec{x}$ away from the observer, with corresponding redshift $z$, as follows:
\begin{equation}
    T_{21}(\vec{x}) \simeq \frac{1}{1+z}[T_S(\vec{x})-T_{\gamma}(z)]\left[1-e^{-\tau(\vec{x})}\right] \, ,
\end{equation}
where $T_S(\vec{x})$ and $\tau(\vec{x})$ are the spin temperature of the neutral hydrogen gas and the optical depth of 21-cm photons at $\vec{x}$ respectively~\cite{Barkana:2022hko}. 
The brightness temperature at points corresponding to redshift $z$ is measured by observing in the radio band at energy $\omega(z) = \omega_{21}/(1+z)$, where $\omega_{21}$ is the energy of the 21-cm transition. 
$T_S$ is set by interactions between HI atoms with other atoms and surrounding CMB and Lyman-$\alpha$ photons, and can be thought of here as characterizing the thermal properties of the gas~\cite{Pritchard:2011xb,Barkana:2016nyr}.

Experiments seek to measure either the global (sky-averaged) brightness temperature as a function of redshift~\cite{Monsalve:2016xbk,Bowman:2018yin,zotero-item-2534,Singh:2017syr} or the power spectrum of $T_{21}$~\cite{DeBoer:2016tnn,Bowman:2012ef,LOFAR:2013jil,Braun:2019gdo}. 
$\gamma \to A'$ conversions impact both kinds of signals. 
In $\Lambda$CDM cosmology, $T_S \simeq T_k$, the baryon kinetic temperature, once the first stars formed and started emitting Lyman-$\alpha$ photons, and therefore $T_{21}^{\Lambda \rm CDM}<0$ during cosmic dawn, prior to significant photoheating occurring.  
$\gamma\to A'$ conversions also lead to a disappearance of photons relative to the CMB backlight, appearing to the observer as a negative contribution to $T_{21}$ as well. 
However, unlike the 21-cm signal, where the value of $T_{21}(\vec{x})$ depends on the local properties of the gas at $\vec{x}$, the $\gamma \to A'$ contribution depends on the integrated conversion probability along the entire line of sight, $P^{\rm tot}_{\gamma \to A'}(\hat{x})$. 
Assuming that $P^{\rm tot}_{\gamma \to A'}(\hat{x}) \ll 1$, i.e.\ that the CMB backlight brightness temperature is always approximately $T_{\gamma,0}$, the observed 21-cm brightness temperature is, to first order, the sum of these two effects:
\begin{equation}
    T_{21}(\vec{x}) = T_{21}^{\rm \Lambda CDM}(\vec{x}) - T_{\gamma,0} P_{\gamma\to A'}^{\rm tot}(\hat{x}) \, .
\end{equation}
The sky-averaged signal is then 
\begin{equation}
    \label{eq:21cm_global}
    \langle T_{21} \rangle(z) = \langle T_{21}^{\rm \Lambda CDM} \rangle(z) - T_{\gamma,0}\langle P_{\gamma\to A'}^{\rm tot}\rangle \, .
\end{equation}
This is the $\gamma \to A'$ signature observable in a 21-cm global signal measurement. 
We first discuss the details related to the global signal in the next section, before commenting on the 21-cm power spectrum signal from $\gamma \to A'$ conversions.

\subsection{21-cm Global Signal}
\label{sec:21cm_global}

We begin by estimating the expected size of the $\gamma \to A'$ signal in the global 21-cm signal. 
To simplify matters, let us assume that $m_{\gamma}^2$ is homogeneous and equal to its cosmic mean value $\overline{m_{\gamma}^2}(z)$ at all redshifts, and only consider conversions occurring after reionization is complete, where $\overline{m_\gamma^2}(z)$ decreases monotonically. 
In this limit, the sky-averaged probability of conversion can be estimated from Eq.~\eqref{eq:conversion_probability} as
\begin{alignat}{1}
    \langle P_{\gamma \to A'}^{\rm tot} \rangle \sim \frac{\pi m_{A'}^2 \epsilon^2}{\omega_{21} (1 + z_{\rm obs})^{-1} (1 + z_*)^2 H(z_*)} \left| \frac{d \ln \overline{m_\gamma^2}}{dz} \right|^{-1}_{z=z_*} \!\!\!\!\!\,, 
\end{alignat}
where $m_{A'}^2 = 4 \pi \alpha_{\rm EM} n_{b,0} (1 + z_*)^3 / m_e$ defines the resonance redshift $z_*$, and $\omega_{21} (1 + z_{\rm obs})^{-1}$ is the present-day photon energy of a 21-cm photon at redshift $z_{\rm obs}$. 
At $z_{\rm obs} = 17$, the 21-cm brightness temperature is roughly on the order of \qty{-100}{\milli\kelvin} in $\Lambda$CDM cosmology. 
We can therefore obtain a rough estimate of the sensitivity by setting $T_{\gamma,0} \langle P_{\gamma \to A'}^{\rm tot} \rangle \gtrsim \qty{100}{\milli\kelvin}$, $z_{\rm obs} = 17$ and assuming matter domination, giving
\begin{alignat}{1}
    \epsilon \lesssim 2 \times 10^{-7} \left(\frac{\qty{5e-14}{\eV}}{m_{A'}}\right)^{1/6} \,.
    \label{eq:global_signal_estimate}
\end{alignat}
This simple estimate suggests that the 21-cm global signal can be highly sensitive to $\gamma \to A'$ conversions in the IGM after reionization. 

To obtain the expected signal more accurately, we numerically model the $\gamma \to A'$ conversions that impact the 21-cm global signal using the formalism described in Appendix~\ref{sec:mean_prob_analytic}, following the work of Refs.~\cite{Caputo:2020bdy,Caputo:2020rnx} closely. 
References~\cite{Caputo:2020bdy,Caputo:2020rnx} showed that the sky-averaged probability of conversion $\langle P_{\gamma \to A'} \rangle$ could be analytically computed, given the one-point PDF of $m_{\gamma}^2$ as a function of $z$. 
Following Ref.~\cite{Caputo:2020rnx}, we assume that $m_\gamma^2$ follows a lognormal distribution given by Eq.~\eqref{eq:lognormal_mgamma_pdf} in the redshift range $0.005 \lesssim z \lesssim 375$, with standard deviation taken from a fit of the nonlinear baryon power spectrum using hydrodynamical simulations. 
We ignore conversions from $5\lesssim z \lesssim 35$ because of the uncertainties associated with reionization. 
For $375 \lesssim z \lesssim 1000$, we take $f_1(m_{\gamma}^2;z) = \delta_D(m_{\gamma}^2 - \overline{m_{\gamma}^2}(z))$, since the Universe is essentially homogeneous at high redshifts; this simplifies the numerical integration required.  

\begin{figure}[t!]
    \centering
    \includegraphics[width=\linewidth]{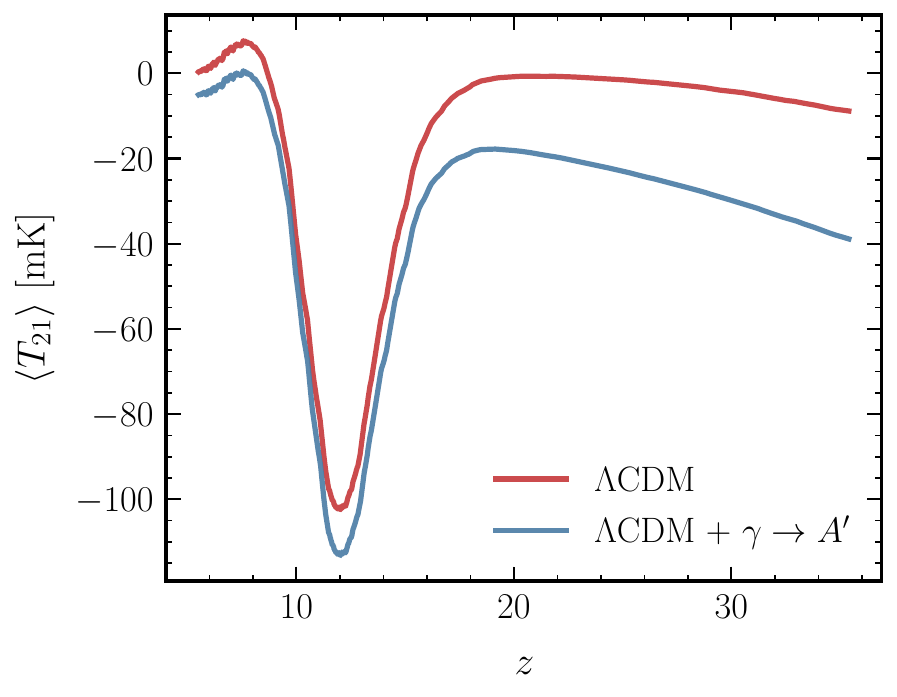}
    \caption{Sky-averaged 21-cm brightness temperature $\langle T_{21}\rangle$ in $\Lambda$CDM (red), and including the effects of $\gamma \to A'$ conversions (blue). Including the effects of dark photons causes a characteristic reduction in $\langle T_{21}\rangle$, which scales as $(1+z)$. Here, the dark photon signal is shown for $m_{A'}=\qty{3e-14}{\eV}$ and $\epsilon=10^{-6}$.~\nblink{Global_Signal}}
    \label{fig:global_signal}
\end{figure}

Figure~\ref{fig:global_signal} shows the global signal for a fiducial choice of astrophysical parameters expected in $\Lambda$CDM cosmology, and the modified signal including the effects of $\gamma \to A'$ conversions for $m_{A'}=\qty{3e-14}{\eV}$ and $\epsilon=10^{-6}$. The increasingly large effect at large $z$ is due to the fact that the probability of conversion scales as $1/\omega$. 

To forecast the sensitivity to dark photons from a 21-cm global signal measurement, we determine the value of $\epsilon$ such that $T_{\gamma, 0}P_{\gamma\to A'}^{\rm tot} > \qty{100}{\milli\kelvin}$ at $z=17$. 
This corresponds to the dark photon signal alone making up approximately two-thirds of the signal expected in $\Lambda$CDM cosmology at this redshift.
Figure~\ref{fig:ska_lims} shows this result as the orange dashed line, which broadly agrees with our rough estimate in Eq.~\eqref{eq:global_signal_estimate} at $m_{A'} \sim \qty{5e-14}{\eV}$, and demonstrates that the 21-cm global signal could be highly sensitive to $\gamma \to A'$ conversions. 
We caution, however, that the exact limit from a global signal experiment would depend heavily on the details of foreground removal, since the $1/\omega$ dependence of the dark photon signal could easily be removed along with conventional foregrounds, which are typically modeled as a smooth function of frequency. 
We leave a detailed analysis of the actual limits from a global signal experiment to future work. 

\subsection{21-cm Power Spectrum}

Given that most of the paper has been dedicated to the detection of fluctuations in the radio sky, it is perhaps surprising that we have not mentioned the 21-cm power spectrum signal at all. 
That is because, although there is a $\gamma \to A'$ signal in the 21-cm power spectrum, the experimental details of 21-cm interferometers such as HERA limit their ability to detect this signal. 
To see why, recall that as a function of frequency, the $\gamma \to A'$ signal only differs by the scaling $1/\omega$. 
The $\gamma \to A'$ signal can therefore be thought of as ``inverse point sources,'' i.e. dark patches on the sky that are highly correlated along the line of sight. 
This is conceptually similar to point sources, a major foreground for 21-cm experiments: because of the smooth spectral dependence of both the dark photon signal and point sources, their contributions to the 21-cm signal are confined to the smallest $k_\parallel$ modes, as opposed to the 21-cm signal, which is expected to exhibit large fluctuations as a function of redshift, and therefore to have power in larger $k_\parallel$ modes. 
Experimental effects additionally cause the foregrounds, and therefore also the $\gamma \to A'$ signal, to leak into higher $k_\parallel$ modes, especially at large $k_\perp$. 

This point substantially limits the ability of 21-cm interferometry experiments such as HERA to search for dark photons. 
Because of this leakage effect, experiments such as HERA excise the so-called ``wedge,'' which is the region of the 2D power spectrum where $k_\parallel \leq m k_\perp + a$ where $m$ and $a$ are set by the experimental details \cite{Liu:2019awk}. 
This approach is advantageous for these experiments since the wedge is exactly the region of the $k_\parallel$-$k_\perp$ plane that is highly contaminated by foregrounds and does not contain the 21-cm signal.
However, the wedge is also the region of the power spectrum in which our signal is concentrated.
Therefore, it appears difficult to search for the $\gamma \to A'$ signal using these experiments because the most relevant portion of the data is removed before further analysis.
However, the data in the wedge are still recorded, and it is possible that with a clever analysis strategy, it could still be possible to extract the $\gamma \to A'$ signal from e.g.\ HERA data. 
We leave an exploration of such a possibility to future work. 

\section{Conclusion}
\label{sec:conclusion}
In this paper, we have provided a detailed explanation of the modeling necessary to study $\gamma \to A'$ conversions in the radio sky as well as the analysis pipeline that we use to forecast the sensitivity of SKA to $\gamma \to A'$ conversions. 
This modeling falls into two categories: modeling of the signal and of mock observations.
First, we considered three models for the dark photon signal. 
In environment A, we adopted the methods in Refs.~\cite{Pirvu:2023lch,McCarthy:2024ozh} to consider $\gamma \to A'$ conversions in dark matter halos in the late Universe and the correlation of these conversions with the \textit{unWISE} galaxy survey.
We simplify the derivation of the power spectra and correct an error in the halo mass function that was used in those works.
This environment is the focus of~\citetalias{Baker:2025zhf}.

In environment B, we use the semianalytic code \textsc{21cmFAST} to compute the signal due to $\gamma \to A'$ conversions in the IGM during the EoR. 
Additionally, we use this code to model high-redshift galaxies that are observable with the upcoming High Latitude Survey on the Nancy Grace Roman Space Telescope. 

Finally, in environment C, we apply the formalism developed in Refs.~\cite{Caputo:2020rnx,Caputo:2022keo} and apply a simple lognormal Ansatz for the PDF of $m_{\gamma}^2$ to estimate the $\gamma \to A'$ signal due to IGM conversions in the late Universe.

We also model the foregrounds that obscure the $\gamma \to A'$ signal. 
In radio frequencies, galactic synchrotron, free-free radiation, and spinning dust foregrounds, as well as extragalactic point source foregrounds are significant and complicate the search for any cosmological signal. 
Our forecast analysis pipeline demonstrates that radio experiments can overcome these challenges. 
In a real experiment, radio maps at different frequencies can be combined to extract the dark photon signal using the needlet ILC algorithm. 
This algorithm constructs the linear combination of the radio maps with the minimum variance while maintaining the strength of the dark photon signal.
This widely used algorithm enables experiments to extract signals that are orders of magnitude fainter than the foregrounds.

We use our pipeline to process our mock null-signal maps, which we then use to forecast upper limits on the sensitivity of SKA. 
In~\citetalias{Baker:2025zhf}, we find that SKA is most sensitive to conversions in environment A for $\qty{e-13}{\eV} \lesssim m_{A'}\lesssim \qty{5e-13}{\eV}$, when cross-correlated with a low-redshift galaxy survey (Fig.~\ref{fig:ska_lims}, red), finding a potential improvement over the $\textit{Planck}$-\textit{unWISE} cross-correlation from Ref.~\cite{McCarthy:2024ozh} by a factor of 4.
We found that this excellent sensitivity is largely due to the smaller beams used in SKA, which give access to more small-scale information and allow for better point source subtraction.
These sensitivities are shown in Fig.~\ref{fig:ska_lims} for the variety of models that we consider. 

In this paper, we also find that SKA cross-correlated with a high-redshift galaxy survey (environment B) has moderately better sensitivity to $\gamma \to A'$ conversions in the EoR IGM as compared to a search for $\gamma \to A'$ spectral distortions in FIRAS (Fig.~\ref{fig:ska_lims}, blue), at a lower mass range centered at \qty{2e-14}{\eV}.  
Finally, based on our preliminary estimate of the $\gamma \to A'$ signal in environment C, we find that SKA is likely to be sensitive to these conversions as well, slightly outperforming the FIRAS limits, and motivating a more detailed study of this signal using hydrodynamical simulations.

Beyond SKA, we also consider the outlook for observing $\gamma \to A'$ conversions in EoR 21-cm experiments.
We find that these conversions leave a strong signal in a 21-cm global signal experiment (Fig.~\ref{fig:ska_lims}, orange), provided it can be disentangled from other foregrounds.
We also note that, as detailed above, it appears to be much more difficult to search for $\gamma \to A'$ conversions in 21-cm power spectrum experiments. 

To conclude, in this work and~\citetalias{Baker:2025zhf}, we present the first search for $\gamma \to A'$ conversions in the radio sky. 
We find that upcoming radio experiments are highly sensitive to these conversions. 
Unique aspects of radio experiments enable this sensitivity and many of the arguments we have made here apply to other searches for new physics, highlighting the unique promise of upcoming radio experiments in the continuing search for physics beyond the Standard Model. 

\section{Acknowledgments}
We would like to thank Colin Hill, Junwu Huang, Cristina Mondino, and Julian Mu{\~n}oz for helpful conversations. The work in this paper makes extensive use of the \textsc{numpy}~\cite{Harris:2020xlr}, \textsc{scipy}~\cite{Virtanen:2019joe}, \textsc{matplotlib}~\cite{Hunter:2007ouj}, \textsc{astropy}~\cite{Astropy:2013muo,Astropy:2018wqo,Astropy:2022ucr}, \textsc{halomod}~\cite{Murray:2013qza,Murray:2020dcd}, \textsc{healpy}, and \textsc{healpix} packages~\cite{Gorski:2004by}. We are also pleased to acknowledge that the computational work reported on in this paper was performed on the Shared Computing Cluster which is administered by Boston University's Research Computing Services.
E.B. was supported by the Boston University Dean's Fellowship program. H.L. is supported by the U.S. Department of Energy under Grant No. DE-SC0026297, as well as the Cecile K. Dalton Career Development Professorship, endowed by Boston University trustee Nathaniel Dalton and Amy Gottlieb Dalton. 

\section{Data Availability}
The code and simulation data that support the findings of this article are available in Ref.~\cite{Baker_code_repo}.

\clearpage 

\appendix
\begin{widetext}

\section{The Halo Model}
\label{sec:halo_model_der}
In this appendix, we derive the autopower spectrum of $P_{\gamma\to A'}$ and its cross-correlation with galaxies described by the HOD model in Ref.~\cite{Kusiak:2022xkt}. 
This derivation proceeds in several steps. 
First, we derive the halo model predictions of the power spectra of two general cosmological fields that trace the distribution of halos,  $U_1(\chi \hat{n})$ and $U_2(\chi \hat{n})$, each defined at some comoving distance $\chi$ and direction $\hat{n}$. 
Since we are interested in the angular power spectrum of these observables, we will ultimately consider the integral of $U_i(\chi \hat{n})$ over the line of sight, as follows:
\begin{equation}
    \tilde{U}_i(\hat{n}) = \int d\chi \, U_i(\chi\hat{n})\, . 
\end{equation}
On our path to deriving the power spectra of these observables, we first compute the ensemble-averaged value of $\tilde{U}_i(\hat{n})$, denoted $\langle \tilde{U}_i \rangle$, and then consider fluctuations in these fields. 
From these fluctuations, we then derive a function known as the \textit{multipole kernel}, which completely defines the power spectrum.
Deriving this multipole kernel for our observables of interest will be the main focus of our modeling. 

After this derivation for general fields $U_i$, we will specify $U_i(\chi \hat{n})$ to be either $-T_{\gamma, 0}dP_{\gamma\to A'}/d\chi$ at a point $\chi\hat{n}$ or $d\delta_g/d\chi$, where $\delta_g$ is the galaxy overdensity field, and finally derive their associated multipole kernels and power spectra. 

To begin, we suppose that, inside each halo, we can write $U_1(\chi_i \hat{n}) = u_1(\hat{n}\cdot\hat{\alpha}; z,m)$ where the halo center is at $\chi \hat{\alpha}$.
In other words, we express $U_1$ as a function of the angle between the line of sight and the halo center and the halo mass $m$ and redshift. 
This follows from the spherical symmetry of the halo. 
The same expression is true for $U_2$, but we specify $U_1$ here for clarity. 

Next, we divide space at this redshift into sufficiently small chunks $\Delta V_i$, each containing $N_i$ halo centers where $N_i=0$ or 1. 
Then, the total value of $U_1(\chi_1\hat{n})$ is given by summing $u_1$ over all halos, as follows:
\begin{equation}
    \label{eq:p_per_los}
    U_1(\chi_1\hat{n}) = \sum_i N_i u_1(\hat{n}\cdot\hat{\alpha_i}; z_1, m_i) \, ,
\end{equation}
where $\hat{\alpha}_i$ is the direction to the center of the halo $i$, if $N_i = 1$. 

To compute the average value of $\tilde{U}_1$ over the entire sky, we first take the ensemble average of Eq.~\eqref{eq:p_per_los} and integrate over space.
In the halo model~\cite{vandenBosch:2012nq}, this ensemble average over halo mass, denoted by $\langle \cdots \rangle_m$, is defined by
\begin{equation}
    \langle N_i u_1\rangle_m (\hat{n}\cdot \hat{\alpha}_i, z) =\int dm \, n(z, m)\Delta V_i u_1(\hat{n}\cdot \hat{\alpha}_i; z, m_i) \, , \label{eq:halo_ensemble}
\end{equation}
where $n(z, m)$ is the halo mass function. 
This gives 
\begin{equation}
    \left\langle U_1(\hat{n}, z)\right\rangle_m = \int dm \, n(z, m) \sum_i \Delta V_i u_1(\hat{n}\cdot \hat{\alpha}_i; z, m_i) \, . 
\end{equation}
Then, we average over all lines of sight by replacing 
\begin{equation}
    \label{eq:halo_sum_to_int}
    \sum_i \Delta V_i \to \int d^3\chi \, ,
\end{equation}
which amounts to averaging over all angles and integrating the signal along each line of sight to get $\tilde{U}_1$ averaged over each line of sight, as follows:
\begin{equation}
     \langle \tilde{U}_1 \rangle = 2\pi \int d\chi \, \chi^2 \int d(\hat{n} \cdot \hat{\alpha})\, \int dm \, n(z,m) u_1(\hat{n}\cdot \hat{\alpha}; z, m) \, . 
\end{equation}
Here, we have assumed isotropy to perform the $\phi$ integral.

\subsection{Fluctuations in the Halo Model}
\label{sec:halo_auto_Cl}

Next, we compute the cross-correlation of the two tracers $U_1$ and $U_2$. 
First, consider $U_1(\chi_1 \hat{n})U_2(\chi_2\hat{n}')$. In the halo model, we have
\begin{equation}
    U_1(\chi_1\hat{n})U_2(\chi_2\hat{n}') =\sum_i N_i u_1(\hat{n}\cdot \hat{\alpha}_i; z_1, m_i) \sum_j N_j u_2(\hat{n}'\cdot \hat{\alpha}_j; z_2, m_j)\, ,
\end{equation}
where $\hat{\alpha}_i$ is the direction to the center of the halo $i$. 
In order to simplify this expression, we treat two cases separately: the one-halo term, where $i=j$, and the two-halo term, where $i\neq j$.

The one-halo term is
\begin{equation}
\left[U_1(\chi\hat{n}) U_2(\chi\hat{n}')\right]_{\rm 1h} = \sum_i N_i u_1(\hat{n}\cdot \hat{\alpha}_i; z_i, m_i) u_2(\hat{n}'\cdot \hat{\alpha}_i; z_i, m_i)\, ,
\end{equation}
where we have used that $N_i^2=N_i$ since it is either 0 or 1. Then, as we did for the sky-averaged signal, we take the ensemble average over mass using Eq.~\eqref{eq:halo_ensemble} to get 
\begin{equation}
\left\langle U_1 U_2\right\rangle_{m,\rm 1h} = \int dm \, n(z,m) \sum_i \Delta V_i u_1(\hat{n}\cdot\hat{\alpha}_i; z, m_i)  u_2(\hat{n}'\cdot\hat{\alpha}_i; z, m_i)\, .
\end{equation}

At this point, $\langle U_1 U_2 \rangle_{m, \rm 1h}$ is still a function of $\hat{n}$, $\hat{n}'$, and all the $\alpha_i$ for each halo.  
We can now compute the two-point cross-correlation function of the integrated tracers,
\begin{equation}
    \xi^{u_1 u_2, {\rm 1h}}(\hat{n} \cdot \hat{n}') \equiv \left\langle \tilde{U}_1(\hat{n})\tilde{U}_2(\hat{n}') \right\rangle
\end{equation}
by making the same replacement as Eq.~\eqref{eq:halo_sum_to_int} and integrating over all angles $\hat{\alpha}$, as follows:
\begin{equation}
    \xi^{u_1 u_2, {\rm 1h}}(\hat{n} \cdot \hat{n}') = \int d\chi\, \chi^2  \int dm\, n(m, z)\int d\Omega_\alpha \, u_1(\hat{n}\cdot \hat{\alpha}; z, m)  u_2(\hat{n}'\cdot \hat{\alpha}; z, m)\, .
\end{equation}
Note that $\xi^{u_1,u_2,\rm 1h}$ is only a function of $\hat{n}\cdot \hat{n}'$ due to isotropy.

To make further headway with this expression, we expand $u$ in terms of Legendre polynomials $P_\ell$, simplify the expression using spherical harmonic identities, and perform the angular integral, as follows:
\begin{align}
    \int d\Omega_\alpha \, u_1(\hat{n}\cdot \hat{\alpha}; z, m) u_2(\hat{n}'\cdot \hat{\alpha}; z, m) &= \int d\Omega_\alpha \, \sum_\ell u_{1,\ell} P_\ell(\hat{n}\cdot \hat{\alpha})
    \sum_{\ell'} u_{2,\ell'} P_{\ell'}(\hat{n}'\cdot \hat{\alpha}) \nonumber \\
    &= \int d\Omega_\alpha \, \sum_\ell u_{1,\ell} \sum_{m} \frac{4\pi}{2\ell+1}Y_{\ell m}(\hat{n})Y^*_{\ell m}(\hat{\alpha})
    \sum_{\ell'} u_{2, \ell'} \sum_{m'} \frac{4\pi}{2\ell'+1}Y_{\ell' m'}(\hat{n}')Y^*_{\ell' m'}(\hat{\alpha}) \nonumber \\
    &= \sum_\ell u_{1,\ell} u_{2,\ell}\sum_{m} \left(\frac{4\pi}{2\ell+1}\right)^2Y_{\ell m}(\hat{n})Y^*_{\ell m}(\hat{n}') \nonumber \\
    &= \sum_\ell u_{1, \ell}u_{2,\ell} \frac{4\pi}{2\ell+1} P_\ell (\hat{n}\cdot \hat{n'}) \, ,
\end{align}
where 
\begin{equation}
u_{i,\ell} = \frac{2\ell+1}{2}\int_{-1}^1d(\hat{n} \cdot \hat{\alpha})\, u_i(\hat{n} \cdot \hat{\alpha})P_\ell(\hat{n} \cdot \hat{\alpha}) \, ,
\end{equation}
which we define as the multipole kernel of $U_i$ that was mentioned above, and is a function of $z$ and $m$. 
The one-halo term of the correlation function is therefore
\begin{equation}
    \xi^{u_1, u_2,{\rm 1h}}(\hat{n}\cdot\hat{n}') =\int d\chi\, \chi^2 \int dm \, n(z,m) \sum_\ell u_{1, \ell}u_{2, \ell} \frac{4\pi}{2\ell+1} P_\ell (\hat{n}\cdot \hat{n}') \,,
\end{equation}
or in terms of the angular power spectrum, using Eq.~\eqref{eq:xi_and_Cl_relation},
\begin{equation}
    C_\ell^{u_1, u_2,{\rm 1h}} = \left(\frac{4\pi}{2\ell+1}\right)^2\int d\chi \, \chi^2 \int dm \, n(z,m) u_{1, \ell} u_{2, \ell} \, .
    \label{eq:gen_Cl_1halo}
\end{equation}
$C_\ell$ is determined entirely by the multipole kernel.
Therefore, to compute the angular power spectrum of some signal in the halo model, we must simply determine the multipole kernel of that signal.
This is computed in Appendix \ref{sec:dark_photon_signal_halo_model} following Ref.~\cite{Pirvu:2023lch} and the galaxy model from Ref.~\cite{Kusiak:2022xkt}.

Next, we use a similar approach to derive the two-halo term in terms of the multipole kernels. 
Once again, we have 
\begin{equation}
    U_1(\chi_1\hat{n})U_2(\chi_2\hat{n}') = \sum_i N_i u_1(\hat{n}\cdot \hat{\alpha}_i; z_1, m_i)\sum_j N_j u_2(\hat{n}'\cdot\hat{\alpha}_j; z_2, m_j)\,.
\end{equation}
Next, we take the ensemble average over halo mass, 
\begin{alignat}{3}
    \label{eq:2halo_ensemble}
    & \langle N_i u_1(\hat{n}'\cdot\hat{\alpha}_i, z_1, m)N_ju_2(\hat{n}\cdot\hat{\alpha}_j, z_2, m') \rangle^{\rm 2h}_{m} && = &&\int dm \, n(z_1, m) \Delta V_i u_1(\hat{n}'\cdot\hat{\alpha}_i; z_1, m)  \int dm' \, n(z_2, m') \Delta V_j u_2(\hat{n}'\cdot\hat{\alpha}_j; z_2, m') \nonumber \\
    & && && \times [1+\xi^{\rm hh}(\hat{\alpha}_i \cdot \hat{\alpha}_j; z_1, m, z_2, m')]\, ,
\end{alignat}
where the halo autocorrelation function $\xi^{\rm hh}$, defined in e.g.\ Ref.~\cite{vandenBosch:2012nq}, accounts for halo clustering, and is given to leading order by 
\begin{equation}
    \label{eq:xi_hh}
    \xi^{\rm hh}(\hat{\alpha}_i\cdot\hat{\alpha}_j; z_1, m, z_2, m') \simeq b(z_1, m) b(z_2, m')\xi^{\rm lin}(\hat{\alpha}_i \cdot \hat{\alpha}_j, z_1, z_2)\, .
\end{equation}
Here, $b$ is the bias function from Ref.~\cite{Tinker:2010my} and $\xi^{\rm lin}$ is the linear matter two-point correlation function between two points at redshifts $z_1$ and $z_2$, along lines of sight with an opening angle of $\hat{\alpha}_i \cdot \hat{\alpha}_j$. $\xi^{\rm lin}$ is related to the linear matter power spectrum by a Fourier transform,
\begin{equation}
    \label{eq:xi_lin_to_Cl}
    \xi^{\rm lin}(\hat{\alpha}_i \cdot \hat{\alpha}_j; z_1, z_2) = \int \frac{d^3 \vec{k}}{(2\pi)^3} e^{i \vec{k} \cdot (\chi_1 \hat{\alpha}_i -\chi_2 \hat{\alpha}_j)}\, P^{\rm lin}(k, z_1, z_2)\, ,
\end{equation}
where $\chi_1$ and $\chi_2$ are the comoving distances to redshifts $z_1$ and $z_2$, respectively, and we take $P^{\rm lin}(k, z_1, z_2) = \sqrt{P^{\rm lin}(k, z_1)P^{\rm lin}(k, z_2)}$~\cite{Pirvu:2023lch}. 

Returning to the two-halo term, after taking the ensemble average over halo mass, we have
\begin{equation}
    \left\langle U_1 U_2\right\rangle_{m}^{\rm 2h} = \int dm \, n(z, m) \sum_i \Delta V_i u_1(\hat{n}'\cdot\hat{\alpha}_i; z_1, m)  \int dm' \, n(z_2, m') \sum_j \Delta V_j u_2(\hat{n}'\cdot\hat{\alpha}_j; z_2, m') [1+\xi^{\rm hh}(\hat{\alpha}\cdot\hat{\alpha}'; z_1,m, z_2, m')]\, ,
\end{equation}
and once again replacing the sums with integrals over all space, we can get the two-point correlation function $\xi^{u_1, u_2, {\rm 2h}}(\hat{n}\cdot\hat{n}')$, as follows:
\begin{alignat}{3}
    & \xi^{u_1, u_2, {\rm 2h}}(\hat{n} \cdot \hat{n}') &&= &&\int d\chi_1 \, \chi_1^2 \int d\chi_2 \, \chi_2^2 \int dm_1 \, n(z_1, m_1) \int dm_2 \, n(z_2, m_2) \nonumber \\ 
    & && &&\times \int d\Omega_{\alpha_1} \, \int d\Omega_{\alpha_2} \, u_1(\hat{n}\cdot\hat{\alpha}_1;z_1, m_1)u_2(\hat{n}'\cdot\hat{\alpha}_2;z_2, m_2)[1+\xi^{\rm hh}(\hat{\alpha}\cdot\hat{\alpha}'; z_1,m_1, z_2, m_2)]\, . \label{eq:2halo_corr_func}
\end{alignat}

Again, we can simplify the angular integrals using Legendre polynomials using the same identities as before, which gives 
\begin{equation}
    \int d\Omega_{\alpha_1} \, \int d\Omega_{\alpha_2} \,u_1(\hat{n}\cdot\hat{\alpha}_1;z_1, m_1)u_2(\hat{n}'\cdot\hat{\alpha}_2;z_2, m_2)[1+\xi^{\rm hh}(\hat{\alpha}\cdot\hat{\alpha}_2; z_1,m_1, z_2, m_2)] =\sum_{\ell} u_{1,\ell} u_{2, \ell} h_\ell \left(\frac{4\pi}{2\ell+1}\right)^2P_\ell(\hat{n}\cdot\hat{n}')\, ,
\end{equation}
where 
\begin{equation}
    h_\ell = \frac{2\ell+1}{2}\int_{-1}^1d\cos\theta\,\xi^{\rm hh}(\theta)P_\ell(\cos\theta) \, .
\end{equation}
Note that the part of the angular integral not including $\xi^{\rm hh}$ factorizes into $\langle \tilde{U}_1 \rangle \langle \tilde{U}_2 \rangle$, so it only contributes to the monopole. 
We are interested in fluctuations here, so we neglect this term.
Substituting these angular results in, we find that, up to a constant, 
\begin{equation}
    \xi^{u_1, u_2, {\rm 2h}} =\int d\chi_1 \, \chi_1^2 \int d\chi_2 \, \chi_2^2 \int dm_1 \, n(z_1, m_1) \int dm_2 \, n(z_2, m_2) \sum_{\ell} u_{1,\ell} u_{2, \ell} h_\ell \left(\frac{4\pi}{2\ell+1}\right)^2P_\ell(\hat{n}\cdot\hat{n}')\, . \label{eq:2halo_corr_func_post_sub}
\end{equation}
Then, we again read off the angular power spectrum (for $\ell \neq 0$) using Eq.~\eqref{eq:xi_and_Cl_relation}, as follows:

\begin{alignat}{3}
   & C_\ell^{u_1, u_2, {\rm 2h}} &&= &&\left(\frac{4\pi}{2\ell+1}\right)^3\int d\chi_1\, \chi_1^2 \int d\chi_2 \,\chi_2^2 \int dm_1 \, n(z_1, m_1) \int dm_2 \, n(z_2, m_2) u_{1,\ell} (z_1, m_1) \nonumber \\
   & && &&\times u_{2, \ell}(z_2, m_2) h_\ell(z_1, z_2, m_1, m_2) \, .
   \label{eq:gen_Cl_2halo}
\end{alignat}
The final observed power spectrum is then 
\begin{align}
    C_\ell^{u_1, u_2} = C_\ell^{u_1, u_2, {\rm 1h}} + C_\ell^{u_1, u_2, {\rm 2h}}\, .
    \label{eq:gen_Cl}
\end{align}

\subsection{Dark Photons in the Halo Model}
\label{sec:dark_photon_signal_halo_model}

To compute $C_\ell^{\rm TT}$ and $C_\ell^{\rm Tg}$ using the formalism described above, we must first compute the relevant multipole kernels. 
We first consider $u^{\gamma\to A'}_\ell(z, m)$, the multipole kernel of the temperature deficit from $\gamma \to A'$ conversions.
In order to compute $u^{\gamma\to A'}_\ell(z,m)$, we follow the derivation in Ref.~\cite{Pirvu:2023lch}, although we make substantial simplifications to the presentation of the formalism there.

We first assume that halos at a given redshift $z$ and mass $m$ are identical and spherically symmetric, with the total matter distribution following an NFW profile~\cite{Navarro:1995iw}. 
Furthermore, all gas in halos is assumed to be fully ionized. 
Therefore, $m_\gamma^2$ in \unit{\eV} is directly proportional to the gas profile $\rho_{\rm gas}(r; z, m)$, as follows:
\begin{equation}
    m_\gamma^2 = \frac{4\pi \alpha_{\rm EM}}{m_e} \frac{\rho_{\rm gas}(r; z, m)}{m_p} \equiv  \kappa \rho_{\rm gas} (r; z, m)\, ,
\end{equation}
where $r$ is the radial distance to the halo center and $\kappa = \qty{5.7e-38}{\eV\squared\per\msol\mega\parsec\cubed}$~\cite{Pirvu:2023lch}. 
We adopt the Active Galactic Nucleus (AGN) feedback gas profile from Ref.~\cite{Battaglia:2016xbi}\footnote{Note that we correct a typo in the exponent of $\rho_{\rm gas}$ from Ref.~\cite{Battaglia:2016xbi} to match the usual expression for a generalized NFW profile from Ref.~\cite{Zhao:1995cp}. Additionally, in Ref.~\cite{Pirvu:2023lch}, the authors use the prefactor $\Omega_b/\Omega_{cdm}$ in their expression to get the fraction of halo mass that is baryonic matter. However, this should be $\Omega_b/\Omega_m$ since the halo mass includes both baryonic and dark matter.}, as follows:
\begin{equation}
    \rho_{\rm gas}(r; z,m) = \frac{\Omega_b}{\Omega_m} \rho_c(z)\rho_0(z,m)\left(\frac{x}{x_c}\right)^\gamma \left[1+\left(\frac{x}{x_c}\right)^{\alpha(z,m)}\right]^{\frac{-\beta(z,m)+\gamma}{\alpha(z,m)}}\!\!\!\!\!\!\!\!, 
\end{equation}
with $x \equiv r/R_{200}(z,m)$. 
Here, $\Omega_b$ is the baryon density parameter, $\Omega_m$ is the total matter density parameter, $\rho_c$ is the critical density of a flat Friedmann--Lema\^{\i}tre--Robertson--Walker (FLRW) universe, and $R_{200}$ is the radius from the center of the halo at which the total matter density is equal to $200\rho_c$. 
In our adopted AGN feedback model, the core scale $x_c=0.5$ and $\gamma=-0.2$ are fixed; the variables $\rho_0$, $\alpha$ and $\beta$, generically labeled $A$ below, are fit with power laws according to
\begin{equation}
    A = a \left(\frac{m_{200}}{\qty{e14}{\msol}}\right)^b (1+z)^c, \quad m_{200} = \frac{4\pi}{3}R_{200}^3(200\rho_c)\, .
\end{equation}
Here, $\{a,b, c\}$ for each quantity are fit from simulation data and are given in Table 2 of Ref.~\cite{Battaglia:2016xbi}.  

We now need to understand the geometry of $\gamma \to A'$ conversions within halos. 
Figure~\ref{fig:halo_cartoon} illustrates the setup of the problem. 
In terms of $\rho_{\rm gas}$, under the assumption that the gas is fully ionized, we have
\begin{equation}
    \frac{d \ln m_\gamma^2}{dt} = \frac{1}{\rho_{\rm gas}}\frac{d\rho_{\rm gas}}{ds}\, ,
\end{equation}
where $s$ parametrizes a chord that passes through a halo along a line of sight $\hat{n}$. 
For a halo with center at  $\chi\hat{\alpha}$, we can express the physical perpendicular distance from the chord to the center of the halo as $\chi\theta/(1+z)$, where $\cos\theta = \hat{n} \cdot \hat{\alpha}$. 
Then, $s$ is defined by 
\begin{equation}
    s^2 = r^2 - \frac{\chi^2\theta^2}{(1+z)^2}\, .
\end{equation}

This expression allows us to write $|d\ln m_\gamma^2/dt|^{-1}$ at a point of resonance conversion in terms of the halo properties:
\begin{alignat}{1}
    \left| \frac{d \ln m_\gamma^2}{dt}\right|^{-1}_{r=r_{\rm res}} = \frac{m_{A'}^2}{\kappa} \left|\frac{d\rho_{\rm gas}(r; z, m)}{dr}\right|_{r=r_{\rm res}}^{-1} a(\theta; z, m)\Theta(r_{\rm vir}-r_{\rm res})\,,
\end{alignat}
where
\begin{alignat}{1}
    a(\theta; z, m) \equiv \left(1- \frac{\chi^2 \theta^2}{(1+z)^2r_{\rm res}^2} \right)^{-1/2}
\end{alignat}
for a resonance that occurs at distance $r_{\rm res}$ from the center of the halo. 
$r_{\rm res}(z,m)$ can be found by solving for the radius at which the resonance conversion is met in a halo. 
The step function ensures that conversions happen within the halo at a radius less than the virial radius $\rho_{\rm gas}(r_{\rm vir}; z, m) \approx \Delta \rho_c(z)$, where $\Delta = 18\pi^2 + 82x-39x^2$ with $x\equiv \Omega_m(z)-1$~\cite{Bryan:1997dn}.  
Also note that $a(\theta; z,m)$ is only defined for angles less than $\theta_{\rm max} \equiv r_{\rm res}(z,m)(1+z)/\chi$. 
Physically, this corresponds to the maximum angle for which the path of the dark photon passes through the halo. 

Using this result, the temperature deficit due to $\gamma \to A'$ conversions per halo along $\hat{n}$ is 
\begin{equation}
    \label{eq:prob_conv_per_halo}
    u^{\gamma\to A'}(\hat{n} \cdot \hat{\alpha}; z, m) = -T_{\gamma, 0}\frac{2\pi \epsilon^2 m_{A'}^4}{\kappa\omega(z)} \left|\frac{d\rho_{\rm gas}}{dr}\right|_{r_{\rm res}}^{-1} a(\theta; z, m)\Theta(r_{\rm vir}-r_{\rm res})\, ,
\end{equation}
and the factor of 2 appears because there are two conversions per halo, since we assume spherical symmetry. 
The factor of $-T_{\gamma, 0}$ appears relative to the analogous expression in Ref.~\cite{Pirvu:2023lch} because we are considering the temperature deficit due to $\gamma \to A'$ conversions, not just the  probability of conversion. 

\begin{figure}[t!]
    \centering
    \includegraphics[width=0.4\linewidth]{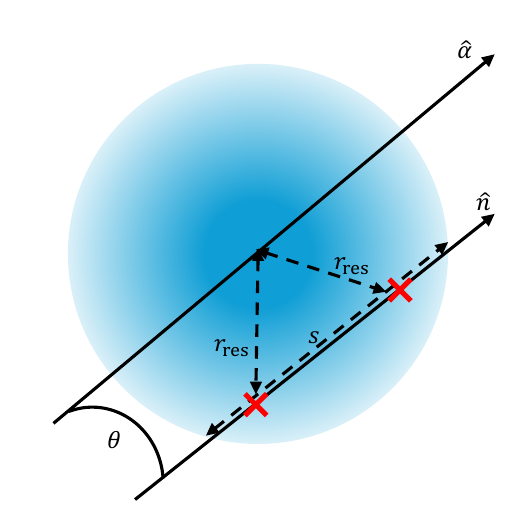}
    \caption{Geometry of $\gamma \to A'$ conversions inside a dark matter halo, where the shading schematically represents the gas density of the halo. The halo center is located along $\hat{\alpha}$. The conversions occur at the points marked with x's at a radius $r_{\rm res}$ along a chord with length $s$.}
    \label{fig:halo_cartoon}
\end{figure}

To get $u^{\gamma\to A'}_\ell$, we integrate $u^{\gamma\to A'}(\theta; z, m)$ against a Legendre polynomial. Ignoring the prefactors to isolate the angular part of Eq.~\eqref{eq:prob_conv_per_halo} and applying the small angle approximation, we have

\begin{equation}
    \label{eq:al_def}
    a_\ell = \frac{2\ell+1 }{2}\int_0^{\theta_{\rm max}}d\theta \, \theta \left(1-\frac{\theta^2}{\theta_{\rm max}^2}\right)^{-1/2} P_\ell(\cos \theta) \, .
\end{equation}
Note that the integral is cut off at $\theta_{\rm max}$. 
We can evaluate this integral explicitly by using the small angle approximation and a particularly convenient series expansion of the Legendre polynomial, as follows:

\begin{align}
    a_\ell &= \frac{2\ell+1 }{2}\int_0^{\theta_{\rm max}}d\theta \, \theta \left(1-\frac{\theta^2}{\theta_{\rm max}^2}\right)^{-1/2} P_\ell(1-\theta^2/2) \nonumber \\
    &= \frac{2\ell+1 }{2}\sum_{k=0}^\ell \frac{\ell!}{k!(\ell - k)!}\frac{(\ell +k)!}{\ell! k!} \int_0^{\theta_{\rm max}}d\theta \, \theta \left(1-\frac{\theta^2}{\theta_{\rm max}^2}\right)^{-1/2} \left(\frac{\theta^2}{4}\right)^k \nonumber \\
    &= \frac{\sqrt{\pi}}{2}\theta_{\rm max}^{3/2}(4-\theta_{\rm max}^2)^{1/4}P_\ell^{-1/2}(1-\theta_{\rm max}^2/2)\, ,
\end{align}
where $P_\ell^{-1/2}$ is the associated Legendre polynomial of degree $\ell$ and order $-1/2$. 
This expression is particularly useful for avoiding expensive numerical integrals over high-degree Legendre polynomials. 
Therefore, we have the multipole kernel for $\xi^{\rm TT}$ and $C_\ell^{\rm TT}$, as follows:
\begin{equation}
    u^{\gamma\to A'}_\ell = -T_{\gamma,0} \frac{2\pi \epsilon^2 m_{A'}^4}{\kappa\omega(z)} \left|\frac{d\rho_{\rm gas}}{dr}\right|_{r_{\rm res}}^{-1}(z,m) a_\ell(z, m)\Theta(r_{\rm vir}-r_{\rm res})\, .
\end{equation}

\subsection{Halo Occupation Distribution}
\label{sec:hod}
Because $\gamma \to A'$ conversions trace the distribution of dark matter halos, which galaxies also trace, there is a nonzero cross-correlation between $\gamma \to A'$ conversions in halos and the galaxies that occupy them. 
To compute this cross-correlation, we make use of a HOD model, which is an extension of the halo model and describes how galaxies are distributed within halos. 

In this section, we first outline the HOD model that we adopt, which is the best-fit model from Ref.~\cite{Kusiak:2022xkt} and describes the \textit{unWISE} ``blue'' galaxy sample from Ref.~\cite{Krolewski:2019yrv,Krolewski:2021yqy}. 
The \textit{unWISE} blue sample is a low-redshift subset of the full \textit{unWISE} catalog with a mean redshift of $0.6$.

In our adopted HOD model, a halo can host a central galaxy located at its core and some number of satellite galaxies distributed throughout the halo. 
The mean number of central galaxies is given by 
\begin{equation}
    \label{eq:Nc_def}
    N_c(m) = \frac{1}{2}\left[1+\erf\left(\frac{\log m- \log m_{\rm min}}{\sigma_{\log m}}\right)\right]\, ,
\end{equation}
where $m_{\rm min}$ is the minimum halo mass that can host a central galaxy. 
The mean number of satellite galaxies is taken to follow a power law given by
\begin{equation}
    \label{eq:Ns_def}
    N_s(m) = N_c(m)\left(\frac{m}{m_\star}\right)^{\alpha_s} \, .
\end{equation}

From these expressions, we can write the average number density of galaxies at a redshift $z$ as~\cite{vandenBosch:2012nq}
\begin{equation}
    \overline{n}_g(z) = \int dm \ n(z, m) [N_c(m) + N_s(m)] \, .
\end{equation}

Using the halo model, the number density of galaxies at a point is 
\begin{equation}
    n_g(\chi \hat{n}) = \sum_i N_{i} u_g(\hat{n}\cdot\hat{\alpha}_i, z, m_i) \,,
\end{equation}
where $N_i$ is the number of halos in the volume $\Delta V_i$, as before, and $u_g$ describes the distribution of the galaxies as a function of angle from the halo center. 

In the HOD model, we assume that the central galaxy is located at the center of the halo and that the satellite galaxies follow the same distribution as the dark matter itself. 
We take this to be the NFW distribution. 
Therefore,
\begin{equation}
   u_g(\hat{n}\cdot\hat{\alpha}_i, z, m) = A(z)\left[N_{c,i}\delta_D(\chi \hat{n} - \chi \hat{\alpha}_i) + N_{s,i} u_{\rm NFW}(\hat{n}\cdot \hat{\alpha}_i, z, m) \right]\, ,
\end{equation}
where $N_{c,i}$ ($N_{s,i}$) is the number of central (satellite) galaxies in the halo with center in $\Delta V_i$~\cite{vandenBosch:2012nq}, and $u_{\rm NFW}$ is a probability distribution function proportional to an NFW profile for a halo of mass $m$ that integrates to 1 over all space. 
Specifically, the Fourier transform of $u_{\rm NFW}$ is
\begin{alignat}{3}
   & \tilde{u}_{\rm NFW}(k; z,m) &&= &&f(\lambda_{\rm NFW}c)\left[\sin(q)\left(\Si(\tilde{q})- \Si(q)\right) +\cos(q)\left(\Ci(\tilde{q}) - \Ci(q)\right) - \frac{\sin(\tilde{q}-q)}{\tilde{q}}\right]\, , \label{eq:nfw_ft} \\
   & f(\lambda_{\rm NFW}c) &&\equiv &&\left[\ln \left(1+\lambda_{\rm NFW}c\right)-\frac{\lambda_{\rm NFW}c}{1+\lambda_{\rm NFW}c} \right]^{-1} \! \!\!\!\! ,
\end{alignat}
where
\begin{equation}
    q = \frac{k R_{200}}{c}, \qquad \tilde{q} = (1+\lambda_{\rm NFW}c)q\, \, ,
\end{equation}
$\lambda_{\rm NFW}R_{200}$ is the NFW truncation radius, and $c$ is the concentration from Ref.~\cite{Bhattacharya:2011vr}. $\Si$ and $\Ci$ are the sine and cosine integrals, respectively. 
Finally, $A(z)$ is a normalization factor that we derive below to ensure that the redshift distribution of the galaxies matches that of the \textit{unWISE} catalog. 

\begin{table}[t!]
    \centering
    \begin{tabular}{l S[table-format=2.3]}
    \toprule \toprule
    {Parameter} & {Value} \\ 
    \midrule
    $\alpha_s$                            & 1.304  \\
    $\sigma_{\log m}$                     & 0.687 \\
    $\log_{10}(m_{\rm min}/\unit{\msol})$ & 11.97 \\
    $\log_{10}(m_\star/\unit{\msol})$     & 12.87 \\
    $\lambda_{\rm NFW}$                   & 1.087 \\
    \bottomrule \bottomrule
    \end{tabular}
    \caption{Parameters of the best fit HOD model from Ref.~\cite{Kusiak:2022xkt} that we adopt in this work.}
    \label{tab:K22_params}
\end{table}

From this expression, we can compute the galaxy multipole kernel $u^{g}_\ell$. To make progress, we express $u^g(\hat{n}\cdot \hat{\alpha}_i)$ using its Fourier transform and simplify using the plane wave expansion and the properties of spherical harmonics, as follows:
\begin{alignat}{3}
    & u^{g}(\hat{n}\cdot \hat{\alpha_i}, z, m) &&= &&A(z) \int \frac{d^3k}{(2\pi)^3} e^{i k\cdot( \chi-\chi_i)} [N_{c,i} +N_{s,i}\tilde{u}_{\rm NFW}(k,z, m)] \\
    & &&= &&A(z) \frac{2}{\pi}\sum_{\ell m}\sum_{\ell' m'}\int d^3k\, i^{\ell -\ell'} j_{\ell}(k \chi)j_{\ell'}(k\chi_i) Y_{\ell m}(\hat{k})Y^*_{\ell m}(\hat{n})Y^*_{\ell' m'}(\hat{k})Y_{\ell' m'}(\hat{\alpha}_i) [N_{c,i} +N_{s,i}\tilde{u}_{\rm NFW}(k,z, m)] \nonumber \\
    & &&= &&A(z) \frac{2}{\pi}\sum_{\ell m}\int dk\, k^2 j_{\ell}(k \chi)j_{\ell}(k\chi_i) Y^*_{\ell m}(\hat{n})Y_{\ell' m'}(\hat{\alpha}_i) [N_{c,i} +N_{s,i}\tilde{u}_{\rm NFW}(k,z, m)] \nonumber \\
    & &&= &&A(z) \frac{2}{\pi}\sum_{\ell} \frac{2\ell+1}{4\pi} \int dk\, k^2 j_{\ell}(k \chi)j_{\ell}(k\chi_i) [N_{c,i} +N_{s,i}\tilde{u}_{\rm NFW}(k,z, m)] P_\ell(\hat{n}\cdot \hat{\alpha_i})
\end{alignat}
Next, we can simplify further using the Limber approximation, which is $k^2 j_\ell(k \chi)j_{\ell}(k \chi') \approx (\pi/2)\delta_D(k-(\ell+1/2)/\chi)\delta_D(\chi-\chi')/\chi^2$~\cite{Limber:1954zz,LoVerde:2008re}. Then,

\begin{alignat}{3}
    & u^{g}(\hat{n}\cdot \hat{\alpha_i}, z, m_i) &&= &&A(z) \frac{2}{\pi}\sum_{\ell} \frac{2\ell+1}{4\pi} \int dk\, \frac{\pi}{2}\delta_D\left(k-\frac{\ell+\frac{1}{2}}{\chi}\right)\frac{1}{\chi^2}\delta_D(\chi-\chi') [N_{c,i} +N_{s,i}\tilde{u}_{\rm NFW}(k,z, m_i)] P_\ell(\hat{n}\cdot \hat{\alpha_i})\nonumber \\
    & &&= &&A(z) \sum_{\ell} \frac{2\ell+1}{4\pi} \frac{1}{\chi^2}\delta_D(\chi-\chi_i) \left[N_{c,i} +N_{s,i}\tilde{u}_{\rm NFW}\left(\frac{\ell+\frac{1}{2}}{\chi}, z, m_i\right)\right] P_\ell(\hat{n}\cdot \hat{\alpha_i})\, .
\end{alignat}
From here, we read off $u_\ell^g$, as follows:
\begin{equation}
    u_{\ell,i}^g = A(z) \frac{2\ell+1}{4\pi} \frac{1}{\chi^2}\delta_D(\chi-\chi_i) \left[ N_{c,i} +N_{s,i}\tilde{u}_{\rm NFW}\left(\frac{\ell+\frac{1}{2}}{\chi}, z, m_i\right)\right]
\end{equation}
Finally, we integrate over each halo volume (which eliminates the delta function) and take the ensemble average to get 
\begin{equation}
    \label{eq:gal_multipole_kernel}
    u_{\ell}^g = A(z) \frac{2\ell+1}{4\pi} \left[N_c(m)  + N_s(m)\tilde{u}_{\rm NFW}\left(\frac{\ell+\frac{1}{2}}{\chi}; z, m\right)\right] \, .
\end{equation}

Now, we can derive $A(z)$. First, we define the redshift distribution of the galaxies in the catalog, $dN_g/dz$, which has been normalized such that $\int dz \, dN_g/dz=1.$ The HOD must match this distribution, so we need
\begin{alignat}{3}
    & \int dz \frac{dN_g}{dz} &&= &&\int d^3 \, \chi A(z) \int dm \, n(z,m) u^g(\theta, z, m) \nonumber \\
    & &&= &&\int dz \, \frac{\chi^2 A(z)}{H(z)}  \int d\Omega \int dm \, n(z,m) u^g(\theta, z, m) \nonumber \\
    & &&= &&\int dz \, \frac{\chi^2 A(z)}{H(z)} \overline{n}_g(z) \, ,
\end{alignat}
where the last line follows by construction in the halo model. Therefore, by comparing integrands, we see that
\begin{equation}
    A(z) = \frac{H(z)}{\chi^2}\frac{dN_g}{dz} \overline{n}^{-1}_g \, ,
\end{equation}
which fully determines the galaxy multipole kernel.

In this model, there are five free parameters, whose values were computed in Ref.~\cite{Kusiak:2022xkt} from fits to the \textit{unWISE} data and cross-correlations with \textit{Planck} maps, and are listed in Table~\ref{tab:K22_params}.

\subsection{Modeling Uncertainties in the Halo Model}
\label{sec:halo_modeling_choices}
In the halo model, we made several modeling choices about the gas profile and HOD.
These choices introduce some degree of systematic uncertainty, which we evaluate in this section.

First, in our fiducial analysis, we adopt the AGN feedback model from Ref.~\cite{Battaglia:2016xbi} to describe the baryon distribution inside the halo. 
This profile was fit to simulation data that is meant to capture AGN feedback mechanisms which affect the distribution of baryons near the halo center. 
The AGN feedback reduces the density of baryons near the center of the halo relative to the density of dark matter, thereby imposing a cutoff on $m_{\gamma}^2(r)$ in the halo.
This cutoff in turn sets an upper limit on $m_{A'}$ for which $\gamma \to A'$ conversions can occur inside the halo.

To explore the effect of this modeling choice, we also considered two alternate models for the gas profile.
The derived sensitivity on $\epsilon$ that results from these various modeling choices is shown in Fig.~\ref{fig:halo_model_comparisons}, left. 

\begin{figure}
    \centering
    \includegraphics[width=0.4\textwidth]{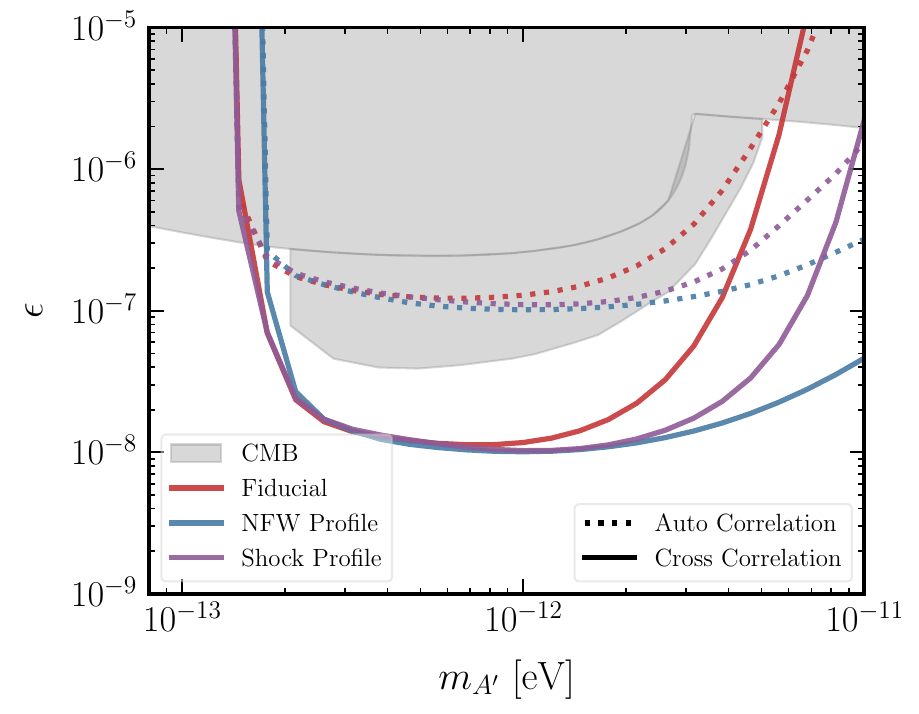}
    \includegraphics[width=0.4\textwidth]{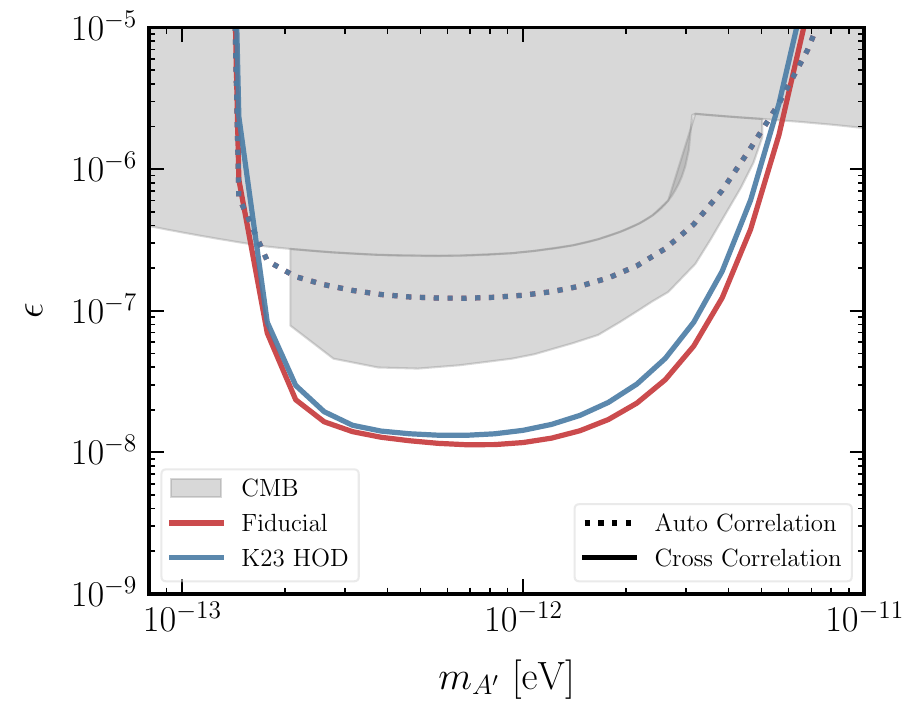}
    \caption{In each panel, we compare the effect of modeling choices on the derived SKA sensitivity to $\epsilon$. \textit{Left:} Two alternative choices for the halo gas profile are compared to the AGN Feedback model from Ref.~\cite{Battaglia:2016xbi} used in our fiducial analysis (red). The shock feedback model from Ref.~\cite{Battaglia:2016xbi} (purple) and an NFW profile (blue) both give similar results to the fiducial model for $\qty{e-13}{\eV}\lesssim m_{A'} \lesssim \qty{e-12}{\eV}$. At larger masses, the sensitivity to dark photons is highly dependent on the gas profile. The fiducial profile gives the most conservative results. \textit{Right:} SKA's sensitivity to dark photons is shown with two choices of HOD model. Our fiducial HOD model from Ref.~\cite{Kusiak:2022xkt} (Red) and the HOD profile from Ref.~\cite{Kusiak:2023hrz} (blue) give similar results.~\nblink{Limits}}
    \label{fig:halo_model_comparisons}
\end{figure}
The first alternative profile we consider is the ``shock heating'' model from Ref.~\cite{Battaglia:2016xbi}.
This model only accounts for heating through gravitational effects and represents less extreme baryonic feedback mechanisms near the halo center.
As a result, the gas density near the center of the halo is generally larger than in the fiducial AGN feedback model.
Finally, we considered the case where the baryons exactly trace the dark matter distribution in the halo and follow an NFW profile. 
Although this scenario is unrealistic, this choice of gas profile allows us to understand the full significance of the baryonic feedback mechanisms in affecting SKA's sensitivity to dark photons. 
The NFW profile is unbounded at small radii, meaning heavier dark photons could convert near the center of halos, unlike in the other two models that we consider.

In the left panel of Fig.~\ref{fig:halo_model_comparisons}, we see that the choice of gas profile makes little difference at small dark photon masses. 
Lighter dark photons preferentially convert on the outskirts of halos, where baryonic feedback mechanisms are largely unimportant, and the gas is expected to track the dark matter.
This in turn means that SKA's sensitivity to dark photons with $\qty{e-13}{\eV}\lesssim m_{A'} \lesssim \qty{e-12}{\eV}$ does not depend precisely on the halo gas profile.

At larger $m_{A'}$, however, SKA's sensitivity does depend greatly on the choice of gas profile. 
In particular, models with less baryonic feedback predict that SKA will have greater sensitivity to dark photons with $m_{A'} \gtrsim \qty{e-12}{\eV}$ than our fiducial analysis.
This is largely a result of the cutoff in gas density mentioned above.
Regardless, we see that the AGN feedback model is the most conservative choice, and thus is the basis of our fiducial analysis. 

Another source of uncertainty not captured by the halo model is the impact of 
small-scale inhomogeneities. 
Our results are largely insensitive to these inhomogeneities as well. 
Specifically, we find that $C_\ell^{\rm Tg}$ is dominated by correlations between $\gamma \to A'$ conversions in a main halo and the satellite galaxies surrounding this halo.
The presence of these satellite galaxies is uncorrelated with any small-scale inhomogeneities in the main gas profile, meaning these perturbations will not affect our results. 

Figure~\ref{fig:one_halo_two_halo} provides evidence for this point. In Fig.~\ref{fig:one_halo_two_halo}, we plot the decomposition of $C_\ell^{\rm Tg}$ into one-halo and two-halo terms for a dark photon with $m_{A'}=\qty{6e-13}{\eV}$ and $\epsilon=5\times 10^{-8}$ at $\omega_0/2\pi = \qty{410}{\mega\hertz}$.
We consider two cases in this figure. 
The solid lines show $C_\ell^{\rm Tg}$ when we consider both central and satellite galaxies in the expression for $u^{g}_\ell$. 
The dashed lines show the case where we exclude satellite galaxies, i.e. letting
\begin{align}
    u_\ell^{ g} =A(z) \frac{2\ell+1}{4\pi}N_c(m)\, .
\end{align}
By comparing these two cases, it is clear that the satellite galaxy contribution dominates $C_\ell^{\rm Tg}$. 
In other words, SKA will mostly be sensitive to correlations between $\gamma \to A'$ conversions and the locations of satellite galaxies. 
However, since these satellite galaxies are dispersed on the outer reaches of halos, their location is uncorrelated with any small-scale perturbations in the main halo and these perturbations will not affect the predicted $C_\ell^{\rm Tg}$. 

\begin{figure}
    \centering
    \includegraphics[width=0.45\textwidth]{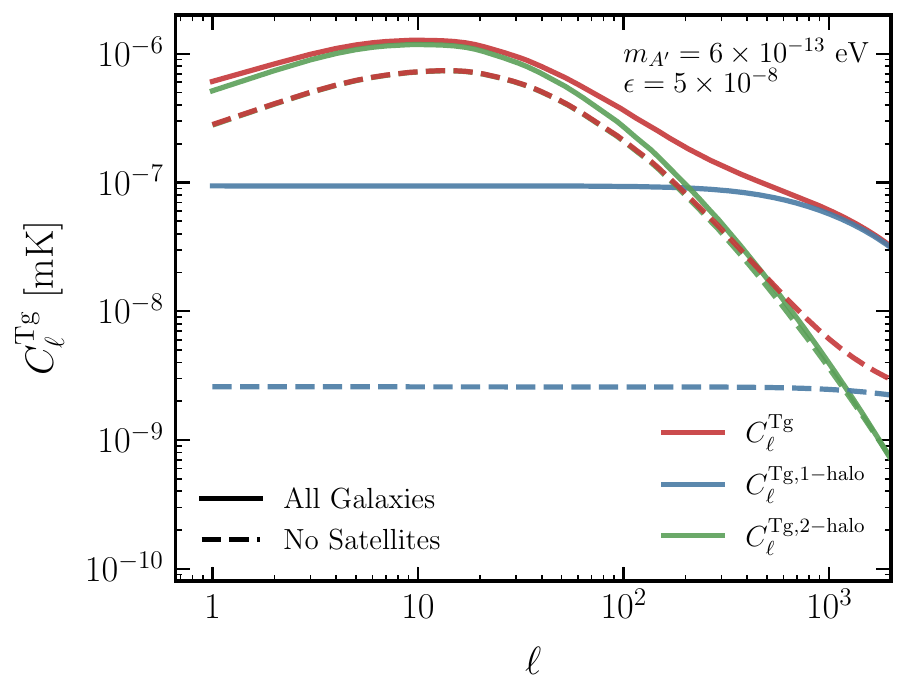}
    \caption{One-halo (blue) and two-halo (green) contributions to $C_\ell^{\rm Tg}$ are shown for a dark photon with $m_{A'} = \qty{6e-13}{\eV}$ and $\epsilon=5\times 10^{-8}$ at $\omega_0/2\pi = \qty{410}{\mega\hertz}$. The solid lines show $C_\ell^{\rm Tg}$ from our main analysis, where we include contributions from satellite and central galaxies in $u_\ell^g$. The dashed lines show $C_\ell^{\rm Tg}$ when satellite galaxies are excluded from $C_\ell^{\rm Tg}$. The amplitude of the one-halo term decreases substantially in the latter case, indicating that most of the signal comes from correlations between satellite galaxies and $\gamma \to A'$ conversions in the main halo.~\nblink{Halo_Model_Decomposition}}
    \label{fig:one_halo_two_halo}
\end{figure}

Finally, the adopted HOD model is another potential source of modeling uncertainty. 
As explained in more detail in Sec.~\ref{sec:galaxy_modeling}, we choose the HOD model from Ref.~\cite{Kusiak:2022xkt} for our fiducial analysis (``K22'').
Alternatively, we also consider the HOD model from Ref.~\cite{Kusiak:2023hrz} (``K23''). 
This HOD model was fit to smaller-scale data than K22 and could therefore better describe the scales that are important to this analysis.
However, this model has some unphysical features as explained more fully in Ref.~\cite{Goldstein:2024mfp} and therefore is not the basis of our fiducial analysis.
Regardless, as shown in the right panel of Fig.~\ref{fig:halo_model_comparisons}, SKA's forecasted sensitivity to dark photons is largely insensitive to the choice of HOD model.

\section{Lognormal Ansatz for Environment C}
\label{sec:lognormal_derivation}

In this appendix, we outline a derivation of $C_\ell^{\rm C, TT}$ under our Ansatz that $m_{\gamma}^2$ is distributed according to lognormal one-point and two-point PDFs. 
This model is adopted from Ref.~\cite{Caputo:2022keo} and this discussion closely follows the derivation presented there. 
We first derive the prediction for the mean probability of conversion in this model, and then discuss fluctuations in the signal and derive $C_\ell^{\rm C, TT}$. 

\subsection{Mean Probability of Conversion}
\label{sec:mean_prob_analytic}

First, we consider the analytic model of the mean probability of conversion described in Refs.~\cite{Caputo:2020bdy,Caputo:2020rnx}. We use this model to understand the impact of $\gamma \to A'$ conversions on the 21-cm global signal (see Sec.~\ref{sec:21cm_global}) and it provides an introduction to the analytic model we use to estimate the magnitude of fluctuations in $P_{\gamma \to A'}$ in the late Universe. 
We model these fluctuations from $0.005 \lesssim z \lesssim 4$ for consistency with our modeling of environment A.

This formalism is similar to the halo model that we use above. 
There, the probability of conversion and its fluctuations were computed using the distribution of halos in space and their properties. 
Here, we consider something more general by computing the PDF of $m_{\gamma}^2$ itself, which in turn directly gives the mean probability of conversion. 

First, we express the differential probability of conversion per unit time as~\cite{Caputo:2020rnx}
\begin{equation}
    \frac{dP_{\gamma\to A'}}{dt}= \frac{\pi m_{A'}^2\epsilon^2}{\omega(t) }\delta_D\left(m_\gamma^2(t)-m_{A'}^2\right) m_{\gamma}^2(t) \, ,
\end{equation}
where $\delta_D$ is the Dirac delta function. The mean differential probability of conversion is found by integrating over the PDF $f(m_{\gamma}^2;t)$ to get
\begin{alignat}{3}
    & \frac{d\langle P_{\gamma\to A'}\rangle}{dt} &&=  &&\frac{\pi m_{A'}^2 \epsilon^2}{\omega(t)} \int dm_{\gamma}^2 \, f(m_\gamma^2;t) \delta_D\left(m_\gamma^2(t)-m_{A'}^2\right) m_{\gamma}^2(t)  \nonumber \\ 
    & &&= &&\frac{\pi m_{A'}^4 \epsilon^2}{\omega(t)} f(m_\gamma^2=m_{A'}^2;t) \,.  
\end{alignat}
The total mean probability of conversion is then given by integrating $d\langle P_{\gamma \to A'} \rangle/dt$ over time:
\begin{equation}
    \label{eq:analytic_avg_P}
    \langle P_{\gamma\to A'}\rangle = \int_{t_1}^{t_2} dt \, \frac{d\langle P_{\gamma\to A'}\rangle}{dt} \, ,
\end{equation} 
where we only consider conversions between $t_1$ and $t_2$. 
Therefore, in order to compute the mean probability of conversion, we only need to specify the PDF of $m_{\gamma}^2$. 
Here, we take $f_1$ to be a lognormal distribution given by 
\begin{equation}
    \label{eq:lognormal_mgamma_pdf}
    f_1(m_{\gamma}^2; z) = \frac{(1+\delta_{\rm b})^{-1}}{\sqrt{2\pi \Sigma^2(z)}}\exp[-\frac{[\ln(1+\delta_{\rm b})+\Sigma^2(z)/2]^2}{2\Sigma^2(z)}] \, ,
\end{equation}
where $\delta_{\rm b}$ is the baryon overdensity and $\Sigma^2(z)\equiv \ln[1+\sigma_b^2(z)]$, where $\sigma^2_b$ is the standard deviation of baryon density fluctuations~\cite{Caputo:2020rnx}. Additionally, we adopt the choice of Ref.~\cite{Caputo:2020rnx} and only consider fluctuations such that $1+\delta_{\rm b} \leq 100$. 

\subsection{Fluctuations in the Probability of Conversion}

Beyond computing the mean probability of conversion using this analytic framework, we can also estimate fluctuations in the probability of conversion to construct $C^{\rm C, TT}_\ell$ and the associated two-point correlation function $\xi^{\rm C, TT}(\theta)$, as described in Ref.~\cite{Caputo:2022keo}. 
In particular, we will use this model to estimate the signal from $\gamma \to A'$ conversions in the IGM from $0.005 \lesssim z \lesssim 4$, i.e. conversions that occur in underdense and mean-density regions. 
In this derivation, we assume that the two-point PDF $f_2$ of $m_\gamma^2$ is a lognormal distribution.
As discussed in the main body, this is an unrealistically simple approximation and a much bolder assumption than assuming that the one-point PDF of $m_{\gamma}^2$ is a lognormal distribution. 
Therefore, the results derived in this subsection should be taken only as a basic estimate of the $\gamma \to A'$ signal that motivates a more sophisticated study.

The full derivation of $C_\ell^{\rm C,TT}$ is provided in Ref.~\cite{Caputo:2022keo}, so we only sketch the main points here. 
First, consider the probability of conversion along a line of sight, generically given by 
\begin{equation}
    P_{\gamma\to A'} = \frac{\pi \epsilon^2 m_{A'}^4}{\omega_0} \int dz \frac{\delta_D(m_{\gamma}^2(\vec{\chi}, z)-m_{A'}^2)}{H(z)(1+z)^2}\, ,
\end{equation}
where $\vec{\chi} = \chi\hat{n}$~\cite{Caputo:2022keo}. 
Then, using Eq.~\eqref{eq:analytic_avg_P}, the fluctuation in the probability of conversion $\delta P(\hat{n})= P_{\gamma\to A'}(\hat{n})-\langle P_{\gamma\to A'}\rangle $ is
\begin{equation}
    \delta P(\hat{n}) = \frac{\pi\epsilon^2 m_{A'}^4}{\omega_0} \int\frac{dz}{H(z)(1+z)^2}[\delta_D(m_\gamma^2(\vec{\chi}, z)-m_{A'}^2) - f_1(m_\gamma^2=m_{A'}^2,z)]\,.
\end{equation} 

At this point, we can use a similar approach to the derivation of the power spectrum in the halo model. 
First, compute the two-point correlation function $\xi^{\rm C, TT}(\theta)$ by averaging $\delta P(\hat{n})\delta P(\hat{n}')$ across all lines of sight.
This is most easily done by expanding the $P(\hat{n})P(\hat{n}')$ in terms of spherical harmonics. 
Then, using Eq.~\eqref{eq:xi_and_Cl_relation}, this expansion gives the power spectrum $C_\ell^{\rm C, TT}$. 
This yields
\begin{equation}
    C_\ell^{\rm C, TT} = T_{\gamma, 0}^2\left[\frac{\pi \epsilon^2 m_{A'}^4}{\omega_0}\right]^2 \int \frac{dz\, W^2(z)H(z)}{\chi^2}\tilde{Q}\left(\frac{\ell}{\chi}, z\right) \, ,
\end{equation}
where $W(z) \equiv [H(z)(1+z)^2]^{-1}$, as follows:
\begin{alignat}{1}
    \tilde{Q}(k, z) = 4\pi \int d\rho \, \rho^2 j_0(k \rho)Q(\rho, z) \,,
\end{alignat}
and 
\begin{alignat}{1}
    Q(\rho, z) = f_2\left(\rho, m_{A'}^2, m_{A'}^2; z\right) - [f_1(m_\gamma^2=m_{A'}^2;z)]^2 \, .
\end{alignat}
$f_2(\rho, m_{A'}^2, m_{A'}^2, z)$ is the two-point PDF of $m_{\gamma}^2$, evaluated at two points at redshift $z$, separated by a distance $\rho$, with $m_{\gamma}^2 = m_{A'}^2$ at both points. 
$Q$ is the two-point correlation function of $m_{\gamma}^2$ at a redshift $z$ at two points $\chi\hat{n}$ and $\chi\hat{n}'$ and is given by the difference between the two-point PDF $f_2$ of $m_{\gamma}^2$ and the one-point PDF from Eq.~\eqref{eq:lognormal_mgamma_pdf}.
As mentioned above, we take $f_2$ to be a lognormal distribution defined by:
\begin{equation}
    \label{eq:2pt_mgamma_pdf}
    f_2\left(\rho, m_\gamma^2(\vec{\chi})=m_{A'}^2, m_\gamma^2(\vec{r'}) \right) = \frac{1}{2\pi \overline{m^2_\gamma}^2(z)\sqrt{\Sigma^4(z)-X^2(\rho; z)}}\exp\left[-\frac{L^2(m_{A'}^2;z)}{\Sigma^2(z)+X(\rho; z)}\right] \frac{1}{(1+g(m_{A'}^2;z))^2} \, .
\end{equation}
Here, $g(m_{A'}^2;z) \equiv m_{A'}^2/\overline{m^2_{\gamma}}(z)-1$, $X(\rho; z)=\ln [1+\xi_{\rm b}(\rho;z)]$, and $L(m_{A'}^2;z)=\ln[1+g(m_{A'}^2; z)] + \Sigma^2(z)/2$. Also, $\xi_{\rm b}$ is the baryon two-point correlation function, which is the Fourier transform of the baryon power spectrum, as follows:
\begin{equation}
    \xi_{\rm b}(\rho; z) = \int \frac{d^3\vec{k}}{(2\pi)^3}j_0(k\rho)P_b(k; z)\, .
\end{equation}

\section{Needlet ILC Algorithm}
\label{sec:needlet_ilc_appendix}
In this appendix, we describe in detail the needlet ILC algorithm that we use to process the mock null-signal foreground maps. 
This algorithm is conceptually similar to the basic ILC procedure outlined in Sec.~\ref{sec:needlet_ilc}, but works in needlet space instead of real space.
Transforming a map to needlet space involves filtering it with both real-space and harmonic-space filters, which allows us to separate out both spatial and angular information. 
Therefore, the algorithm is more effective at extracting signals from highly anisotropic foregrounds.
The end result of this algorithm is still a single post-ILC map that is the optimal linear combination of the input foreground maps such that the resulting map has the minimum possible variance without decreasing the strength of the dark photon signal.
We perform this algorithm with the \textsc{pyilc} code \cite{McCarthy:2023hpa} and our explanation of this algorithm closely follows that of Ref.~\cite{McCarthy:2023hpa}.

First, we begin with the $N_f$ real-space maps $T_i(\hat{n})$, where $i$ indexes the frequency of the maps. 
Then, we filter each map with $N_I$ harmonic space filters $h^I_\ell$. This gives $N_f \times N_I$ maps, $T_i^I(\hat{n})$, given by
\begin{equation}
    T_i^I(\hat{n}) = \sum_{\ell m}h^I_\ell T_{\ell m}^i Y_{\ell m}(\hat{n})\, .
\end{equation}
In other words, $T_i^I(\hat{n})$ is found by first performing a spherical harmonic decomposition on $T_i(\hat{n})$ and then multiplying the resulting $T_{\ell m}^i$ by the needlet filter $h_\ell^I$ to get the filtered map. 
This step separates out information on different angular scales.
In this work, we adopt the choice of needlet filters from Ref.~\cite{McCarthy:2024ozh}, which are defined by
\begin{equation}
    h_\ell^I = 
    \begin{cases}
        \cos\left(\frac{\pi}{2} \frac{\ell_{\rm peak}^I-\ell}{\ell_{\rm peak}^I - \ell_{\rm peak}^{I-1}}\right) & \ell_{\rm peak}^{I-1} \leq \ell < \ell_{\rm peak}^I \\
        \cos\left(\frac{\pi}{2} \frac{\ell - \ell_{\rm peak}^I}{\ell_{\rm peak}^{I+1} - \ell_{\rm peak}^{I}}\right) & \ell_{\rm peak}^{I} \leq \ell < \ell_{\rm peak}^{I+1}  \\
        0 & \rm else \, ,
    \end{cases}
\end{equation}
and are shown in Fig.~\ref{fig:needlets}.

\begin{figure}[t!]
    \centering
    \includegraphics[width=0.5\linewidth]{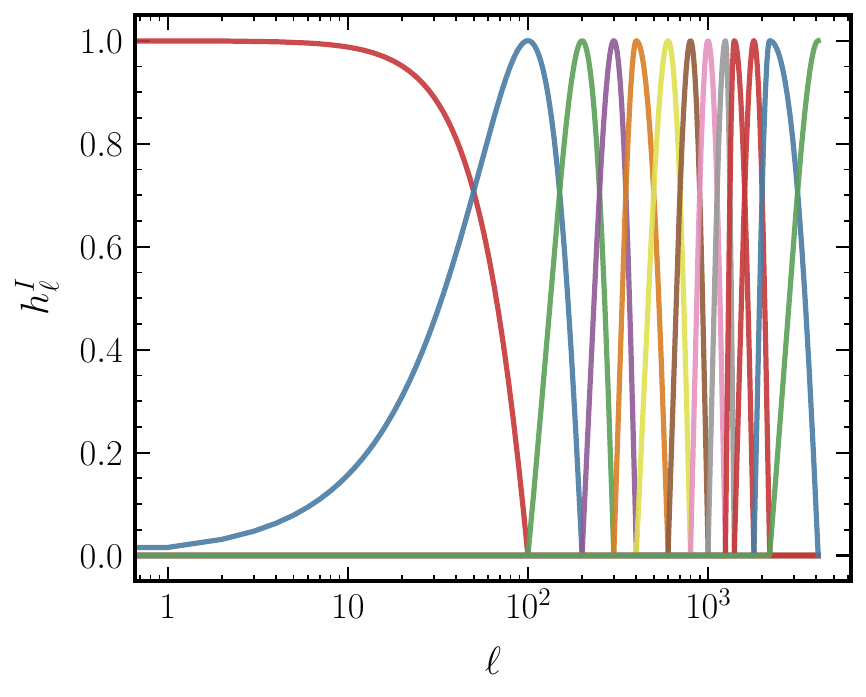}
    \caption{Cosine needlet filters $h_\ell^I$ used in the needlet ILC algorithm in this work, adopted from Ref.~\cite{McCarthy:2024ozh}.~\nblink{Needlet_Plot}}
    \label{fig:needlets}
\end{figure}

Next, we compute the covariance of the needlet filtered maps, $\mathcal{C}_{ij}(I, \hat{n})$.
This covariance is defined at each pixel and needlet scale, meaning it contains both real-space and angular information, unlike in the standard algorithm. 
To compute the covariance matrices, we first smooth each map $T_i^I(\hat{n})$ with a Gaussian kernel with variance~\cite{McCarthy:2023hpa} 
\begin{equation}
    \label{eq:gaussian_real_space_var}
    \sigma^2_{{\rm real}, I} = 2 \left( \frac{N_f-1}{0.01 N^I_{\rm modes}}\right)\, ,
\end{equation}
where $N^{I}_{\rm modes}$ is the number of modes at the needlet scale $I$, weighted by the appropriate sky fraction if a mask has been used. 
The factor of $0.01$ is chosen to reduce the ``ILC bias'' to 1\% of the total signal variance, which comes from a spurious correlation between the signal and noise when we perform the algorithm on a map of finite resolution. 
We define the smoothed maps as $\mathcal{F}_I\left(T_i^I(\hat{n})\right)$. Then, the covariance matrix between maps $T_i^I(\hat{n})$ and $T_j^I(\hat{n})$ is~\cite{McCarthy:2023hpa}

\begin{alignat}{3}
    & \mathcal{C}_{ij}^I(\hat{n}) &&= &&\mathcal{F}_I\left\{\left[T_i^I(\hat{n}) - \mathcal{F}_I\left(T_i^I(\hat{n})\right)\right] \right. \nonumber \\
    & && &&\times \left. \left[T_j^I(\hat{n}) - \mathcal{F}_I\left(T_j^I(\hat{n})\right)\right]\right\}\,.
\end{alignat}

In words, the covariance is found by first smoothing each $T^I_i(\hat{n})$, then taking the difference with the unsmoothed map to get $T^I_i(\hat{n})-\mathcal{F}(T_i^I(\hat{n}))$. 
Next, we compute the product between this and the same quantity at a different frequency $j$. 
Finally, this product is smoothed again. 
In total, this gives $N_I$ covariance matrices that each have size $N_f\times N_f$ at every pixel in the map. 

Next, we calculate the ILC weights $w_i^I(\hat{n})$ according to Eq.~\eqref{eq:clean_coeff}. 
There is one for each pixel, scale, and frequency. 
Then, we form $N_I$ maps by taking the linear combination: 
\begin{equation}
    T_{\rm ILC}^I(\hat{n}) = \sum_i w_i^I(\hat{n})T_i^I(\hat{n})\, .
\end{equation}

From here, we refilter $T_{\rm ILC}^I(\hat{n})$ and add them together to get the final result, the post-ILC map, given by
\begin{equation}
    T_{\rm ILC}(\hat{n})=\sum_I \left[\sum_{\ell m} h_\ell^I T_{{\rm ILC}, \,\ell m}^IY_{\ell m}(\hat{n})\right] \, . 
\end{equation}
This is done by first taking the spherical harmonic transform of each $T_{\rm ILC}^I(\hat{n})$ to get the spherical harmonic coefficients $T^I_{{\rm ILC}, \ell m}$. 
Then, these are multiplied together with the needlet filters $h_\ell^I$ and the inverse spherical harmonic transform is taken to get $N_I$ pixel space maps. 
Finally, these maps are added together to get $T_{\rm ILC}(\hat{n})$.

\section{Radio Bias}
\label{sec:radio_bias}

As we mentioned in the main body, a realistic experiment is likely to observe a positive cross-correlation between the post-ILC map and a galaxy survey such as \textit{unWISE}. 
This results from the radio emission of these galaxies, which cannot be entirely removed by the ILC procedure.
If this bias is not removed, SKA would report artificially strong limits on dark photons.
In Ref.~\cite{McCarthy:2024ozh}, the authors described a data-driven procedure to remove this bias, which they applied successfully to CMB data.
We emphasize that our results in this work correspond to a successful removal of the radio bias at SKA using (for example) the procedure from Ref.~\cite{McCarthy:2024ozh}.
Therefore, our results are not artificially optimistic due to this effect.
Regardless, in this section, we first perform a rudimentary estimate of the magnitude of this bias at SKA in order to understand the size of the potential challenge. 

A proper estimate of this bias would require a model that draws mock samples of extragalactic radio sources that are correlated with the mock galaxy surveys. 
Additionally, we would need to carefully understand the effects of point source masks at SKA, which would remove the brightest radio sources.
Some of these sources are likely also found in optical galaxy survey catalogs like \textit{unWISE}, but the precise nature of the relationship between bright radio sources and \textit{unWISE} galaxies is difficult to evaluate without SKA data.

On account of these challenges, we construct a simplified extragalactic point source map that has identical statistics to the \textit{unWISE} catalog. 
More specifically, we begin with the $\delta_g(\hat{n})$ \textit{unWISE} map. 
Then, we construct an extragalactic point source map $T_{\rm ps,0}(\hat{n})$ at a reference frequency $\nu_0$ by
\begin{equation}
    T_{\rm ps,0}(\hat{n}) = \overline{T}_{\rm ps,0}(\delta_g + 1) \, 
\end{equation}
where $\overline{T}_{\rm ps,0}$ is the mean point source brightness temperature from our fiducial point source model at $\nu_0$.
Then, we rescale this map to other frequencies using a power law at each pixel
\begin{equation}
    T_{\rm ps}(\nu,\hat{n}) = T_{\rm ps, 0}\left(\frac{\nu}{\nu_0}\right)^{-\beta(\hat{n})}\!\!\!\!\!\!\!\!\!\!\! ,
\end{equation}
where $\beta(\hat{n})$ is a spectral index at each pixel drawn from a normal distribution with mean $\overline{\beta} = 2.681$ and standard deviation $\sigma_\beta = 0.5$ \cite{Mittal:2024mzv}.
By drawing the spectral index in each pixel from a distribution, we account for some degree of scatter in the radio sources. 

Next, we proceed with our usual analysis procedure by adding these mock extragalactic point source maps to the galactic foreground maps. 
We then perform the ILC algorithm on these mock foreground maps and cross-correlate the resulting post-ILC map with the mock \textit{unWISE} galaxy catalog. 
The resulting power spectrum is shown in Fig.~\ref{fig:radio_bias_Cls} (left panel, red) and compared to the absolute value of the predicted signal for $m_{A'}=\qty{5.6e-13}{\eV}$ and $\epsilon=5\times 10^{-8}$ at $\omega_0/2\pi = \qty{410}{\mega\hertz}$ (left panel, black). 
We emphasize that the true signal is negative, but we show its absolute value here for easier comparison. 

\begin{figure}[h ]
    \centering
    \includegraphics[width=\textwidth]{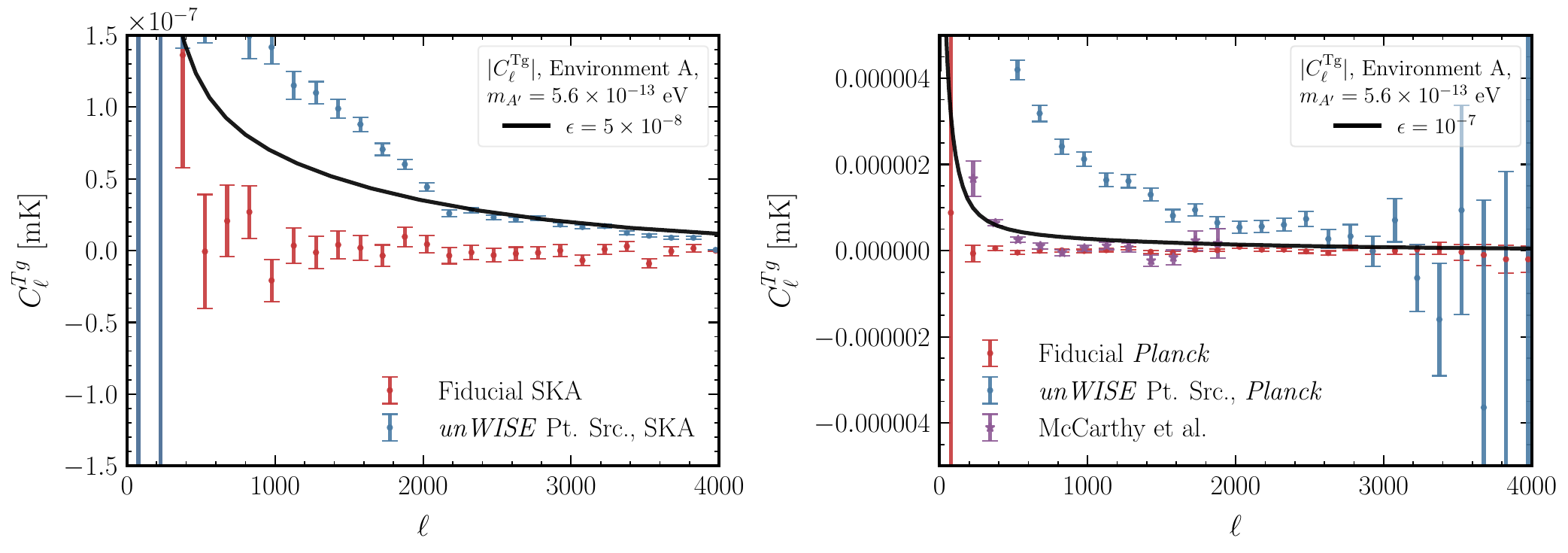}
    \caption{Cross-power spectrum of the post-ILC map constructed with the mock low-redshift galaxy survey is shown. \textit{Left:} $C_\ell^{\rm Tg, obs}$ between the post-ILC map constructed from the fiducial SKA mock foregrounds is shown in red and the power spectrum constructed from the alternative point source model at SKA frequencies described in Appendix~\ref{sec:radio_bias} is shown in blue. The absolute value of the predicted signal for a dark photon with $m_{A'}=\qty{5.6e-13}{\eV}$ and $\epsilon=5\times 10^{-8}$ is also plotted with $\omega_0/2\pi = \qty{410}{\mega\hertz}$. \textit{Right:} Cross-power spectrum between the post-ILC map constructed from \textit{Planck} maps and the mock low-redshift galaxy survey is shown in red. $C_\ell^{\rm Tg, obs}$ between the post-ILC \textit{Planck} map and the alternative point source model is plotted in blue. These are compared to the absolute value of the predicted signal for a dark photon with $m_{A'}=\qty{5.6e-13}{\eV}$ and $\epsilon= 10^{-7}$ with $\omega_0/2\pi = \qty{410}{\mega\hertz}$. Additionally, for comparison $C_\ell^{\rm Tg, obs}$ from Ref.~\cite{McCarthy:2024ozh} is plotted in purple. This is constructed from \textit{Planck} maps and the actual \textit{unWISE} catalog.~\nblink{Radio_Bias}}
    \label{fig:radio_bias_Cls}
\end{figure}

To validate this procedure, we also create mock point source maps at \textit{Planck} frequencies. Then, we analyze these maps in the same manner as described above and compute $C_\ell^{\rm Tg}$. This power spectrum is also shown in the right panel of Fig.~\ref{fig:radio_bias_Cls} (red). We see that this method overestimates the radio bias by a factor of $\lesssim 10$ compared to the results of Ref.~\cite{McCarthy:2024ozh} (purple), which indicates that our method is likely a conservative estimate for the radio bias.
This is somewhat expected; while the radio source population and the \textit{unWISE} galaxies are correlated, it is unreasonable to expect that they would have identical statistics.

\end{widetext}

\bibliographystyle{apsrev4-1}
\bibliography{sources}

\end{document}